\documentclass[a4paper,11pt]{article}
\pdfoutput=1 % if your are submitting a pdflatex (i.e. if you have
\usepackage{jcappub} % for details on the use of the package, please
\usepackage[T1]{fontenc} % if needed

\usepackage{url}                     
\usepackage{graphicx}
\usepackage{hyperref}
\usepackage{aas_macros}
\usepackage{booktabs}
\usepackage{multirow}
\usepackage{subcaption}
\usepackage{array}
\usepackage{comment}
\usepackage{parskip}
\usepackage{float}
\usepackage[dvipsnames]{xcolor}
\usepackage{soul}
\usepackage{amsmath}
\usepackage{amssymb}
\usepackage{siunitx}
\usepackage{physics}
\usepackage{gensymb}
\usepackage{adjustbox}
\usepackage{bm}

\newcommand{\HI}{\textsc{Hi}}

\newcommand{\ClPCA}{C_{\ell}^{\rm PCA}}
\newcommand{\ClCosmo}{C_{\ell}^{\rm cosmo}}
\newcommand{\ClRes}{C_{\ell}^{\rm Res}}

\title{\textbf{Needlets and foreground removal for SKAO hydrogen intensity maps}}

\author[a,1]{Bianca De Caro,\note{Corresponding author.}}
\author[b,c]{Isabella P.\ Carucci,}
\author[d, e, f]{Stefano Camera,}
\author[h]{Mathieu Remazeilles,}
\author[a]{Carmelita Carbone}

\affiliation[a]{INAF-IASF Milano, Via Alfonso Corti 12, 20133 Milano, Italy}
\affiliation[b]{INAF - Osservatorio Astronomico di Trieste, Via G.B. Tiepolo 11, 34131 Trieste, Italy}
\affiliation[c]{IFPU - Institute for Fundamental Physics of the Universe, Via Beirut 2, 34151 Trieste, Italy}
\affiliation[d]{Dipartimento di Fisica, Università degli Studi di Torino, Via P.
Giuria 1, 10125 Torino, Italy}
\affiliation[e]{INFN-Sezione di Torino, Via P. Giuria 1, 10125 Torino, Italy}
\affiliation[f]{INAF-Osservatorio Astrofisico di Torino, Via Osservatorio 20,
10025 Pino Torinese (TO), Italy}
\affiliation[h]{Instituto de Fisica de Cantabria (CSIC-UC), Avenida de los Castros s/n, 39005 Santander, Spain}

\emailAdd{bianca.decaro@inaf.it}

\abstract{Intensity Mapping (IM) of the 21-cm line of the neutral hydrogen (\HI) has become a compelling new technique to map the large-scale structure of the Universe. One of the main challenges is the presence of strong foreground emissions of several orders of magnitude larger than the \HI~signal. Here, we implement a version of the Principal Component Analysis, a blind component-separation technique, based on a kind of spherical wavelets called needlets. These functions exploit double localization both in real and in harmonic space. We test Need-PCA performances on a set of maps that simulates the SKA MID radio telescope in the AA4 configuration. We compare our results with other component separation methods such as Generalised Morphological Component Analysis (GMCA) and Generalized Needlet Internal Linear Combination (GNILC). All the methods have comparable results, recovering the \HI~signal within 10\% accuracy across the frequency channels, in the multipole range 30 $\lesssim \ell \lesssim$ 136. We also test our pipeline in the presence of systematics such as polarization leakage. We find that the cleaning methods are insensitive to the presence of such systematic, yielding the same results as in the leakage-free case. Finally, under the assumption of a realistic telescope beam with sidelobes, we find that standard PCA and GMCA fails to recover the \HI~signal at larger scales, while the Need-PCA and Need-GMCA are less affected. GNILC tends to over-clean, yielding to a loss of the signal.}


\begin{document}
\maketitle
\flushbottom

\section{Introduction}
\label{sec:faq}
Detecting the 21-cm line emission from atomic neutral hydrogen (\HI) has emerged as a powerful observable in cosmology. 
\HI~permeates the Universe from the Dark Ages following the recombination of the first atoms, through the epoch of reionization when the first stars formed, and into later times, where hydrogen is found in the gas clouds within galaxies \cite{Pritchard-2008}. At those late epochs, \HI~is a valuable tracer of the galaxy distribution.
However, the 21-cm - or, the frequency $\nu_{21}=$ 1420 MHz - line, emitted by the hyperfine transition of \HI, is particularly weak: 
detecting a sufficiently large number of galaxies for cosmology is time- and resource-consuming, and becomes prohibitive at redshift larger than $z\gtrsim0.5$ \cite{Fernandez-2016}. In the last two decades, Intensity Mapping (IM) has been developed as a new technique to efficiently map large portions of the sky at a relatively low cost. %using emission lines. 
Instead of resolving individual galaxies, IM measures the sky’s intensity by integrating the 21-cm line emission over a large area \cite{Bharadwaj2001,Chang2008,Loeb2008}. This approach yields to a modest angular resolution in favour of high redshift resolution, made possible by the redshifting of the line.
Similarly to the cosmic microwave background (CMB), the fluctuations in the observed temperature intensity of \HI~emission act as a (biased) tracer of the underlying matter density fluctuations. Thus, IM of the \HI~emission allows us to reconstruct matter density and constrain the matter power spectrum, which provides statistical information about the field on large scales and at different redshifts. The future of \HI~cosmology is particularly promising, thanks to a growing number of large-scale 21-cm surveys that are are being conducted \citep{chang2010,wolz2017,Anderson2018,Cunnington-2023,Paul2023,Chime2023,uGMRT2024,Carucci-2025} or being proposed  \citep{tianlai,hirax,bingo,fast}. Among the latter, we stress how the SKA Observatory (SKAO)\footnote{\url{https://www.skao.int/}.} will become a game-changer in the field, in terms of sensitivity, speed, and access to higher redshift ranges \cite{Santos-2015}.\\
This work focuses on \textit{single-dish} intensity mapping, which relies on the auto-correlation data of individual telescope dishes \citep{Battye-2013, Bull-2015}, as opposed to standard radio interferometry that would require extremely short baselines to probe large-scale modes. To date, the \HI~signal has only been detected via cross-correlation with galaxy surveys \citep{Masui-2013,Wolz-2022,Cunnington-2023, Carucci-2025,Meerklass-2025}, which helps mitigate systematics.\\
IM observations are affected by strong astrophysical foregrounds, which are $3$--$5$ orders of magnitude brighter than the \HI~signal. Systematics and calibration errors further complicate foreground removal \citep{Shaw2015,Carucci-2020,Wang2022}. The main sources of foreground in the radio range are the \textit{Galactic synchrotron} (cosmic-ray electrons accelerated by the magnetic field in the galactic medium), \textit{free-free emission} (sourced by free electrons scattering off ions in our Galaxy), and extragalactic \textit{point sources} (e.g. active galactic nuclei and radio galaxies in general). While these foregrounds are stronger in amplitude, their spectra are typically smooth in frequency, unlike the \HI~signal, which has near-Gaussian oscillations \cite{Cunnington-2021}. Thus, in principle, it is possible to separate the foreground contributions from the cosmological signal using removal methods. In practice, instrumental effects complicate this separation. Of particular concern is the leakage of polarized emission into the intensity signal due to instrumental miscalibration \citep{Shaw2015,Carucci-2020}. If the instrument is partially miscalibrated, a fraction of the polarized synchrotron emission can leak into the observation and dominate the \HI~signal. Moreover, polarization leakage has a non-smooth frequency dependence, making it hard to distinguish from the cosmological signal.\\  
Several foreground cleaning methods have been proposed, differing primarily in whether they assume a model for the foregrounds. 
However, there are no attempts in the literature to apply simulation or model-based methods to actual observations, since the field is at an early stage and foreground and instrumental effects models are missing or too uncertain to retrieve the signal confidently. Instead, current observational analyses have opted for the  so-called \textit{blind} methods, which do not make any strong assumption on the foregrounds and work on statistical grounds \citep{Masui-2013,Switzer-2013,wolz2017,Cunnington-2023,Carucci-2025,Meerklass-2025}.
Common \textit{blind} methods include the the Principal Component Analysis \cite{Alonso-2014}, Generalized Morphological Component Analysis \cite{Carucci-2020}, Generalized Internal Linear Combination \citep{Remazeilles-2011, Olivari-2016}, Independent Component Analysis \cite{Cunningtion-2019}. For example, PCA has been applied in the analysis of MeerKAT intensity maps cross-correlated with optical galaxies from the WiggleZ Dark Energy Survey \citep{Cunnington-2023, Carucci-2025}.\\
In this study, we optimize a foreground cleaning pipeline to recover the \HI~signal for SKAO-MID in the few years away AA4 configuration \citep{seethapuram_sridhar_2025_16951020}. We use simulated maps of both the cosmological signal and foregrounds, including instrumental effects such as the thermal noise and the telescope beam smoothing. Our method is based on a version of PCA implemented with a kind of spherical wavelets called needlets \citep{Marinucci-2008, Baldi-2006, Pietrobon-2006, Marinucci-2011}. Needlets enjoy some features which are shared by other wavelets, such as double localization in pixel and harmonic space, simple reconstruction formula after decomposition, uncorrelation properties which make them useful when dealing with incomplete data.\\
Previous works have already exploited the needlets for recovering the 21-cm signal from 2D maps. Foreground cleaning methods with needlets such as GNILC has been applied to BINGO-like simulations (\citep{Marins-2022, deMericia-2023}), in comparison with other methods such as GMCA and FastICA (Fast Independent Component Analysis).
In \cite{Dai-2025}, a modified version of GNILC has been tested also in SKAO-MID like simulations, testing the effect of the telescope beam on the foreground cleaning methods.
As a further step, in this work we assess the performance of different foreground cleaning methods in the presence of polarization leakage. We analyse its impact on signal recovery for SKAO MID-like simulations and evaluate the robustness of the pipeline under such systematics.\\
The effect of the polarization leakage in the foreground cleaning performance has been studied also in \cite{Podczerwinski-2024}. They applied a needlet-based foreground cleaning technique called Needlet Karhunen-Loève (NKL) to a set of simulated maps at high redshift 1.84 < z < 2.55). In this work, we focus to the redshift range relevant for a SKAO-MID-like survey.\\
The paper is structured as follows. In Sec.~\ref{sec:comp_sep_method} we describe the formalism of the different foreground cleaning methods and the assumptions assumed by PCA, GMCA and GNILC. In Sec.~\ref{sec:simulations} we describe the simulations and the full data-cube we use in our analysis. In Sec.~\ref{sec:needlets} we summarize the properties of the needlets. In Sec.~\ref{sec:pipeline} we describe the application of the different cleaning methods to the simulated data and assess their performances. Sec.~\ref{sec:results} shows the obtained results and the discussion. We conclude in Section \ref{sec:conclusions}.
%%%%%%%%%%%%%%%%%%%%%%%%%%%%%%%%%%%%%%%%%%%%%%%%%%
\section{Foreground cleaning methods}
\label{sec:comp_sep_method}
In this section we review some of the most used and studied methods for the 21-cm foreground cleaning. All methods rely on the assumptions that the foreground emissions are slowly varying with frequency, i.e. they are highly correlated, and are order of magnitude larger than the \HI~signal, which, in contrast, is almost uncorrelated in frequency. Thus, we can identify the dominant foreground contributions as a set of smooth functions of frequency to be subtracted from the data in order to isolate the \HI~signal. We start by defining the sky brightness temperature along a given direction in the sky $\hat{\bm{n}}$ and at frequency $\nu$ as \cite{Alonso-2015}
\begin{equation}
\label{eq:brightness_temp}
    T(\nu, \hat{\bm{n}}) = \sum^{N_{\rm fg}}_{k=1}f_{k}(\nu)S_{k}(\hat{\bm{n}})+T_{\rm cosmo}(\nu, \hat{\bm{n}})+T_{\rm noise}(\nu, \hat{\bm{n}})\,,
\end{equation}
where $N_{\rm fg}$ is the number of foreground sources to subtract, $f_{k}(\nu)$ are the smooth functions of frequency, $S_{k}(\hat{\bm{n}})$ are the foreground sky maps, $T_{\rm cosmo}$ is the cosmological signal and $T_{\rm noise}$ is the instrumental noise contribution. Thus, Eq.~\eqref{eq:brightness_temp} can be written as a set of linear equation computed at $N_{\nu}$ frequencies:
\begin{equation}
\label{eq:linear_problem_fg_cleaning}
    \mathsf{X}= \mathsf{A} \mathsf{S} + \mathsf{r}\,,
\end{equation}
where $X_{i}=T(\nu_{i}, \hat{\bm{n}})$, $A_{ik}=f_{k}(\nu_{i})$, and $r_{i}$ is the sum of the contribution from the cosmological signal and the instrumental noise $r_{i}=T_{\rm cosmo}(\nu_{i}, \hat{\bm{n}})+T_{\rm noise}(\nu_{i}, \hat{\bm{n}})$. Being largely uncorrelated across frequencies, the \HI~signal and the instrumental noise are coupled and recovered together.\\
The scope of foreground cleaning is to determine $\mathsf{A}$, which is called the mixing matrix, and $\mathsf{S}$ in order to accurately recover the target signal as $\mathsf{r}=\mathsf{X}-\mathsf{A}\mathsf{S}$.\\
Foreground removal techniques can be categorized as either \textit{blind}, when they rely only on general foreground characteristics such as spectral smoothness, or \textit{non-blind}, when they require a more detailed and specific model of the foregrounds or the cosmological signal.
\paragraph{Principle Component analysis (PCA)}
The Principal Component Analysis (PCA) is one of the most used \textit{blind} foreground removal techniques. In fact, PCA exploits the main properties of the foregrounds, i.e. the smoothness in frequency and the large amplitude, to compute both the source components $\mathsf{S}$ and the mixing matrix $\mathsf{A}$. \\
It can be demonstrated that the covariance between the frequency channels of a highly correlated data-set has a peculiar eigensystem: most of the information is stored in a few number of very large eigenvalues, called principal components, while the other one are negligibly smaller. Thus, we can attempt to subtract the foreground system from the data-set by looking at the largest N$_{\rm fg}$ eigenvalues and eliminating the corresponding eigenvectors of the covariance matrix. The PCA algorithm can be summarized as follow:
\begin{itemize}
    \item Compute the centred covariance matrix for different frequency channels by averaging over all the available line of sight, corresponding to the number of pixels N$_{\rm pix}$ of the observed temperature map $\mathsf{X}$:
    \begin{equation}
    {\mathsf{ Cov}_{\rm ch, ch'} = \frac{1}{\rm N_{pix}} \sum_{i=1}^{\rm N_{\rm pix}} \rm X^{ch'}_{i} X^{ch}_{i}};
    \nonumber
    \end{equation}
    \item Compute the eigenvalues and eigenvectors of ${\rm Cov}_{\rm ch, \, ch'}$ and diagonalize the matrix: 
    \begin{equation}
    \hat{\mathsf{U}}^{T}\mathsf{Cov}\hat{\mathsf{U}}=\hat{\mathsf{\Lambda}} \equiv {\rm diag}(\lambda_{1},\, ... \,,\,\lambda_{N_{\nu}})  
        \nonumber
    \end{equation}
    \item We identify the N$_{\rm fg}$ eigenvalues corresponding to the foregrounds which are larger than the rest. We build the matrix $\hat{\mathsf{U}}_{\rm fg}$ which columns are the eigenvectors corresponding to the N$_{\rm fg}$ largest eigenvalues, i.e. the principal components. The foreground components $\mathsf{S}$ are found by projecting the observations $\mathsf{X}$ on $\hat{\mathsf{U}}_{\rm fg}$, yielding
    \begin{equation}
    \mathsf{S} = \hat{\mathsf{U}}_{\rm fg}^{T} \mathsf{X}
        \nonumber
    \end{equation}
    and, combined with Eq.~\eqref{eq:linear_problem_fg_cleaning}, we estimate the foreground from the data as
        \begin{equation}
    \mathsf{X}_{\rm fg} = \hat{\mathsf{U}}_{\rm fg}\hat{\mathsf{U}}_{\rm fg}^{T} \mathsf{X}\,.
        \nonumber
    \end{equation}
    The cosmological signal is then recovered as:
    \begin{equation}
    \mathsf{X}_{\HI} = \mathsf{X}- \mathsf{X}_{\rm fg}\,.
        \nonumber
    \end{equation}
\end{itemize}
The choice of N$_{\rm fg}$ is the key to optimize the estimation of the foreground from the data and thus remove it. As discussed above, due to their spectral properties the foreground emissions can be represented by few number of eigenvalues. On the contrary, the \HI~cannot be compressed to a few number of components, thus a strong cleaning process, i.e. a large N$_{\rm fg}$, can remove part of the cosmological signal together with the foreground, typically large-scale line of sight modes \cite{Cunnington-2021}. Of course, if N$_{\rm fg}$ is too small, fraction of the foreground emission will not be removed, dominating the \HI~signal. Thus, a fine balance on the choice of the number of components to remove is critical.\\
In a simulation-based approach, working with pure-\HI~and pure-foreground maps, we can calculate the contribution of these components left on the residuals after the cleaning process.
This leakage in the residuals are computed by projecting the mixing matrix onto the pure-\HI~and pure-foreground as
\begin{align}
    \label{eq:HI_fg_leak}
    \mathsf{X}_{\rm leak\ fg} &= \mathsf{X}_{\rm fg} - \mathsf{U}_{\rm fg}\mathsf{U}_{\rm fg}^T \mathsf{X}_{\rm fg} \\
    \mathsf{X}_{\rm leak \ \HI} &=\rm  \mathsf{U}_{\rm fg}\mathsf{U}_{\rm fg}^T \mathsf{X}_{\HI}\,.
\end{align}
The first equation defines the foreground that leaks into the cosmological signal plus noise, while the second one defines the cosmological plus noise signal that leaks into the estimated foreground. The cleaning process is successful when the power spectra of the foreground and \HI~ plus noise leakage are negligible compared to that of the cosmological signal \cite{Carucci-2020}.
\paragraph{Generalized Morphological Component Analysis (GMCA)}
GMCA is a blind source separation algorithm that assumes that the target components can be represented sparsely in a suitable signal domain, such as the Fourier or wavelet domains \cite{Bobin-2007}. A signal is considered sparse if most of its coefficients are zero.
For instance, periodic signals are typically sparse in the Fourier domain, as they can be described using only a few coefficients. This sparsity assumption is crucial, as it significantly enhances the contrast between different components, facilitating their separation.\\
GMCA imposes sparsity by solving the following optimization problem \cite{Carucci-2020}:
\begin{equation}
    \label{eq:GMCA_optim_problem}
    \{ \tilde{\mathsf{A}}, \tilde{\mathsf{S}} \} = \min_{\mathsf{A},\mathsf{S}} \sum^{N_{\rm fg}}_{i=1} \lambda_{i} ||\mathsf{S}_{i}||_{1} + ||\mathsf{X}-\mathsf{AS}||^{2}_{F}\,,
\end{equation}
where the first term is a sparsity constraint term and the second is a data-fidelity term. Indeed, $||\cdot||_{1}$ is the $\ell_{1}$ norm defined by $||\mathsf{Y}||_{1}=\sum_{i,j}|\mathsf{Y}_{i,j}|$. And $||\cdot||_{F}$ is the Frobenius norm defined by $||\mathsf{Y}||^{2}_{F}=\mathrm{Trace}(\mathsf{YY^{T}})$. Specifically, the $\lambda_{i}$ terms act as regularization coefficients — sparsity-thresholds — that are crucial for ensuring robustness against noise in the problem, which is, in our problem, the difference in intensity between the foregrounds and the cosmological signal. GMCA initially estimates these coefficients using the median absolute deviation (MAD) method and gradually reduce them to a final level determined by the noise characteristics.\\
As PCA, GMCA is a \textit{blind} method, as the only input needed is the number of sources N$_{\rm fg}$, while both the mixing matrix $\mathsf{A}$ and the sources $\mathsf{S}$ are not modelled. In the \HI~IM context, the wavelet domain has been demonstrated to be optimal for a sparse description of the astrophysical foregrounds \cite{Carucci-2020}, hence, we perform the analysis also with a wavelet-decomposition of the maps before performing the minimization in Eq.~\eqref{eq:GMCA_optim_problem}. Originally, the wavelet dictionary used is that of the isotropic undecimated wavelet, also known as starlets due to their success in efficiently describe astrophysical images \citep{Starck2007}. Here, we will perform contaminant separations with GMCA either applied on spherical harmonics- (Standard-GMCA) and needlet-decomposed (Need-GMCA) maps. 
\paragraph{Generalized Needlet Internal Linear Combination (GNILC)}
The Internal Linear Combination (ILC) method was first used in the analysis of CMB data from the WMAP mission \cite{Bennett-2003}. It extracts one or more components from observations with known spectral properties by applying a vector of weights $\bm{W}$ that enforces a unit response to the desired signal while minimizing the total variance of the residual foregrounds.\\
The data can be modelled as \cite{Remazeilles-CMB-2011}
\begin{equation}
    \label{eq:ilc_system}
    \bm{x}(\hat{\bm{n}})=\bm{a} s(\hat{\bm{n}}) + \bm{n}(\hat{\bm{n}})\,,
\end{equation}
where $\bm{x}(\hat{\bm{n}})$ is a vector of observed maps for each frequency channels, $s(\hat{\bm{n}})$ is the map of the component of interest, $\bm{a}$ is a known mixing vector which scales \texttt{s} across the frequency channels. The term $\bm{n}(\hat{\bm{n}})$ includes all other components such as noise and foregrounds. 
The ILC estimates \texttt{s} as a linear combination $\hat{s}=\bm{W}^{T}\bm{x}(\hat{\bm{n}})$. A key strength of the ILC method is that it does not require any assumptions about the foreground models. However, it is limited to recovering components that have a fixed frequency scaling across the sky.\\
GNILC aims to reconstruct emission which can not be modelled as a single template scaling in a known way with frequency, e.g. the foreground emission. It is an extension of the ILC method into needlet space that compensates for the lack of information on the frequency dependence of the target signal by some \textit{prior} information on its power spectrum \cite{Remazeilles-2011}. The needlet decomposition allows to constrain the emission on localized areas of the sky and defined angular space.\\
GNILC has been successfully applied to CMB data analysis by the Planck collaboration (\cite{Planck-inter-2016}) and later extended to 21-cm data analysis by \cite{Olivari-2016} and the BINGO collaboration \cite{Fornazier-2022}. \\
The key ingredients for the GNILC algorithm are the theoretical \HI~21-cm power spectrum across the frequency bins and the number of foreground components, N$_{\rm fg}$. The value of N$_{\rm fg}$ is estimated via a constrained PCA using the covariance matrix of the \HI~signal, evaluated from the theoretical \textit{prior}. This estimation is performed independently at each needlet scale using the Akaike Information Criterion (AIC) \cite{Akaike-1974}. For each location of the sky and range of needlet scales, the number N$_{\rm fg}$ is found by minimizing the AIC,
\begin{equation}
    \label{eq:AIC}
    \rm AIC(N_{\rm fg}) = 2 n N_{\rm fg} - 2\log{\mathcal{L}_{max}(N_{\rm fg})}\,,
\end{equation}
where\textit{n} is the number of modes in the needlet domain considered and $\mathcal{L}_{\rm max}$ is the maximum likelihood solution of the data covariance matrix given a model of N$_{\rm fg}$ independent foreground components (see Appendix A of \cite{Olivari-2016} for more details). Once the number N$_{\rm fg}$ is known, one can compute the ILC weight matrix $\mathsf{W}$ at each needlet scale and reconstruct the \HI~signal from the observations.
%%%%%%%%%%%%%%%%%%%%%%%%%%%%%%%%%%%%%%%%%%%%%
\section{Gaussian and log-normal simulations}
\label{sec:simulations}
In this Section we describe the set of simulated data we considered in this work. The simulations are those introduced in \cite{Carucci-2020}.
We simulate a realistic sky with the following components: the cosmological 21-cm signal; the astrophysical foreground emission, both galactic and extragalactic; the polarization leakage, i.e., a fraction of the polarized sky that has leaked in the unpolarized signal due to the receivers' imperfections; and a Gaussian noise contribution due to the temperature of the system.
The simulation consists of a set of 400 full-sky \textsc{HEALPix} maps, one for each frequency channel with a thickness of 1 MHz, generated with $N_{\rm side}=256$, which corresponds to $N_{\rm pix}=12 N_{\rm side}^{2}$. The original set of simulated maps consists of 400 frequency maps of 1 MHz thickness in the range $\nu \in \left[ 900 - 1300 \right]$ MHz, corresponding to the redshift range $z \in \left[ 0.09 - 0.58 \right]$. We consider only a fraction of these range, as described Sec.~\ref{subsec:datacube}.
%%%%%%%%%%%%%%%%%%%%%%%%%%%%%%%%%%%%%
\subsection{Cosmological signal}
\label{subsec:cosmological_signal}
After reionization, \HI~is distributed inside galaxies, self-shielded from the ultra-violet background \cite{Villaescusa-Navarro-2014}. Thus, the \HI~distribution is a (biased) tracer of the underlying dark matter field distribution. The simulated \HI~maps are obtained from the lognormal approximation of the matter field, according to the prescription of \cite{Alonso-2014}, using the code described therein, CRIME. According to this model, the intensity in an observed frequency bin $\delta \nu$ coming from the 21-cm emission of n object at a redshift \textit{z} with neutral hydrogen mass $M_{\HI}$, subtending a solid angle $\delta \Omega$ is given by
\begin{equation}
    \label{eq:intensity_21cm}
    I(\nu, \hat{\bm{n}}) = \frac{3 h_{p} A_{12}}{16 \pi m_{H}} \frac{1}{((1+z)r(z))^{2}} \frac{M_{\HI}}{\delta_{\nu}\,\delta_{\Omega}} \nu_{21}\,,
\end{equation}
where $A_{12}=2.876 \times 10^{-15}$ Hz is the Einstein coefficient corresponding to the emission from the 21-cm hyperfine transition, $h_{p}$ is the Planck's constant and $m_{H}=1.6733 \times 10^{-24}$ g is the hydrogen atom mass. Here, $r(z)$ is the comoving curvature distance $r(z)=c \sinh{H_{0} \sqrt{|\Omega_{k}|}\chi(z)c}/(H_{0} \sqrt{|\Omega_{k}|})$ and $\chi(z)$ is the radial comoving distance $\chi(z)=\int_{0}^{z}\frac{cdz'}{H(z')}$. We can write the intensity $I(\nu, \hat{\bm{n}})$ in terms of a black-body temperature in the Rayleigh-Jeans approximation $T=Ic^{2} / (2k_{B}\nu^{2})$, where $k_{B}$ is the Boltzmann's constant. From this we can obtain the mean brightness temperature coming from redshift \textit{z} and its fluctuations in terms of the neutral hydrogen density:
\begin{equation}
    \label{eq:mean_brigh_Temp}
    T_{21}(z, \hat{\bm{n}}) = (190.55~\mathrm{mK})\frac{\Omega_{\HI} h (1+z)^{2}}{E(z)}(1+\delta_{\HI})\,,
\end{equation}
where $h=H_{0}/100$ km sec$^{-1}$ Mpc$^{-1}$, $E(z)=H(z)/H_{0}$, and $\Omega_{\HI}$ is the \HI~cosmic abundance. Assuming that \HI~density is linearly biased with respect to the matter density, we can write $\delta_{\HI}=b_{\HI}\delta_{M}$.
The \HI~bias and the \HI~cosmic abundance are redshift-dependent and are set to $b_{\HI}(z)=0.3(1+z)+0.6$ in agreement with observations at redshift $z \lesssim 0.8$ \cite{Switzer-2013}, and $\Omega_{\HI}(z)=4(1+z)^{0.6}10^{-4}$, following \cite{Crighton-2015} .\\
The cosmological model is fixed to $\Lambda$CDM with cosmological parameters ${\Omega_{m}, \Omega_{\Lambda}, \Omega_{b}, h}={0.3, 0.7, 0.049, 0.67}$, with an initial cube of side 3 Gpc/h divided in $2048^{3}$. Light-cone effects and redshift-space distortions are included by construction \cite{Carucci-2020}. \\
%%%%%%%%%%%%%%%%%%%%%%%%%%%%%%%%%
\subsection{Contaminants}
\label{subsec:contaminants}
\begin{figure}[ht]
\centering
\setlength{\tabcolsep}{0.01pt}
\begin{tabular}{cc}
\includegraphics[width=0.45\textwidth]{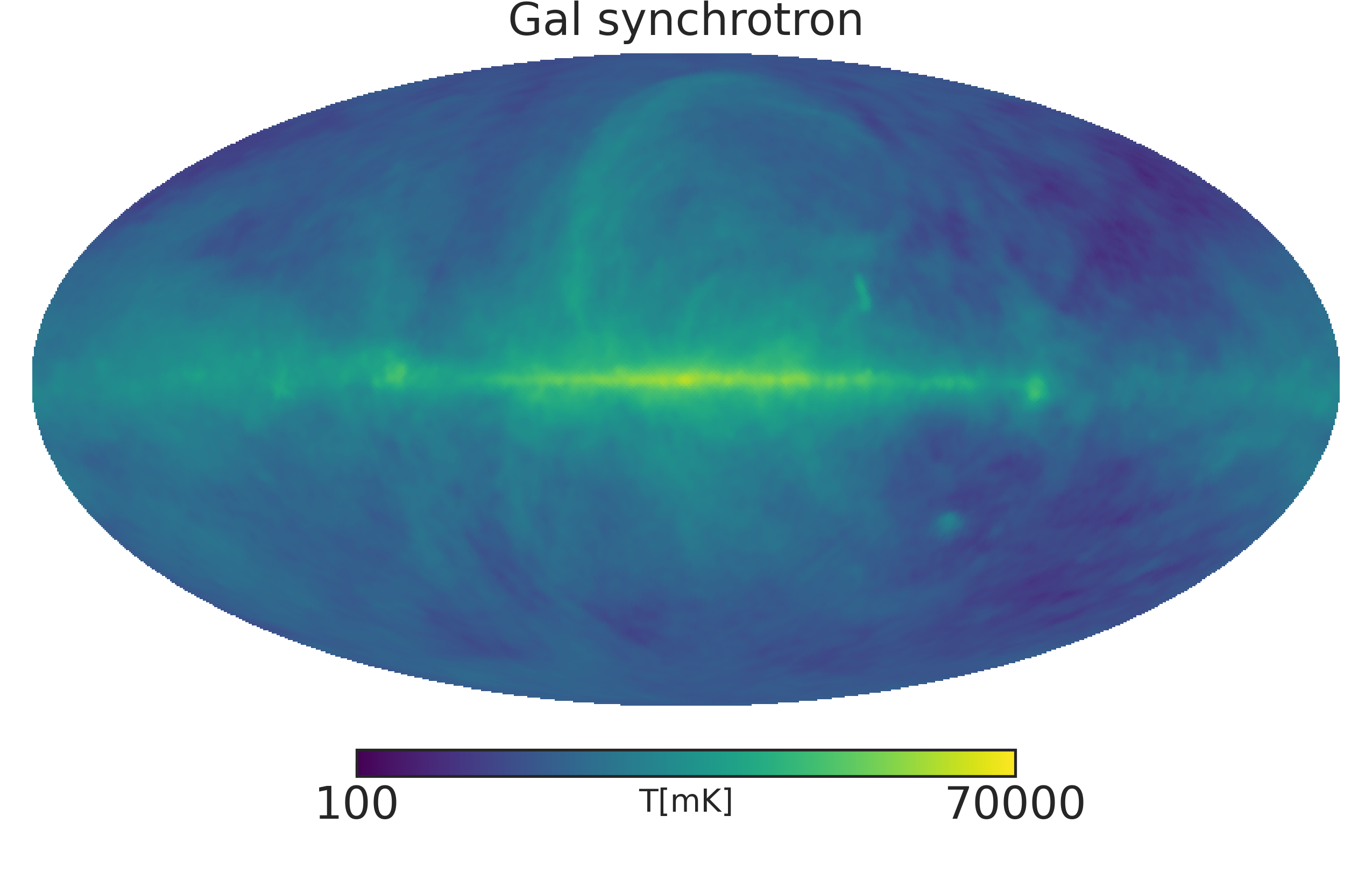}&
\includegraphics[width=0.45\textwidth]{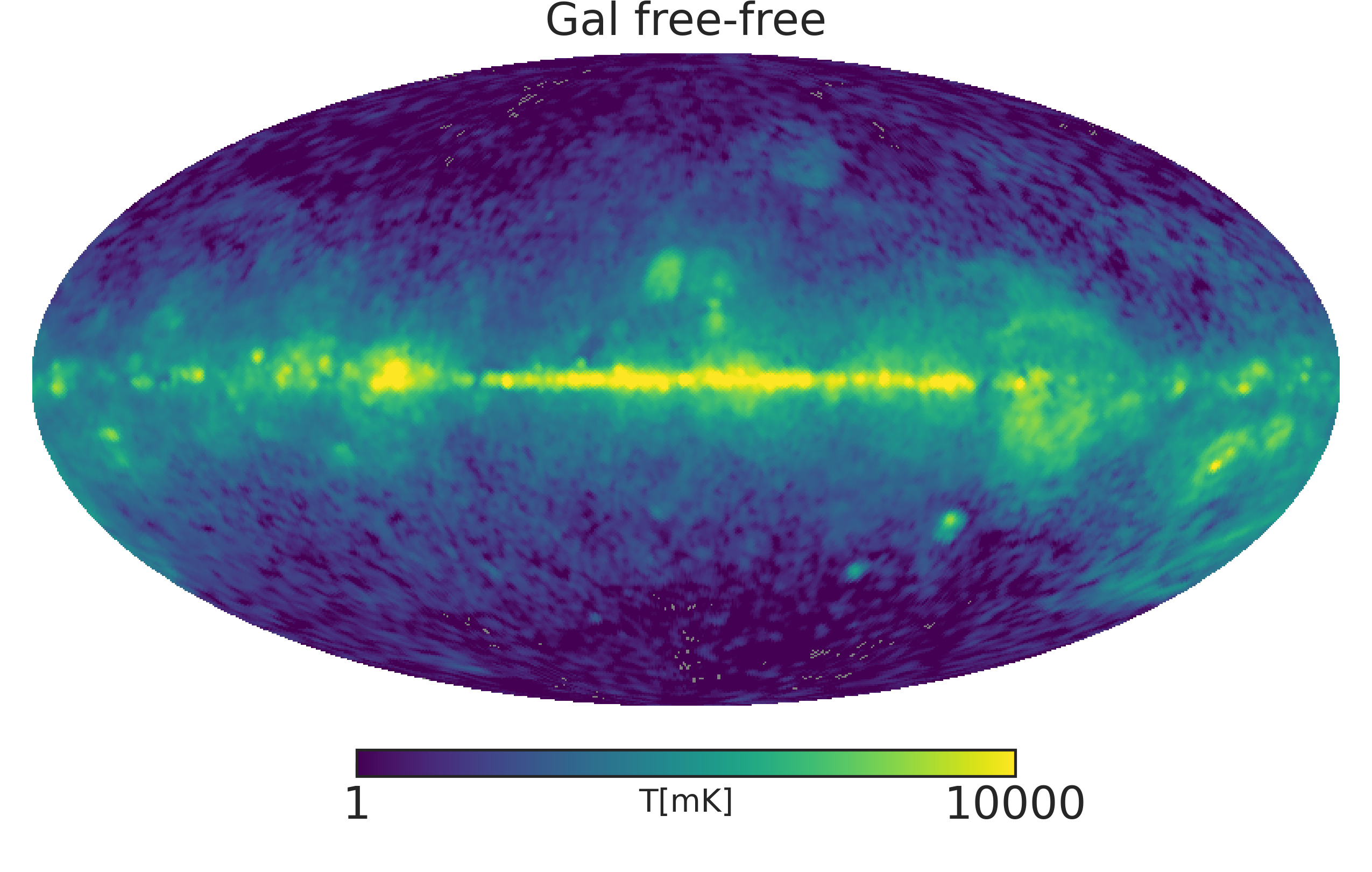}\\
\includegraphics[width=0.45\textwidth]{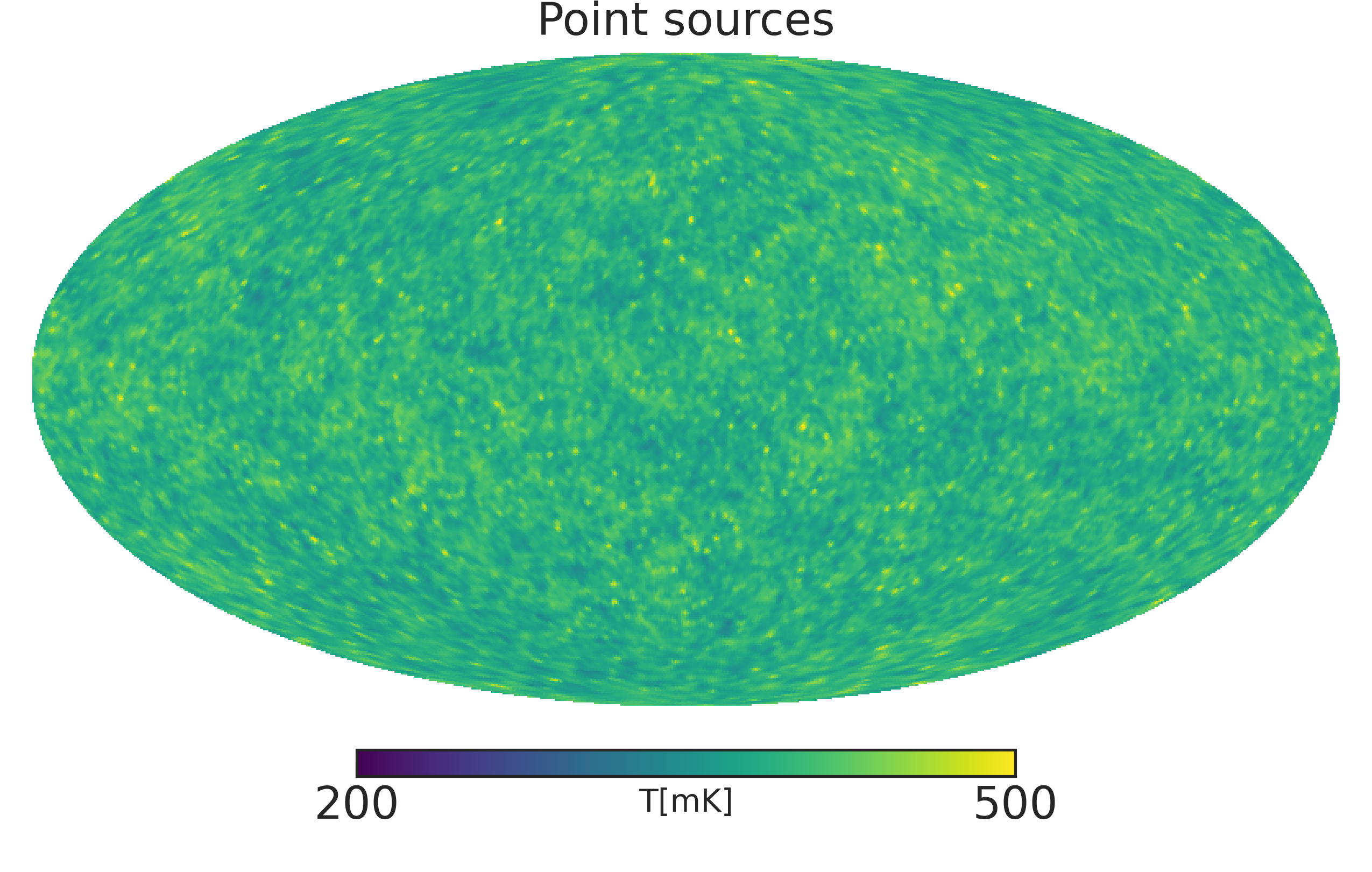}&
\includegraphics[width=0.45\textwidth]{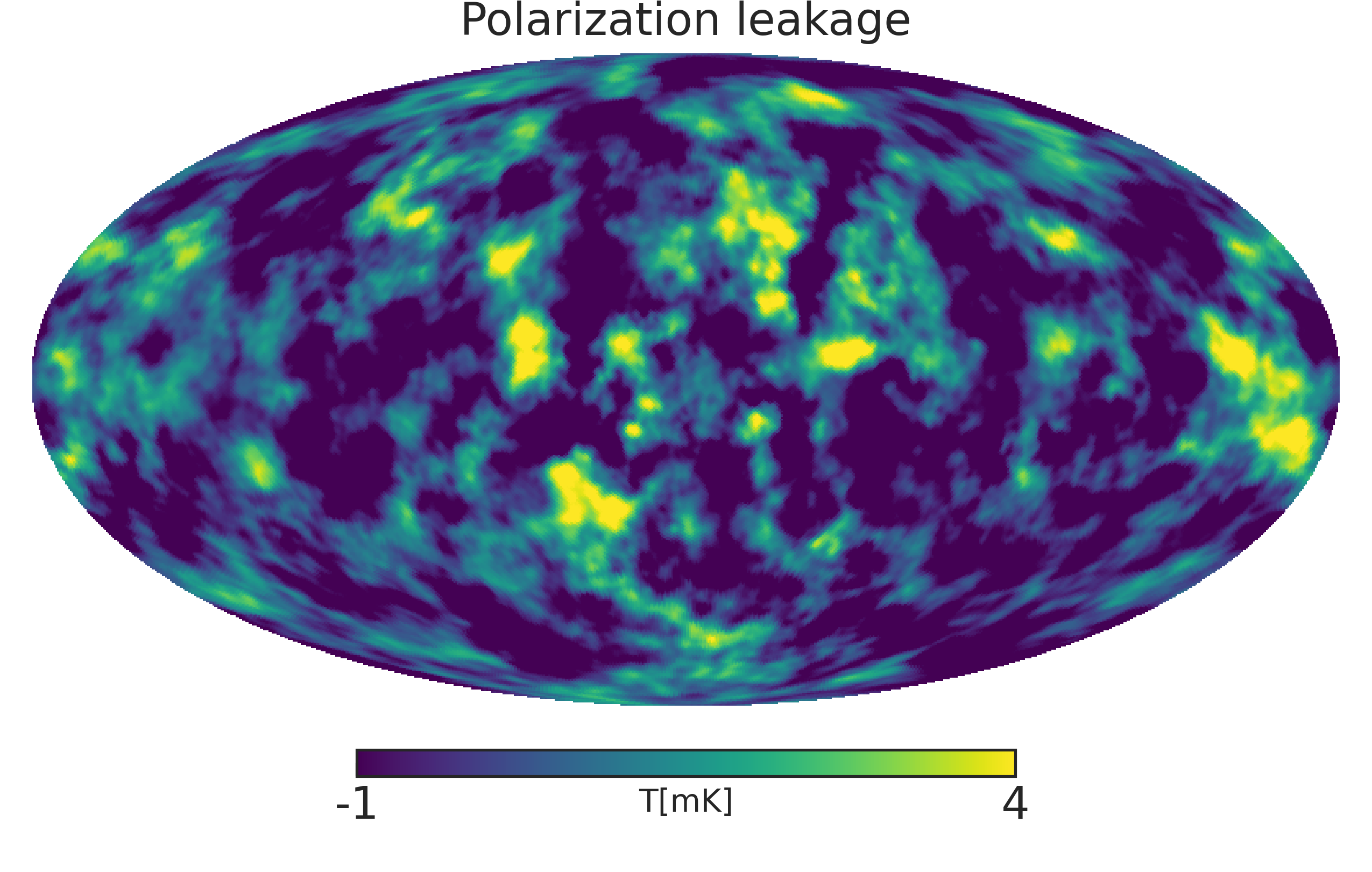}
\end{tabular}
    \caption{Mollweide projection of the intensity of the foreground and systematic contaminants, in the frequency channel 952.5 MHz. From top left and clockwise, the maps show: galactic synchrotron, galactic free-free, polarization leakage and extragalactic point sources. The units are in mK, the scale is logarithmic for the synchrotron and free-free maps and linear for the point sources and the polarization leakage. }
    \label{fig:fg_sims}
\end{figure}

\begin{figure}[ht]
    \centering
    \includegraphics[width=0.8\linewidth]{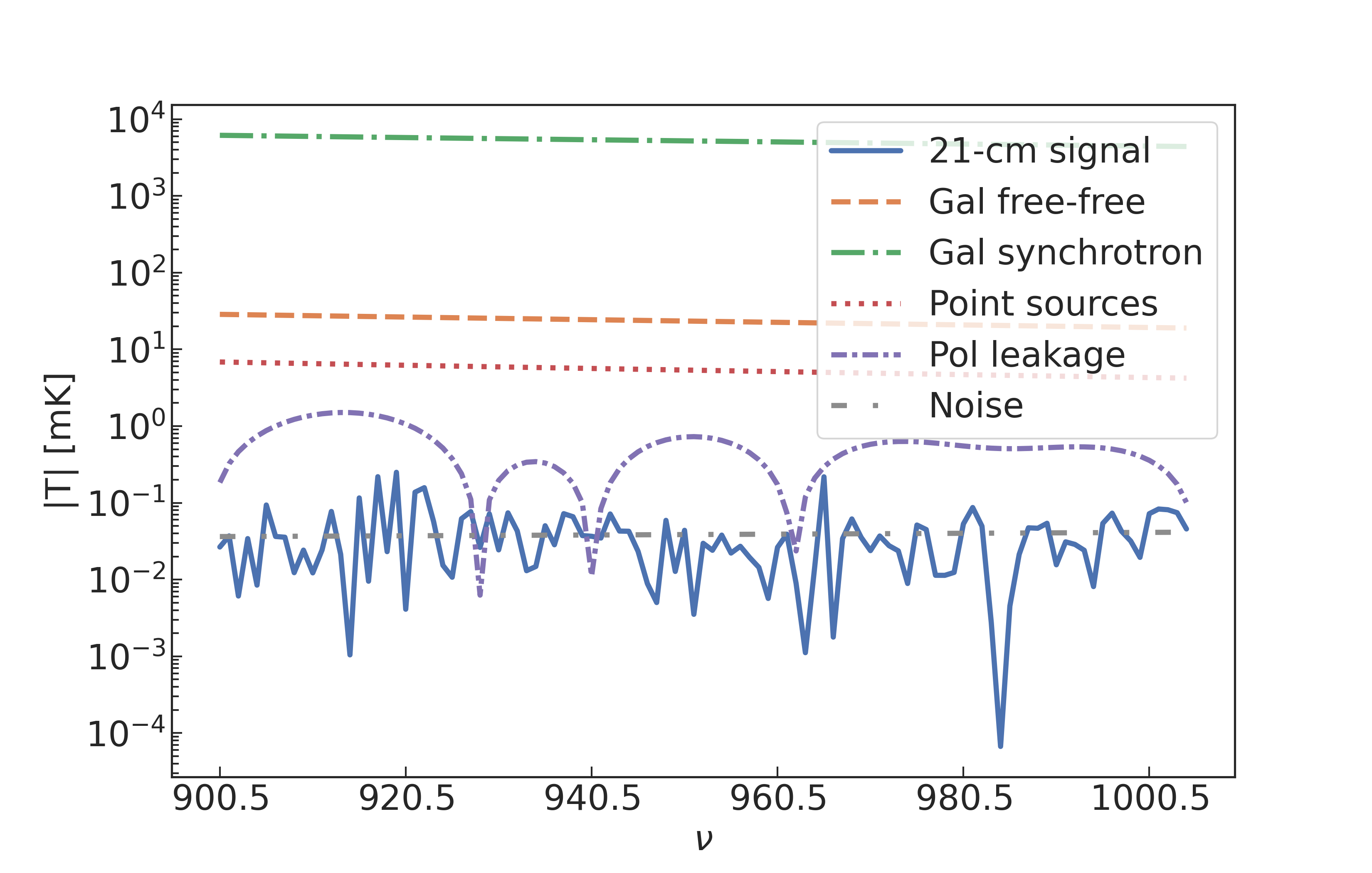}
    \caption{Brightness temperature as a function of the frequency along a line of sight, in the direction of lon=0 deg, lat=-5 deg.}
    \label{fig:brightness_temp}
\end{figure}

The sources that contaminate the 21-cm signal considered in this work are of two types: the astrophysical foregrounds and the polarization leakage. Fig.~\ref{fig:fg_sims} shows the all-sky contaminant components of our simulations.
Fig.~\ref{fig:brightness_temp} shows the brightness temperature of all the contributions, including the foreground emissions, the cosmological signal and the thermal noise.
%%%%%%%%%%%%%%%%%%%%%%%%%%%%
\subsubsection{Astrophysical foregrounds}
\label{susubsec:astro_fg}
The astrophysical foregrounds in the MHz range considered in our simulations are of galactic and extragalactic origins. Here we describe the models that have been implemented to build the simulted maps of the astrophysical emissions. They are described in detail in \cite{Carucci-2020}.\\
\paragraph{Synchrotron emission}
The synchrotron emission in our Galaxy arises from charged electrons accelerating in the magnetic field. It is the brightest emission at the radio frequencies. Its brightness temperature can be modelled as a power-law as a function of frequency, in each pixel \textit{p}, with amplitude $T_{s}$ and spectral index $\beta_{s}$:
\begin{equation}
    \label{eq:power_law_galactic_fg}
    T(\nu, p) = T_{s}\left( \frac{\nu}{\nu_{0}}\right)^{\beta_{s}(\nu, p)}\,.
\end{equation}
The full-sky amplitude as well as the synchrotron spectral index are provided by the Planck Legacy Archive FFP10 simulations. The amplitude is computed from the FFP10 simulations at 217 GHz.\\
The synchrotron spectral index varies across pixels and its map is taken from \cite{Miville-D-2008}. Since the resolution of the maps is around 5 deg, to account for the smaller scales we adapted the small-scale fluctuations from \cite{Santos-2005}. The mean value over the pixels of the spectral index is $\beta_{s}\sim$3.
\paragraph{Free-free emission}
Another diffuse galactic emission is the free-free or Bremsstrahlung. It arises from the interaction of free electrons with ions in the galactic medium. It is mostly concentrated in the Galactic plane, unlike the synchrotron radiation. \\
As for the synchrotron, we model the free-free using a power-law for the brightness temperature, with the spectral amplitude and spectral index provided by the FFP10 simulations. The spectral index is kept fixed across the pixels and frequencies with value $\beta_{s}$=-2.13.
\paragraph{Extragalactic point sources}
Extragalactic point sources—including active galactic nuclei (AGN), radio galaxies, quasars, and others—can be categorized into a diffuse distribution of unresolved continuum sources and a population of bright, isolated point sources.\\
The simulated maps are generated according to the empirical model described in \cite{Battye-2013}. They obtain the differential source count $dN/dS$, which is the number of sources per steradian \textit{N} per unit flux S, from an empirical fit of several continuum surveys at 1.4 GHz. The mean temperature of the continuum populations is given by integrating $dN/dS$ \cite{Olivari-2018}
\begin{equation}
    \label{eq:temp_ps_bk}
    T_{ps} = \Big(\frac{\lambda^{2}}{2 k_{B}} \Big) \int^{S_{\rm max}}_{0} S\frac{dN}{dS}dS \,,
\end{equation}
where $\lambda$ is the wavelength of the observed radiation and $k_B$ is the Boltzmann constant. We assume that all the sources with $S>S_{\rm max}$ are removed from the data, with $S_{\rm max}$= 1 Jy. In practice, we do not model these highest-flux sources and do not go into the details of how this is done in the data, as different strategies are in place depending on the instrumental setup.\\
The contribution of isolated point sources comes from a clustering and a Poisson component; these are calculated from the angular power spectrum from which is generated the pixel map by using the \textsc{HEALPix synfast} routine. Point sources for $S_{\rm max}\lesssim$ 0.01 Jy are randomly injected into the map as fully resolved with mean random temperature \cite{Olivari-2018}
\begin{equation}
    \label{eq:temp_ps}
    T_{ps}(\mathrm{1.4\ GHz},p) = \Big(\frac{\lambda^{2}}{2 k_{B}} \Big) \Omega^{-1}_{\rm pix} \sum^{N}_{i=0} S_{i}\,,
\end{equation}
where $S^{i}$ is the flux of the point sources \textit{i} at 1.4 GHz, \textit{N} is the number of sources in the pixel \textit{p} and $\Omega_{pix}$ is the pixel area. \\
To scale this estimate at different frequencies, we assume a power-law dependence, $T_{ps} \propto \nu^{\alpha}$, where $\alpha$ is randomly chosen for each pixel of the simulated map according to a Gaussian distribution. We choose a mean value of -2.7 a standard deviation of 0.2 \cite{Bigot-Sazy-2015}. 
\subsubsection{Polarization leakage}
\label{subsubsec:pol_leak}
The galactic synchrotron radiation is the most intense diffuse emission in the frequency range considered in this work. Thus, it is fundamental to properly model it as realistic as possible. The linearly polarized part of such emission changes its polarization angle as it crosses the magnetic fields within our Galaxy’s interstellar medium \cite{Alonso-2014}. Unfortunately, the Faraday effect is non-spectrally smooth effect, i.e. it depends on the frequency channels. If any polarization fluctuation leaks into the total intensity for some instrumental issue, it would be difficult to subtract it without losing part of the unpolarized \HI~signal. In fact, a miscalibration of the instrument could yield a leakage of some percentage of the Stokes Q and U parameters of the synchrotron emission into the Stokes I parameter \cite{Alonso-2014}. Thus, the polarization leakage is not constant along the frequencies since the Faraday rotation changed the Stokes Q/U parameters across the channels. \\
In literature, there are two models of the polarization leakage in the frequency range relevant for the 21-cm emission. These models are described in \cite{Alonso-2014} and \cite{Shaw2015}. The two models use the same data-set (\cite{Oppermann-2012}) but produce map with a qualitatively different structure in pixel space, due to the different assumptions \cite{Cunnington-2021}.
We use the model adopted by \cite{Alonso-2014} as it is more conservative than the one proposed by \cite{Shaw2015}, in terms of non-smooth contaminant \cite{Carucci-2020}. This model assumes that all the leakage comes from just the Q Stokes component and not U, which is not totally realistic. They produce with the code CRIME a set of simulated maps of the Stokes Q emission at each frequency. The intensity of the simulated polarization leakage maps is thus:
\begin{equation}
    \label{eq:temperature_pol_leak}
    T_{\rm leak}(\nu, \hat{\bm{n}})=\epsilon_{P} T^{Q}_{\rm synch}(\nu, \hat{\bm{n}})\,,
\end{equation}
where $T^{Q}_{\rm synch}(\nu, \hat{\bm{n}})$ is the Stokes Q maps of the synchrotron emission and $\epsilon_{p}$ is the polarization leakage fraction. We set $\epsilon_{P}=$0.5\% for this study \cite{Carucci-2020}, yielding to a contribution of one order of magnitude higher than the cosmological signal, as seen in Fig.~\ref{fig:brightness_temp}, for regions closer to the Galactic plane.
\subsection{Instrumental effects}
\label{subsec:instrumental_effect}
To mimic a realistic single-dish experiment, we include %as instrumental effect 
the smoothing effect by the telescope beam and the uncorrelated thermal noise\footnote{Regarding the effect of correlated noise, such as 1/f noise, in \cite{Irfan-2024} it is demonstrated that such systematic does not impact the foreground cleaning of the  MeerKAT Large Area Synoptic Survey (MeerKLASS) data, as the 1/f noise affects the very large scales of the angular power spectrum and it can be removed with a proper data calibration.}.\\
We consider a symmetric Gaussian beam with a frequency-dependent width 
\begin{equation}
    \label{eq:theta_fwhm}
    \theta_{\rm FWHM} = \frac{c}{\nu D}\,
\end{equation}
where $c$ is the speed of light and $D$ is the telescope dish diameter. Note that in this case the maps in different frequency range have different resolution. We also test the foreground cleaning strategies with a constant Gaussian beam for all the frequency channels, computing Eq.~\eqref{eq:theta_fwhm} at the minimum frequency (i.e., largest beam).\\
We assume the thermal noise to be Gaussian, with variance per pixel \cite{Spinelli-2021}
\begin{equation}
    \label{eq:sigma_noise}
    \sigma=\frac{T_{\rm sys}}{\sqrt{N_{\rm dish} \tau \Delta \nu} }\,
\end{equation}
where $\tau$ is the integration time in seconds per pixel, $\Delta \nu$ is the width of each frequency channel, $N_{\rm dish}$ is the number of dishes in the array and $T_{\rm sys}$ is the system temperature.
The integration time per pixel is related to the observation time as 
\begin{equation}
    \label{eq:observation_time}
    \rm t_{\rm pix} = \frac{\Omega_{\rm beam}}{4\pi f_{\rm sky}} t_{\rm obs}\,,
\end{equation}
where $\Omega_{\rm beam} = 1.33\theta_{\rm FWHM}^{2}$ is the beam solid angle.\\
The system temperature $T_{\rm sys}$ is calculated as
\begin{equation}
    \label{eq:T_sys}
    T_{\rm sys} = T_{\rm rx}+T_{\rm spl}+T_{\rm CMB}+T_{\rm gal}\, 
\end{equation}
where $T_{\rm spl}\approx 3$ K is the contribution from spill-over, $T_{\rm CMB}\approx 2.7$ K is the temperature of the CMB, $T_{\rm gal} \approx 25 \,\rm K (408 \, \rm MHz / \nu ) ^{2.75}$  is the contribution of our own galaxy, and ${T_{\rm rx}}$ is the receiver noise temperature \cite{SKA-2018}. For SKAO, we assume \cite{SKA-2018}
\begin{equation}
    \label{eq:t_rx_ska}
    T_{\rm rx} = 15\, \rm K\,+\,30\, \rm K\, \left(\frac{\nu}{\rm GHz}-0.75\right)^{2}.
\end{equation}
All the specifications for SKAO-MID AA4 are described in Tab.~\ref{tab:SKA_spec}. Note the the number of dishes N$_{\rm dish}$ includes 133 newly built 15-meter dishes and 64 dishes, with a diameter of 13.5 m,  from the existing MeerKAT telescope. The overall diameter size D$_{\rm dish}$ is computed as a mean between the two different kind of dishes.
\begin{table}[]
\centering
\begin{tabular}{lll}
\hline
\multicolumn{3}{c}{SKAO-MID AA4 specifications}                  \\ \hline
Number of dishes             & N$_{\rm dish}$ & 197       \\
Diameter of the dishes       & D$_{\rm dish}$ & 14.51 m   \\
Observed fraction of the sky & f$_{\rm sky}$    & 50\%      \\
Observation time             & t$_{\rm obs}$    & 10000 hrs \\
Spill-over temperature       & T$_{\rm spill}$  & 3 K       \\
Channel width                   & $\Delta \nu$     & 1 MHz     \\
\hline
\end{tabular}
\caption{Instrumental specifications for computing the thermal noise and the telescope beam. Note that the number of dishes includes both the SKAO and the MeerKAT dishes.}
\label{tab:SKA_spec}
\end{table}
%%%%%%%%%%%%%%%%%%%%%%%%%%%%%%%%%%%%%%%%%%%%%%%%%%
\subsection{Data cube}
\label{subsec:datacube}
\begin{figure}[ht]
    \centering
    \includegraphics[width=0.7\linewidth]{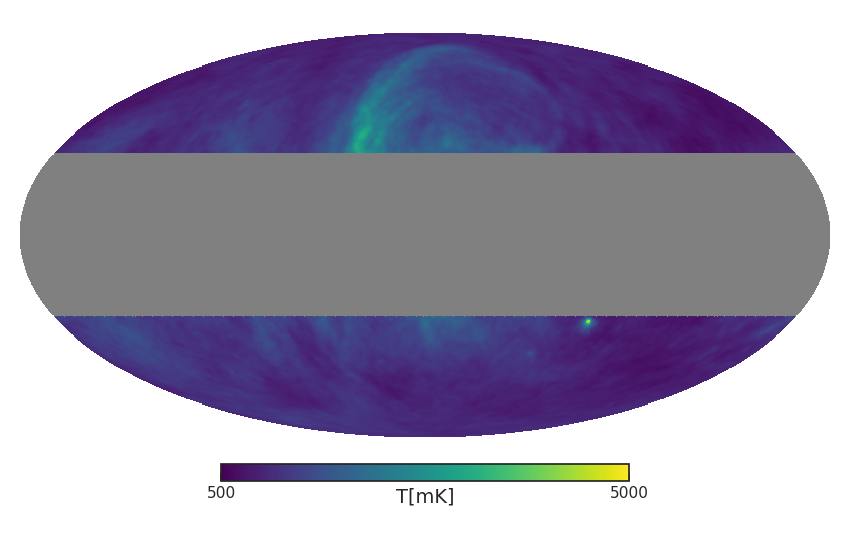}
    \caption{Total temperature map of the observed dataset at channel $\nu=952.5$ MHz. The applied mask covers the 50\% of the pixels in the Galactic plane.}
    \label{fig:obs_data_ch_952.5_mask}
\end{figure}

\begin{figure}[ht]
    \centering
    \includegraphics[width=0.9\linewidth]{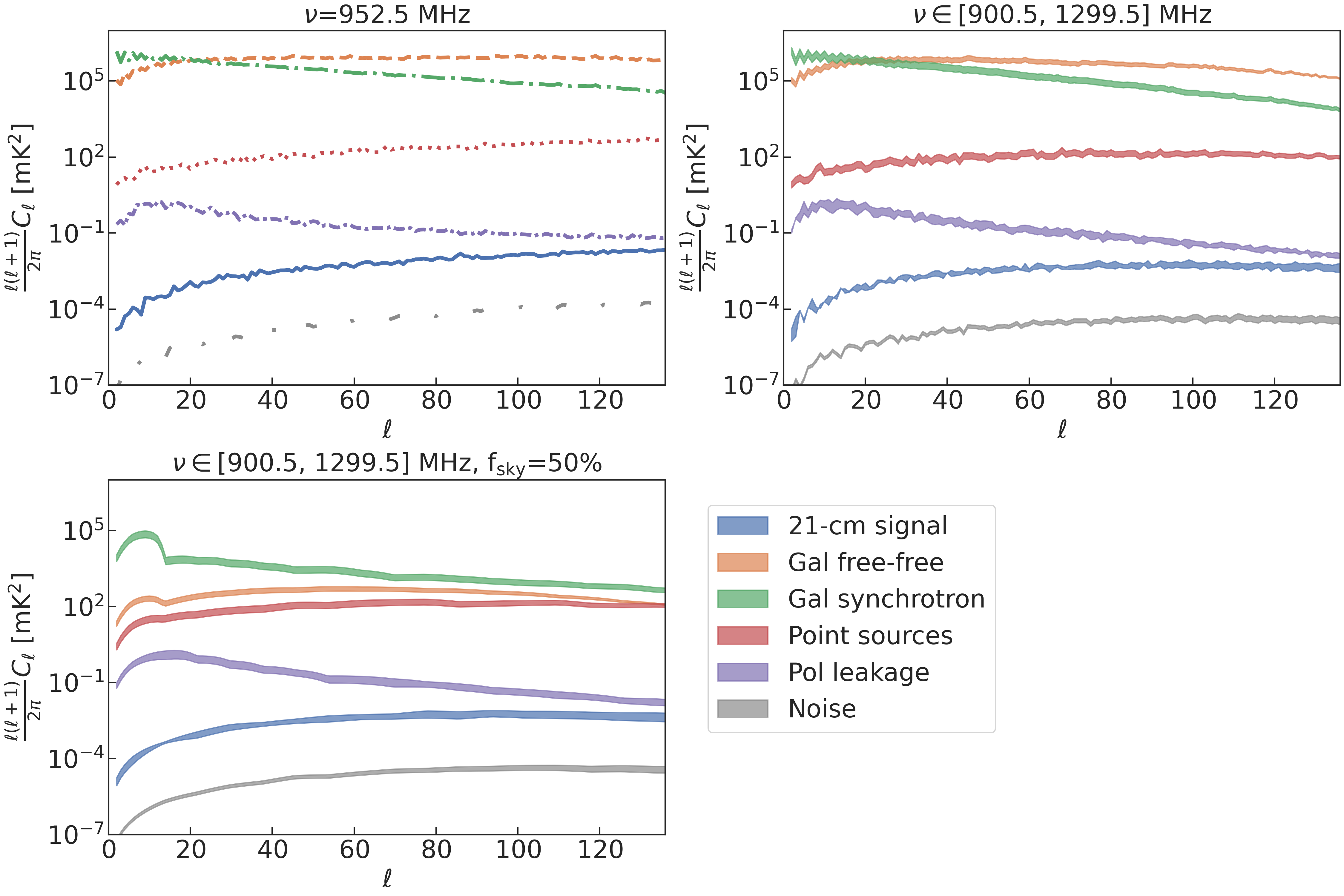}
    \caption{Power spectrum of the simulated components as a function of the angular scales. Left-Top panel: power spectra at the frequency $\nu=$952 MHz. Right-Top panel: power spectra in the frequency range $\nu \in \left[ 900.5 - 1004.5 \right]$ MHz after the convolution with the telescope beam. Bottom panel: angular power spectra of the contaminants when a mask is applied, covering the 50\% of the sky in the Galactic plane.}
    \label{fig:cl_components}
\end{figure}

We sum all the foreground components $\mathsf{F}$ and the contributions from the cosmological signal $\mathsf{C}$ and the thermal noise $\mathsf{N}$ to create the observed data cube, smoothed by the Gaussian beam $\textit{B}$:
\begin{equation}
    \mathsf{X} = \left[\mathsf{F} + \mathsf{C}  \right] \ast\mathsf{B} +\mathsf{N}
\end{equation}
We consider 105 \textsc{HEALPix} maps of 1 MHz thickness in the range $\nu \in \left[ 900.5 - 1004.5 \right]$ MHz,  corresponding to the redshift range $z \in \left[ 0.41 - 0.58 \right]$. According to Eq.~\eqref{eq:theta_fwhm}, the beam width ranges from 1.3 to 1.8 deg. The \textsc{HEALPix} maps have a resolution of $N_{\rm side}=$128 and we consider scales up to $\ell_{\rm max}=3 N_{\rm side}-1$.\\
All simulated maps are masked in the Galactic plane, implementing a mask with $f_{\rm sky}=50\%$, as shown in Fig.~\ref{fig:obs_data_ch_952.5_mask}.\\
The power at the angular scales of the simulated components is plotted in Fig.~\ref{fig:cl_components}, using the estimator later described at the beginning of Section \ref{sec:results}. On the left, each component is plotted at frequency $\nu=$952 MHz: foreground emissions are orders of magnitude higher than the cosmological signal. The right panel shows the effect of the telescope beam in the frequency range $\nu \in \left[ 900.5 - 1004.5 \right]$ MHz: there is a drop in power at $\ell_{\rm beam} \sim \pi/\theta_{\rm FWHM}$. Here, for simplicity, we plot the value at $\theta_{\rm FWHM,\, min}$. The effect of the mask is shown in bottom panel: the masking of the 50\% of the sky in the Galactic plane changes the contributions of the different contaminants.

%%%%%%%%%%%%%%%%%%%%%%%%%%%%%%%%%%%%%%%%%%%%%%%%%%%%%%%%%%%%%%%
\section{Needlets}
\label{sec:needlets}
\begin{figure}[ht]
    \centering
    \includegraphics[width=0.6\linewidth]{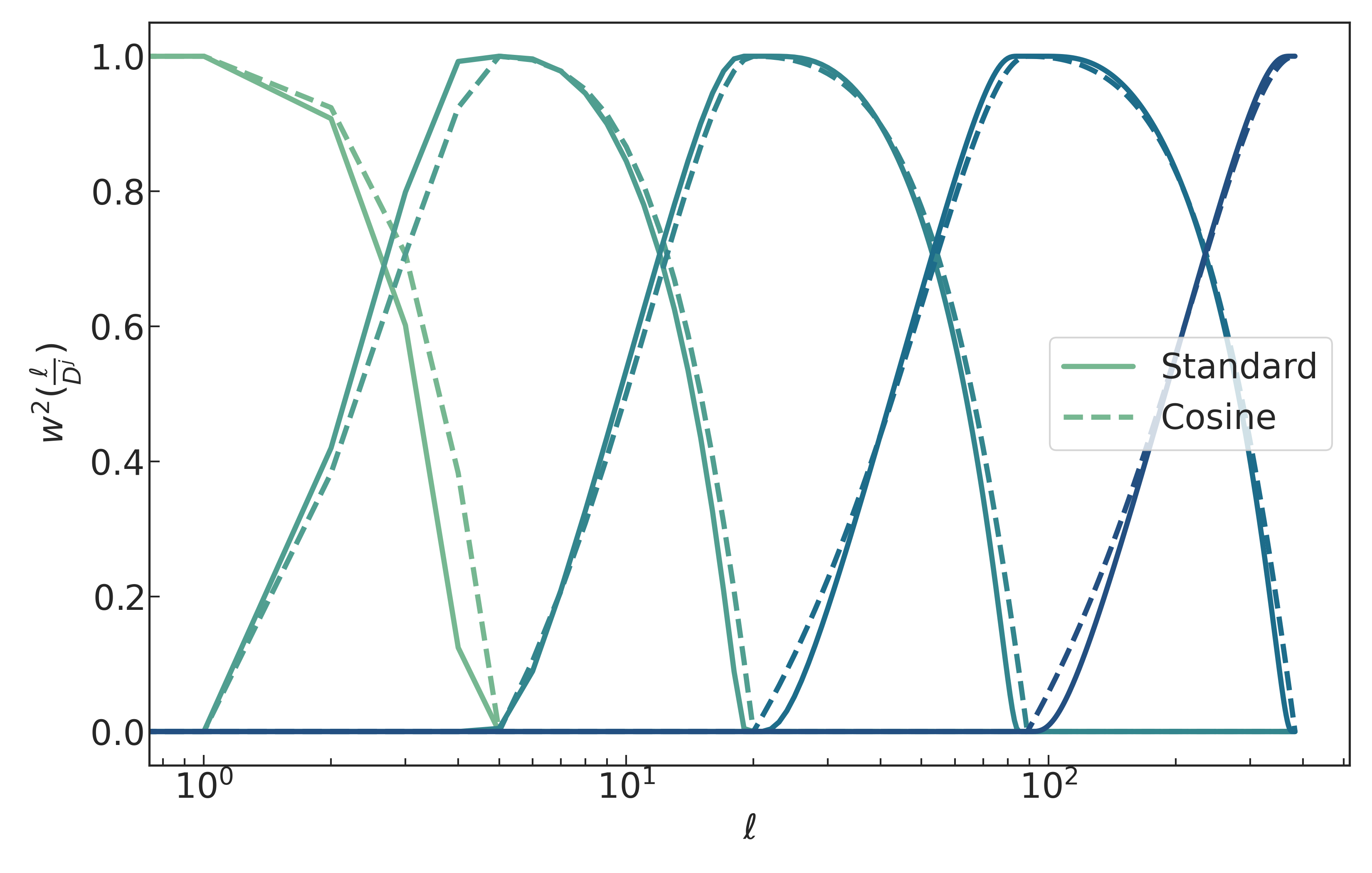}
    \caption{Weight functions for needlets with $j_{\rm max}=4$, $\ell_{\rm max}=383$. The gradient of colors changes from green to blue as the band index \textit{j} increases from $0$ to $4$. The solid line corresponds to the weight function of the standard needlets, the dashed line of the cosine ones.}
    \label{fig:window_cosine_standard_need_jmax4_lmax383}
\end{figure}

Needlets are a kind of spherical wavelets introduced for statistical analysis by \cite{Narcowich-2006};  their properties are discussed in \citep{Baldi-2006, Marinucci-2011}. Their first application in cosmology was for CMB data analysis \citep{Pietrobon-2006,Marinucci-2008}. The main advantage of the needlets is the double localization in the harmonic and in the real space. \\
Needlets exhibit several notable features. They do not rely on any tangent plane approximation and allow for a simple reconstruction formula. Due to their harmonic localization, they are computationally efficient and easily implemented using the standard \textsc{HEALPix} pixelization scheme \cite{HEALPix-2005}. For any fixed angular position, needlet coefficients decay quasi-exponentially in scale space, and it can be shown that the coefficients become asymptotically uncorrelated at fixed angular separation as the scale in harmonic space increases. This property enables them to be treated as approximately independent at small scales.\\
The spherical needlet system $\psi_{\{jk\}}$ is constructed by convolving spherical harmonics $a_{\ell m}$ with a suitable weight function, $w_{\ell}(\cdot)$ and is defined as
\begin{equation}
    \label{eq:psi_need}
    \psi_{jk}(\hat{\bm{n}}) = \sqrt{\lambda_{jk}} \sum^{\left[D^{j+1}\right]}_{\ell=\left[D^{j-1}\right]} w{\Big(\frac{\ell}{D^{j}}\Big)} \sum^{\ell}_{m=-\ell} Y^{*}_{\ell m}(\hat{\bm{n}}) Y_{\ell m}(\xi_{jk})\,,
\end{equation}
where $\left[ \cdot \right]$ denotes the integer part, ${\xi_{jk}}$ and $\lambda_{jk}$ are the cubature points and the weights at frequency \textit{j} and location \textit{k}. The weight functions are usually binned in the harmonic space, with each bin labelled by the index \textit{j}. The parameter \textit{D} controls the bandwidth: smaller values \textit{D} correspond to tighter localization in harmonic space, while larger values improve localization on the sphere.\\
In the \textsc{HEALPix} implementation, $k$ corresponds to a pixel index, the locations $\xi_{jk}$ are identified with the pixel centers, and the cubature weights $\lambda_{jk}$ can be approximated as $4\pi/N_{\rm pix}$, where $\rm N_{\rm pix}$ is the number of pixels.\\
The needlet coefficients $a_{jk}$ of a scalar field $T(\hat{\bm{n}})$ are obtained by projecting the field onto the needlet basis $\psi_{jk}(\hat{\bm{n}})$:
\begin{equation}
    \label{eq:needlets_coeff}
    \begin{split}
          a_{jk}&=\int_{S^2}d\Omega T(\hat{\bm{n}}) \psi_{jk}(\hat{\bm{n}}) \\
          &= \sqrt{\lambda_{jk}} \sum_{\ell m }a_{\ell m} w\Big( \frac{\ell}{D^j}\Big) Y_{\ell m}(\xi_{jk}) \,, 
    \end{split}
\end{equation}
where $a_{\ell m}$ are the spherical harmonic coefficients of the field $T(\hat{\bm{n}})$
\begin{equation}
    \label{eq:alm}
    a_{\ell m} = \int_{S^2}d\Omega T(\hat{\bm{n}})  Y^{*}_{\ell m}(\hat{\bm{n}})\,,
\end{equation}
At each frequency \textit{j}, the needlet coefficients $a_{jk}$ are represented by an \textsc{HEALPix} map.\\
In this work, we implement the so-called \textit{standard} needlets functions. The weight function for standard needlets is shown in Fig.~\ref{fig:window_cosine_standard_need_jmax4_lmax383} for a maximum frequency index $j_{\rm max}=$4. These functions are compactly supported in harmonic space, with support in the range $\frac{\ell}{D^{j}} \in \left[1/D, D\right]$ and involve only a finite number of multipoles. This property guarantees an exact reconstruction of the field after decomposition:
\begin{equation}
    \label{eq:needlets_reconstruction_formula}
    T(\hat{\bm{n}}) = \sum_{jk} a_{jk} \psi_{jk}(\hat{\bm{n}})\,.
\end{equation}
Fig.~\ref{fig:window_cosine_standard_need_jmax4_lmax383} also shows the weight function for \textit{cosine} needlets, which are built on a different weight function, but share the same properties as the standard needlets. The GNILC algorithm uses the \textit{cosine} needlets to decompose the field and calculate the ILC weights. \\
Another class of needlets, known as \textit{Mexican} needlets \cite{Scodeller-2011}, differ in that their weight functions are not compactly supported in harmonic space. Instead, each frequency bin includes contributions from the full range of multipoles, which prevents exact reconstruction of the input field. \\
In this work, we adopt standard needlets with $j_{\rm max}=$4. For completeness, in Appendix~\ref{appendix:std_need_jmax12} we explore the case with $j_{\rm max}=$12 and in Appendix~\ref{appendix:mex_need} we test our pipeline using the \textit{Mexican} needlets.
%%%%%%%%%%%%%%%%%%%%%%%%%%%%%%%%%%%%%%%%%%%%%%%%%%%%%%%%%%%%%%
\section{The pipeline}
\label{sec:pipeline}
This section describes the pipeline we implement to obtain the power spectrum of the \HI~signal.
\subsection{Need-PCA}
\label{subsec:needlets-PCA}
We implement a new version of PCA which, instead of using the spherical harmonics as the \textit{standard} PCA, exploits the decomposition with standard needlets, the \textit{Need}-PCA.\\
We expand the data cube $\mathsf{X}$ in needlet coefficients X$_{jk}$ and we apply PCA to each map corresponding to each coefficient \textit{j}. We compute the covariance matrix as 
\begin{equation}
    \mathsf{Cov}_{j,\,ch,\,ch'}= \frac{1}{\rm N_{\rm pix}} \sum^{N_{\rm pix}}_{k=1} X_{jk}^{\rm ch} X_{jk}^{\rm ch'}\,,
\end{equation}
and its eigenspectrum to identify the principal eigenvalues. Once set the number of components to remove  N$_{\rm fg}$, we compute the foreground residual maps as $\mathsf{X}_{\rm fg,\, j}=\hat{\mathsf{U}}_{\rm fg,\,j}\hat{\mathsf{U}}_{\rm fg,\,j}^{T}\mathsf{X}_{\rm j}$, where $\hat{\mathsf{U}}_{\rm fg,\,j}$ is the mixing matrix at the needlet scale \textit{j}. Thus, the cleaned \HI~maps at scale \textit{j} is recovered as
\begin{equation}
    \label{eq:cleaned_j_map}
    \mathsf{X}_{\HI,\,\rm j} = \mathsf{X}_{j} - \mathsf{X}_{\rm fg,\,j} = \mathsf{X}_{\rm j} - \hat{\mathsf{U}}_{\rm fg,\,j} \hat{\mathsf{U}}_{\rm fg,\,j}^{T} \mathsf{X}_{\rm j}\,.
\end{equation}
As a final step, we reconstruct the \HI~signal by convolving the spherical harmonics coefficients of each X$_{\HI,\,\rm jk}$ maps with the needlet weight functions and then summing up the resulting maps. 
\subsection{Need-GMCA}
\label{subsec:needlets-GMCA}
As for PCA, we implement a version of GMCA which uses needlets (Need-GMCA). As described in Sec.~\ref{subsec:needlets-PCA}, we decompose the data in needlet space and apply GMCA\footnote{\url{https://github.com/isab3lla/gmca4im/tree/master}} to each needlet coefficient. We thus reconstruct the residual maps of the \HI~signal for each frequency channel.
\subsection{GNILC}
\label{subsec:pipeline_GNILC}
As discussed in Sec.~\ref{sec:comp_sep_method}, GNILC require \textit{prior} maps of the \HI~signal as input to evaluate the \HI~dimension in the PCA step. These \HI~maps are generated from the theoretical angular power spectrum calculated in each frequency channel.
We extract from the set of \HI~plus noise simulated maps the angular power spectra using the routine \textsc{anafast} from the \textsc{HEALPix} package. We pass them through a Savitzky–Golay filter\footnote{\url{https://docs.scipy.org/doc/scipy/reference/generated/scipy.signal.savgol_filter.html}} and generate from them a set of maps with the \textsc{synfast} routine of \textsc{HEALPix}. We smooth the maps with the proper telescope beam in each frequency channels, as described in Sec.~\ref{subsec:instrumental_effect}. Finally, we apply the mask to the simulations. \\
We assess the robustness of this choice by comparing the angular power spectrum of the recovered \HI~.\\
In Appendix~\ref{app:gnilc_prior} we also evaluate the impact on the recovery of the \HI~from the data, by comparing the recovered signal from GNILC using a "wrong" choice of the prior.
%%%%%%%%%%%%%%%%%%%%%%%%%%%%%%%%%%%%%%%%%%%%%%%%%%%%%%%%%%%%%%%
\section{Results}
\label{sec:results}
We first test Need-PCA with a set of observations which include the galactic synchrotron and free-free emissions and the extragalactic point sources; later we consider also the contribution from the polarization leakage.\\
Moreover, as a first step we run our pipeline on maps smoothed by a frequency-dependent Gaussian beam, as describe in Sec.~\ref{subsec:instrumental_effect}; then we compare the results with the case of beam kept constant across all the frequency channels. To be more conservative, we choose to calculate the width of the beam (Eq.~\eqref{eq:theta_fwhm}), at the lowest frequency channel, $\nu_{\rm max}=900.5$ MHz, which corresponds to $\theta_{\rm FWHM}=1.3$ deg.
Lastly, we compare Need-PCA performances with other foreground cleaning methods, i.e. the standard-PCA, the Need- and the standard- GMCA and GNILC, described in Sec.~\ref{sec:comp_sep_method}.\\
We assess the performance of the cleaning method by computing the angular power spectrum of the recovered \HI~signal at each frequency channel, $\ClRes$. We extract the angular power spectrum with the software package \textsc{NaMaster}\footnote{\url{https://github.com/LSSTDESC/NaMaster}}, described in \cite{Alonso-2019}. \textsc{NaMaster} is a  pseudo-$C_{\ell}$ estimator that takes into account incomplete sky coverage.\\
For each analysis, we show the recovered $C_{\ell}^{\rm Res}$ in the middle channel of the frequency range, $\nu=$952.5 MHz, and the average $C_{\ell}^{\rm cosmo}$ over all the channels. In all results' plots, we show the input power spectrum in the top panel and in the bottom panel the percent difference $\Delta$ to assess the performance of the cleaning methods, defined as
\begin{equation}
    \label{eq:delta_percentage}
    \Delta :=  C_{\ell}^{\rm Res}/C_{\ell}^{\rm cosmo} -1\,.
\end{equation}
\subsection{Main results}
\label{subsec:results_main}
\begin{figure}[ht]
    \centering
    \includegraphics[width=0.8\linewidth]{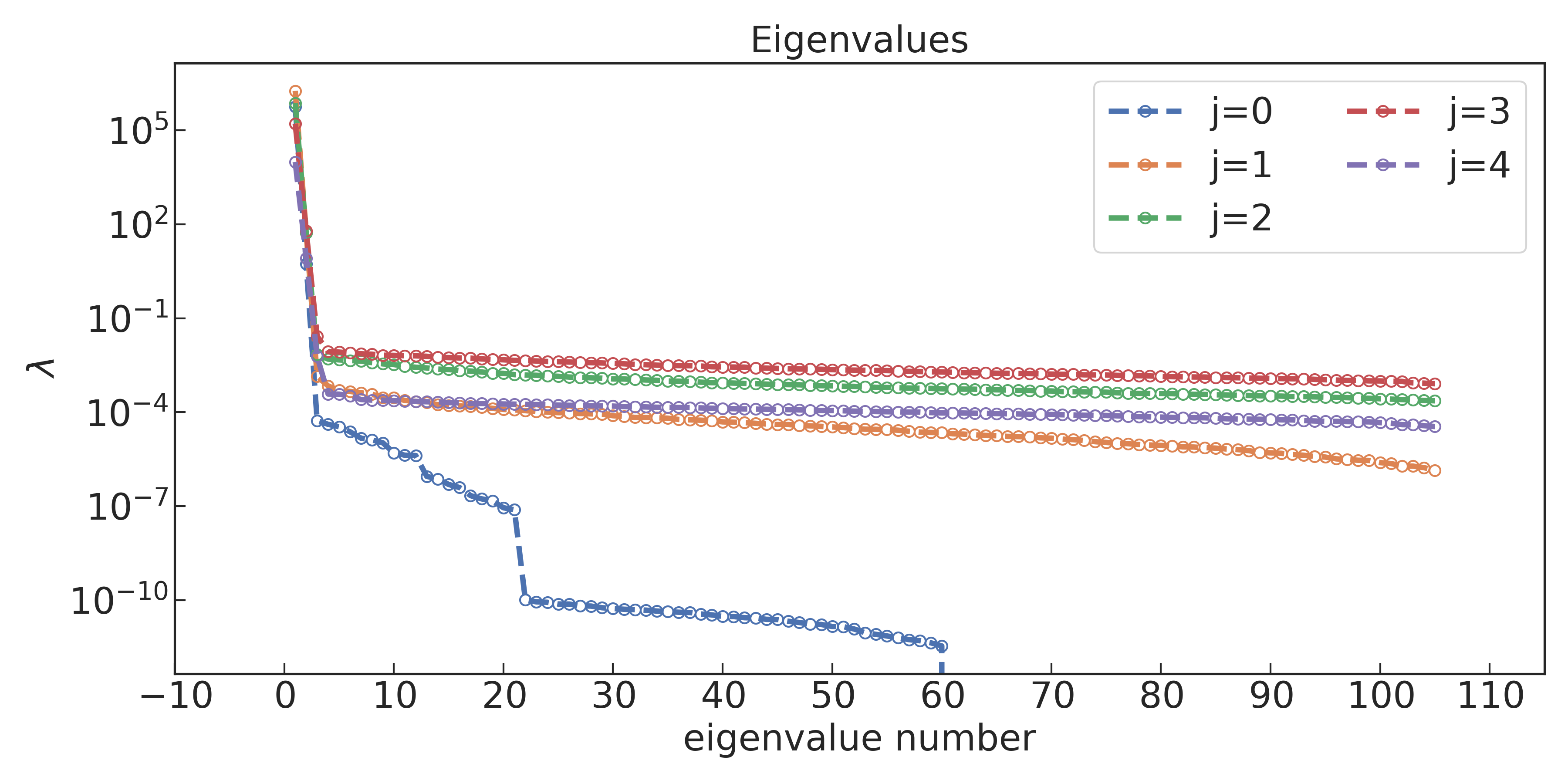}
    \caption{Eigenvalues of the covariance matrix of the data-cube after the needlets decomposition, setting $j_{\rm max}=$4, for each scale. The principal components are identified by the largest eigenvalues.}
    \label{fig:eigenv_synch_ff_ps_SKA_AA4_PCA_need_jmax4_lmax383}
\end{figure}

\begin{figure}[ht]
\centering
\setlength{\tabcolsep}{0.01pt}
\begin{tabular}{ccc}
\includegraphics[width=0.3\textwidth]{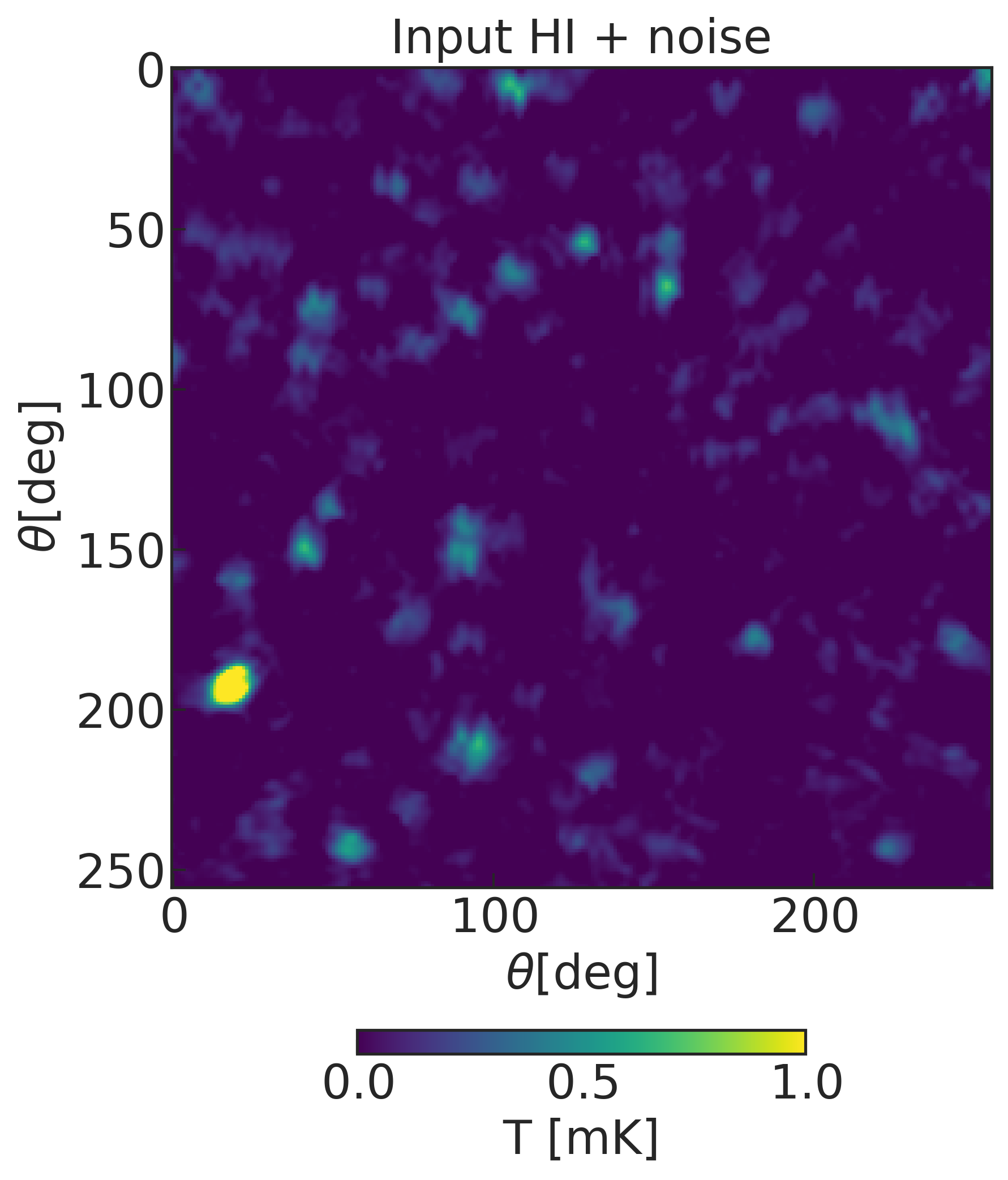}&
\includegraphics[width=0.3\textwidth]{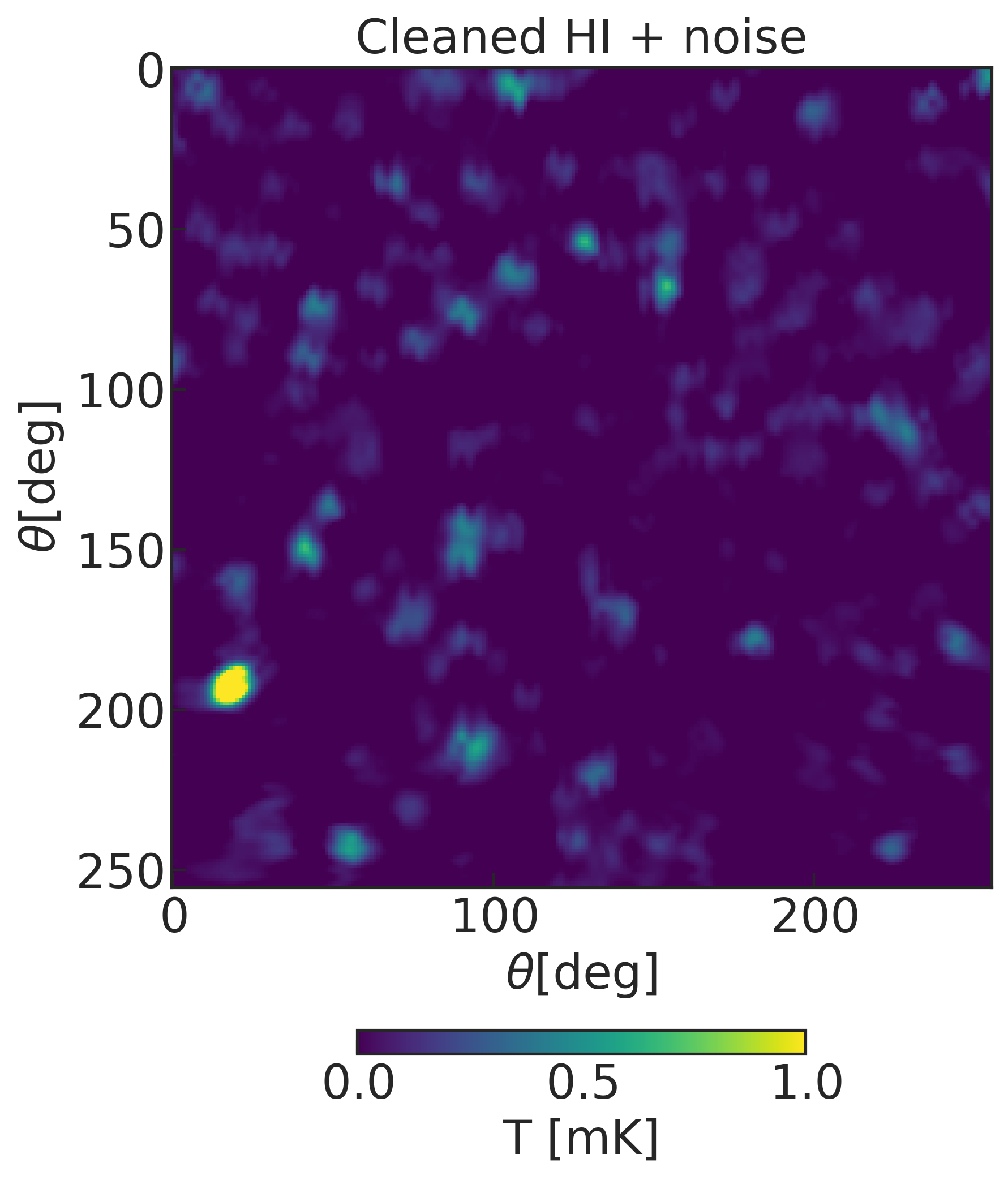}&
\includegraphics[width=0.3\textwidth]{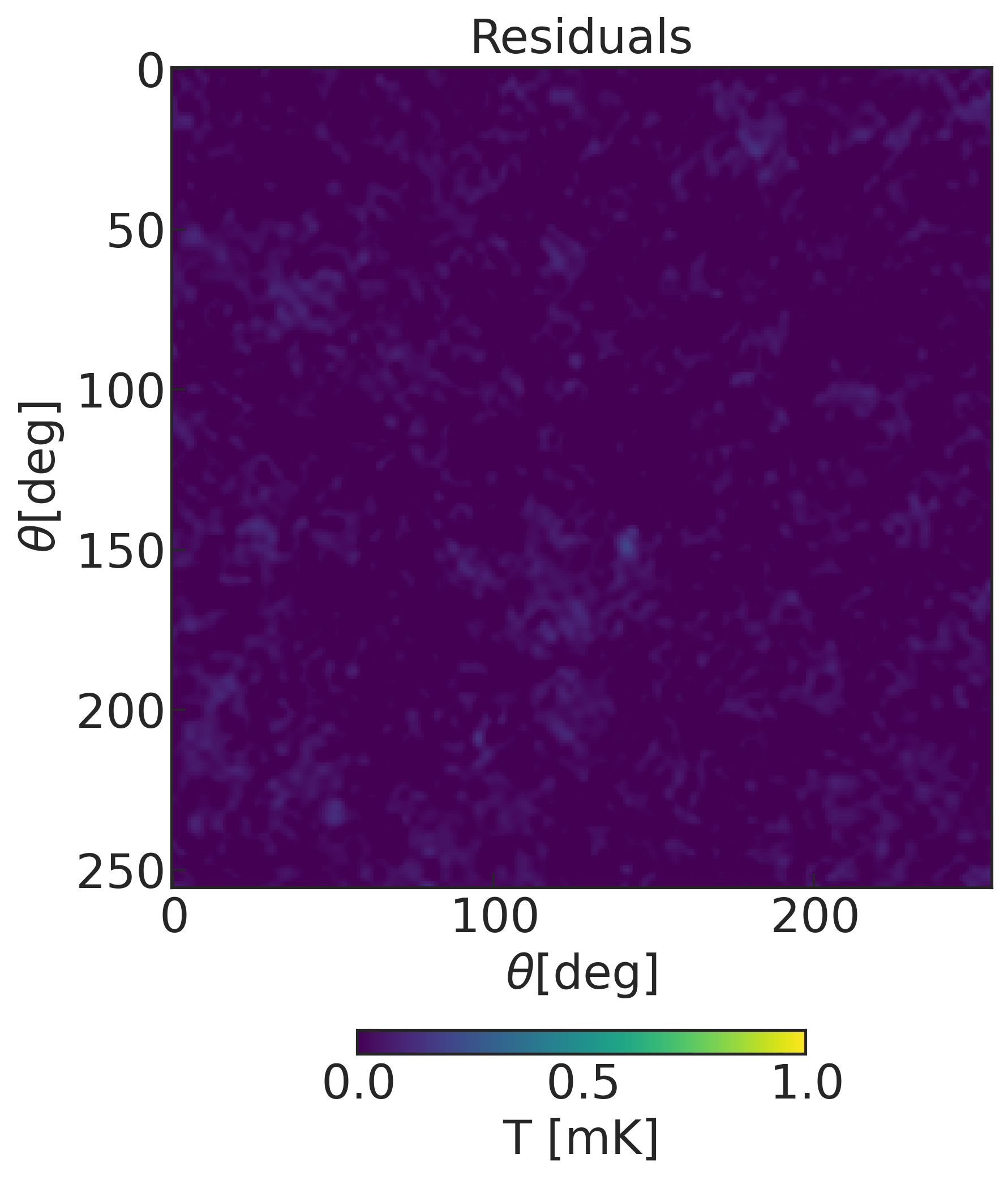}
\end{tabular}
\caption{Gnomonic projection of the input \HI~(left), the cleaned \HI~map after Need-PCA (center) and the residual map, at frequency $\nu=952.5$ MHz. Here we plot a 45.8$\times$45.8 deg$^{2}$ patch centred at the coordinates (lon, lat)=(0,67) deg. In this area, the standard deviation of the residual is 0.034 mK. The telescope beam is Gaussian and varies across the frequency channels.}
\label{fig:gnomview_cleanedHI_SKA_AA4_PCA_need}
\end{figure}

\begin{figure}[ht]
    \centering
    \includegraphics[width=0.7\linewidth]{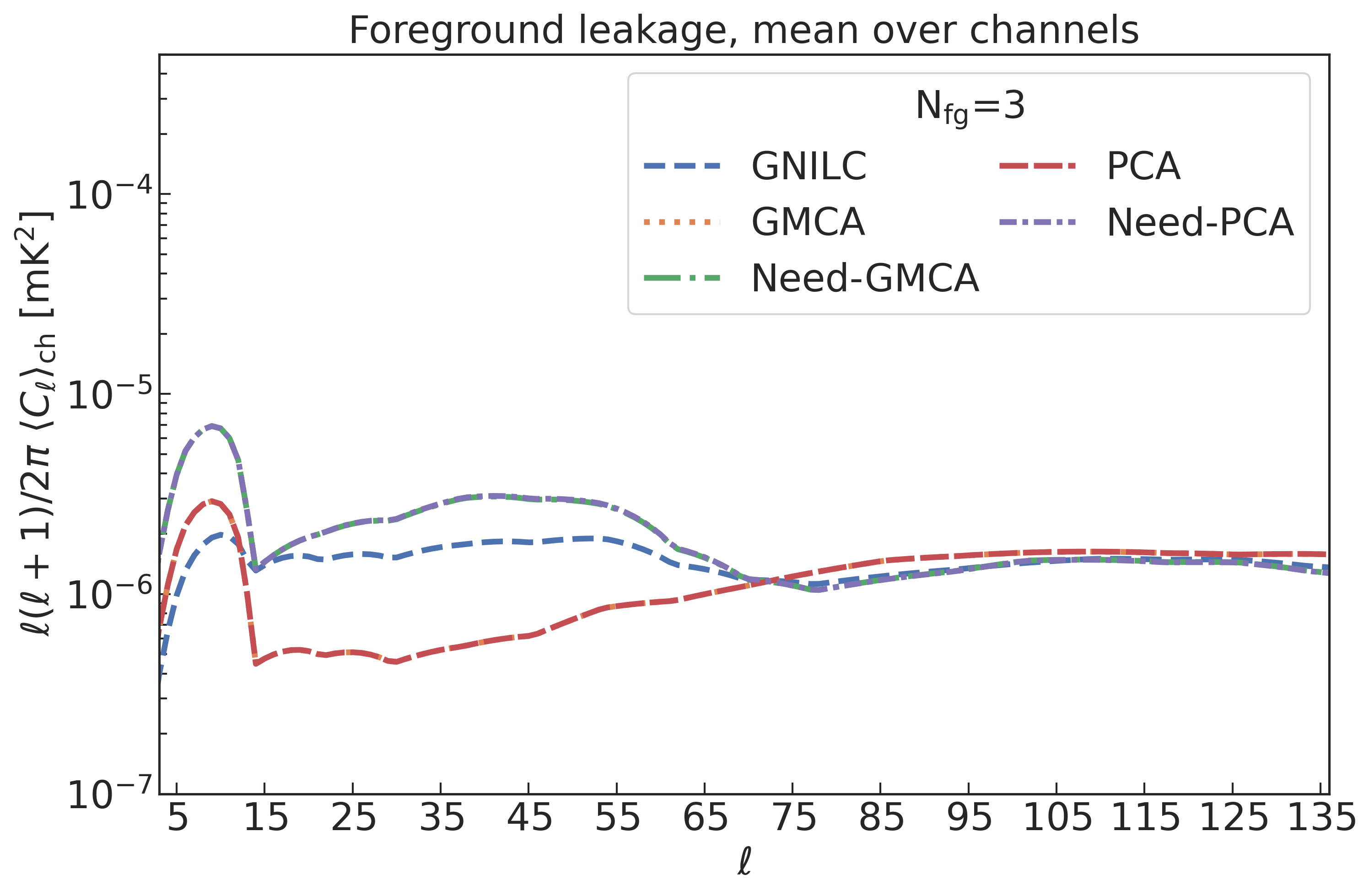}
    \caption{Comparison between the $C_{\ell}$ of the foreground leakage recovered by GNILC (blue line), GMCA (orange line), Need-GMCA (green line), PCA (red line) and Need-PCA (purple line), respectively, obtained by averaging over the frequency channels. The telescope beam is Gaussian and varies across the frequency channels. For low $\ell$, GNILC shows lower foreground leakage than the other methods, since it changes dynamically the number $N_{\rm fg}$ depending on scales and regions. The results of PCA and GMCA, in the standard and needlet versions, almost perfectly overlap.}
    \label{fig:fg_leak_mean_ch_GNILC_PCA_GMCA}
\end{figure}

\begin{figure}[ht]
\centering
\setlength{\tabcolsep}{0.01pt}
\begin{tabular}{cc}
    \includegraphics[width=0.49\linewidth]{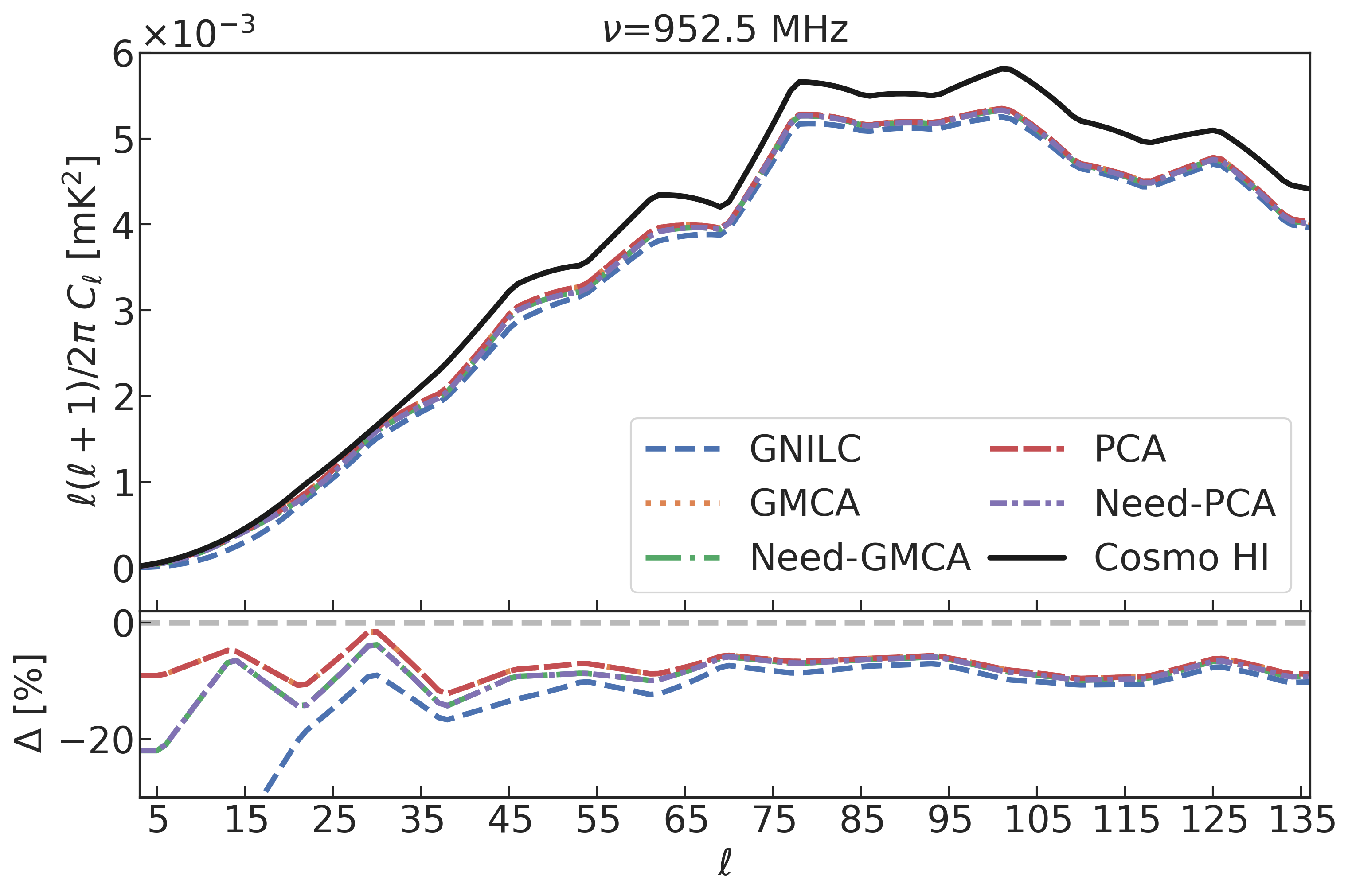}
    \includegraphics[width=0.49\linewidth]{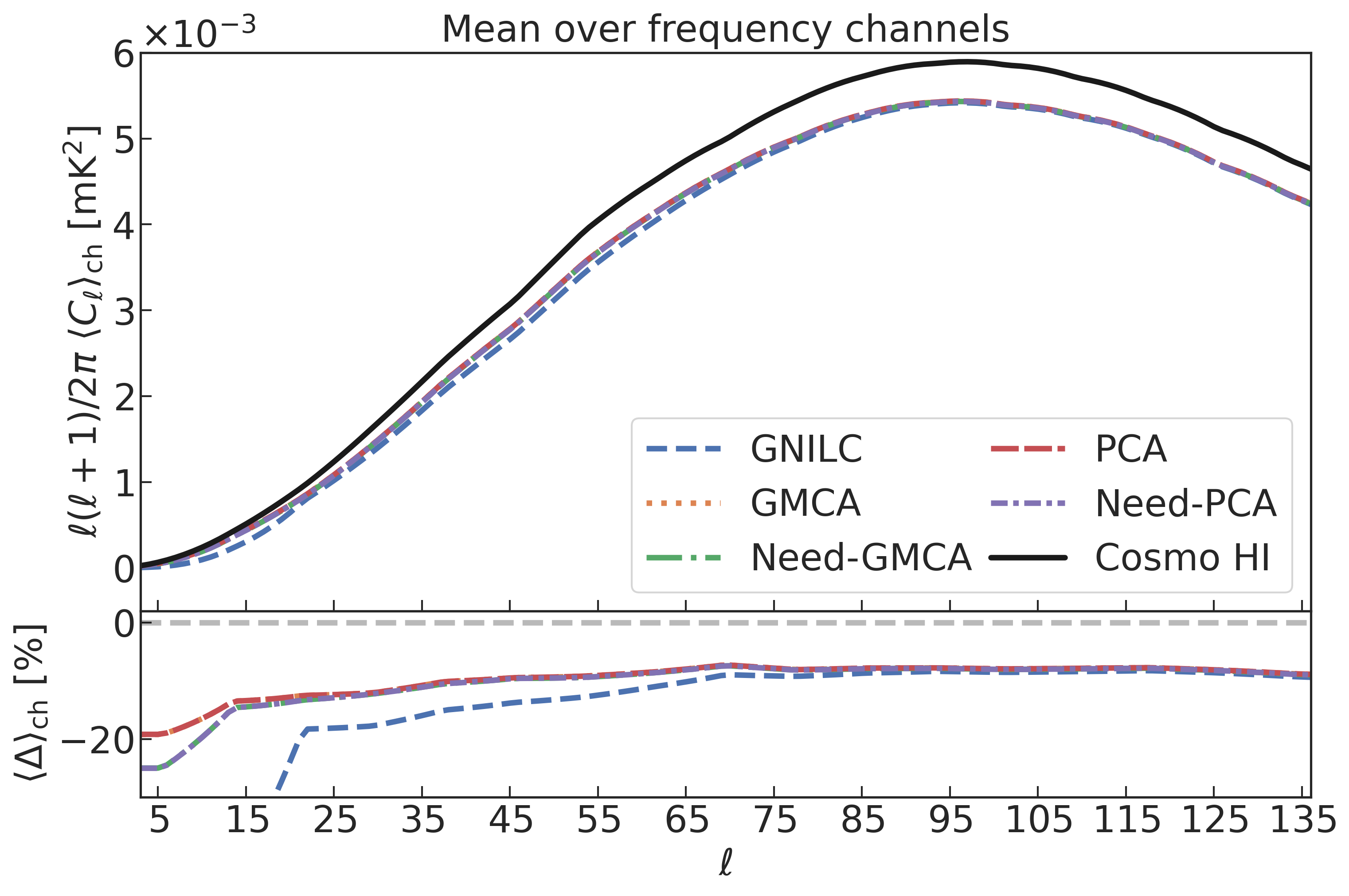}\\
    \includegraphics[width=0.49\linewidth]{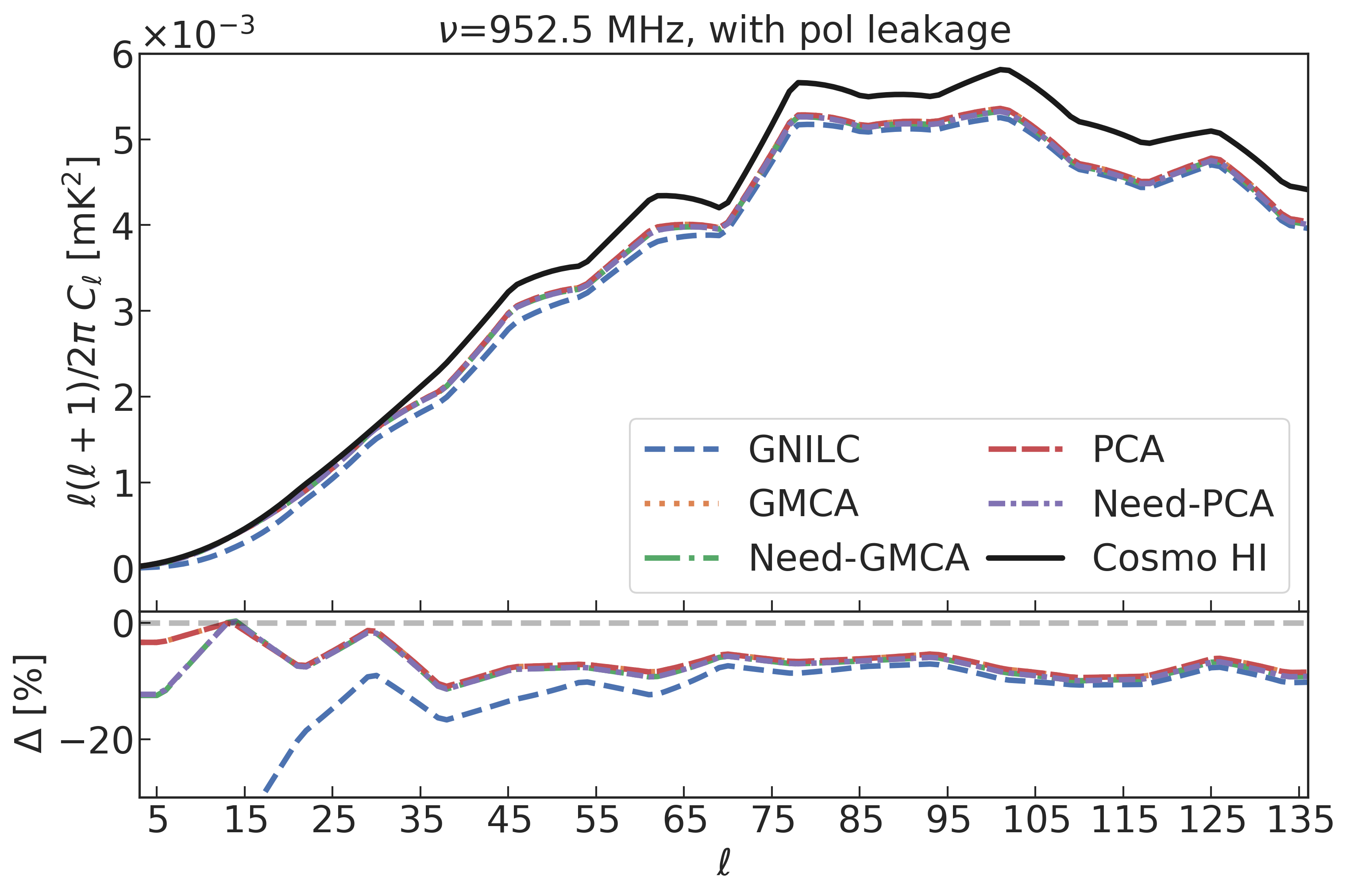}
    \includegraphics[width=0.49\linewidth]  {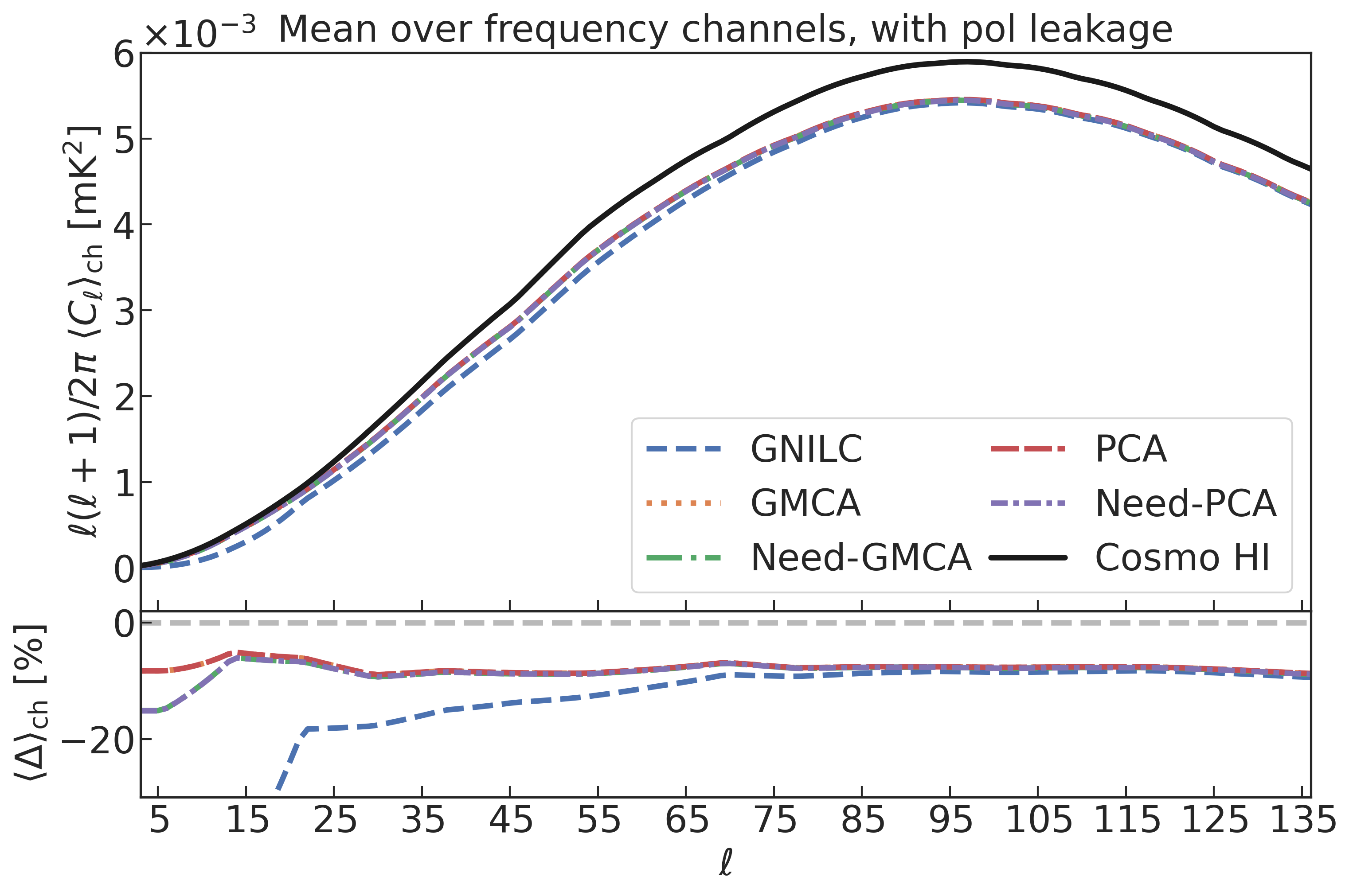}
\end{tabular}
    \caption{Comparison between the $\ClRes$ recovered with GNILC (blue line), GMCA (orange line), Need-GMCA (green line), PCA (red line) and Need-PCA (purple line). The input $\ClCosmo$ is plotted in black. The bottom panel shows the percent difference $\Delta$. The telescope beam is Gaussian and varies across the frequency range. We set N$_{\rm fg}=$3 for PCA, Need-PCA, GMCA and Need-GMCA; GNILC computes N$_{\rm fg}$ for each scales and region of pixel according to the AIC. Left upper panel: $\ClRes$ at $\nu$=952.5 MHz; right upper panel: $\ClRes$ averaged over the frequency channels. In the bottom panel are shown the same quantity adding the polarization leakage to the analysis.}
        \label{fig:panel_summary_SKA_beam}
\end{figure}

To calculate the number of sources to remove, we first look at the eigenvalues of the covariance matrix, as described in Sec.~\ref{subsec:needlets-PCA}. 
As shown in Fig.~\ref{fig:eigenv_synch_ff_ps_SKA_AA4_PCA_need_jmax4_lmax383}, the principal components are identified by the largest eigenvalues, in this case the first three. Thus, we set N$_{\rm fg}=$3. The needlet scale \textit{j}=0, which corresponds to the largest angular scale $\ell \in \left[0,4\right]$, fails to reconstruct the eigenvalues as it takes contributions from a few number of multipoles and is the most affected by the mask.\\
Need-PCA properly reconstructs the cosmological signal, as can be observed from Fig.~\ref{fig:gnomview_cleanedHI_SKA_AA4_PCA_need} for a 45.8 deg $\times$45.8 deg patch of the sky close to the Galactic plane. The left panel is the input \HI~map plus noise, the center panel is the cleaned signal and noise and the right panel shows the residuals. In this region of the sky, the standard deviation of the residuals is 0.034 mK.
To properly address the performance of the cleaning methods at every scale, we calculate the angular power spectrum $\ClPCA$.
Fig.~\ref{fig:panel_summary_SKA_beam} shows the results of the analysis for all the methods.  \\
Overall, all the methods recover the \HI~signal within 10\% for $\ell \gtrsim 30$, averaging over the frequency channels. We stress that the Need- and the standard PCA, as well as the Need- and the standard GMCA\footnote{As demonstrated in \cite{Carucci-2020}, a larger number of frequency channels improves the GMCA reconstruction of the signal. We checked that this is valid also with the SKAO AA4 configuration implemented in this work by increasing the frequency channels (i.e. increasing the redshift range).}, which do not rely on some prior information about the target signal, obtain the same results as GNILC, which instead need a prior in \HI~as input. In fact, while PCA and GMCA are \textit{blind} cleaning methods, GNILC is called \textit{semi-blind}. Another difference among these methods is that GNILC computes the number of sources to remove minimizing the AIC at each needlet scale and location of the sky. On the contrary, in PCA and GMCA methods one needs to set $N_{\rm fg}$ a priori. However, GNILC tends to remove order $\sim10$ foreground components at the largest scales, i.e. small \textit{j}, while it estimates $N_{\rm fg}=3-4$ at smaller scales.\\
At the larger scales (i.e. $\ell \lesssim 30$), GNILC seems to underestimates the $\ClRes$ with respect to the other methods. This is due to an higher foreground contamination that affects the results of PCA and GMCA and their needlet implementations, leading to an apparent power gain at those scales. In fact, GNILC achieves lower foreground residuals than the other methods, as it dynamically adapts the number of foreground components $N_{\rm fg}$ depending on scale and sky region, whereas the other methods rely on a fixed $N_{\rm fg}$. This is shown in Fig.~\ref{fig:fg_leak_mean_ch_GNILC_PCA_GMCA}, where we plot the foreground leakage (see Eq.~\eqref{eq:HI_fg_leak}) for each cleaning method.\footnote{This result was also pointed out by \cite{Olivari-2016}. However, we find different results from \cite{Olivari-2016} on the comparison between PCA and GNILC. They obtain that GNILC recover the $\ClRes$ better than PCA for $\ell \gtrsim 30$, setting a telescope beam with $\theta_{\rm FMWH}=$40 arcmin and a Galactic mask with f$_{\rm sky}\sim$40 \%. Here we obtain the same results from PCA and GNILC at those scales.}
%%%%%%%%%%%%%%%%%%%%%%%%%%%%%%%%%%%%%%%%%%%%%%%%%%%%%%%%%%%%%%%%%%%%%%
\subsection{Adding polarization leakage}
\label{subsec:results_pol_leakage}
\begin{figure}[ht]
\centering
\setlength{\tabcolsep}{0.01pt}
\begin{tabular}{cc}
    \includegraphics[width=0.49\linewidth]{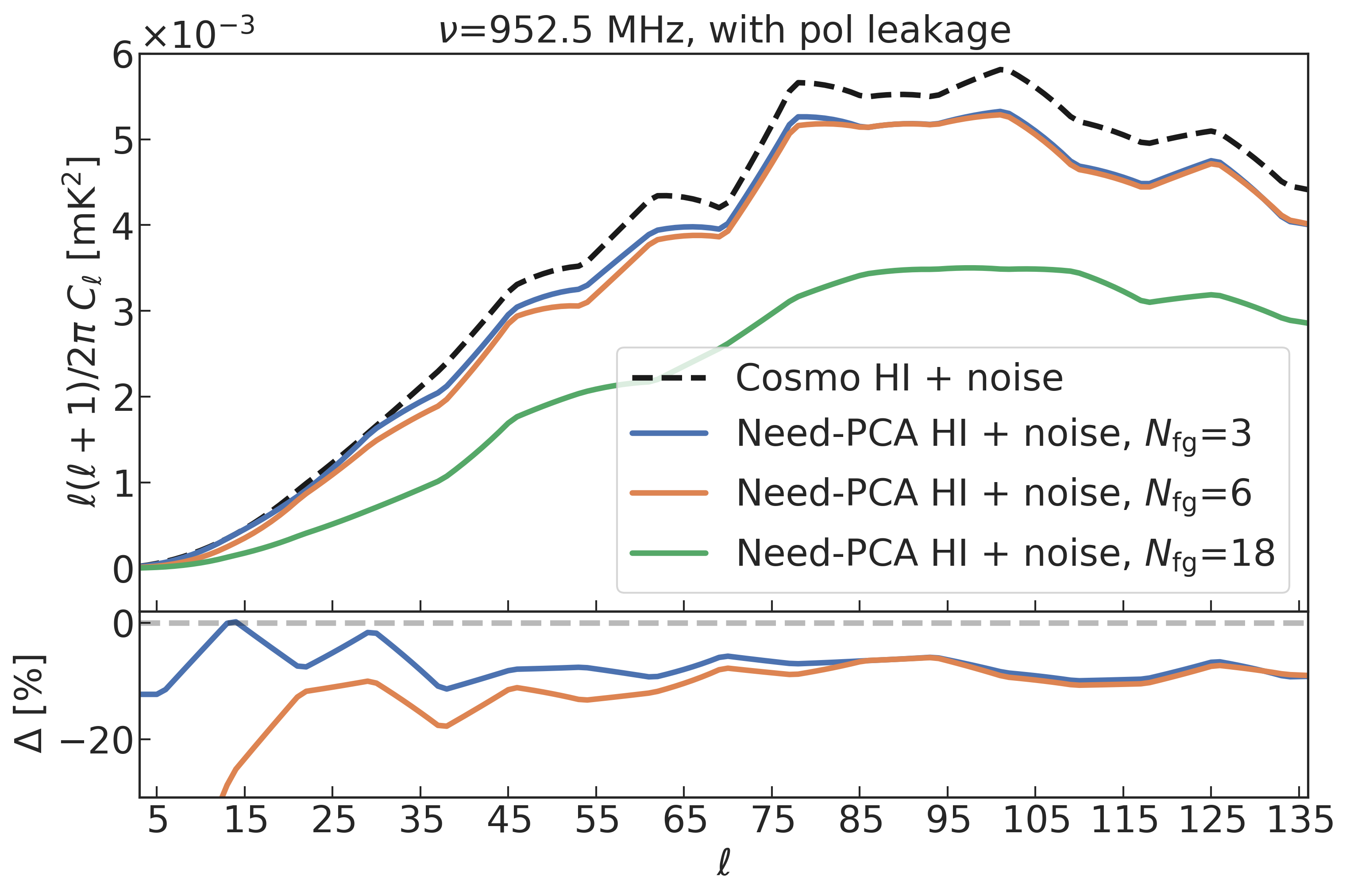}
    \includegraphics[width=0.49\linewidth]{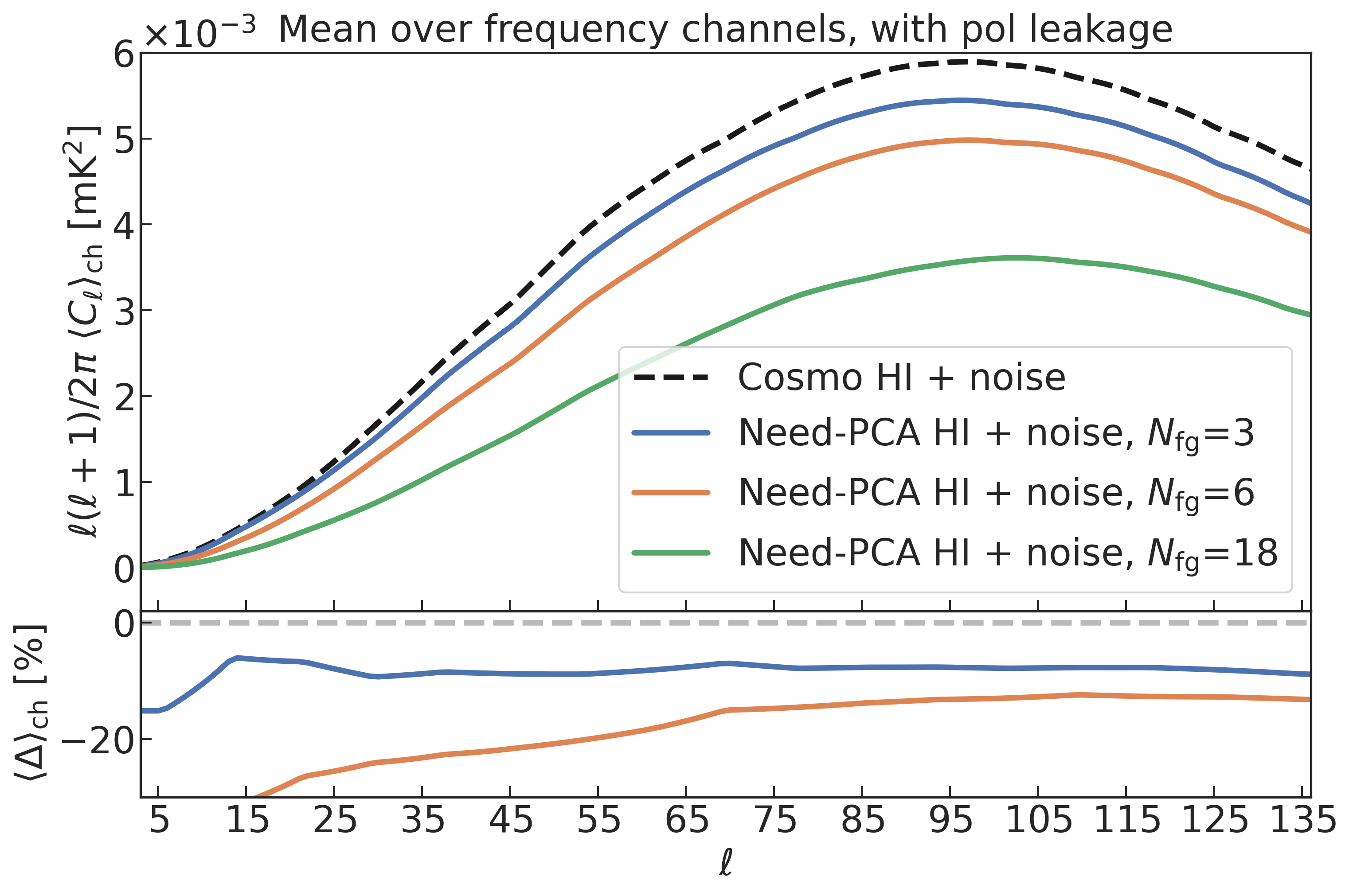}
\end{tabular}
    \caption{Angular power spectrum of the \HI~signal recovered by Need-PCA when adding the polarization leakage and for different values of $N_{\rm fg}$; the dotted black line is the input power spectrum, the blue line is the recovered one. The telescope beam is Gaussian and varies across the frequency channels.} Left: $\ClRes$ at frequency $\nu=$952.5 MHz, right: mean over frequency channels.
    \label{fig:cl_need_pca_sync_ff_ps_pol_Nfg_3_6_18}
\end{figure}

\begin{figure}[ht]
    \centering
    \includegraphics[width=0.7\linewidth]{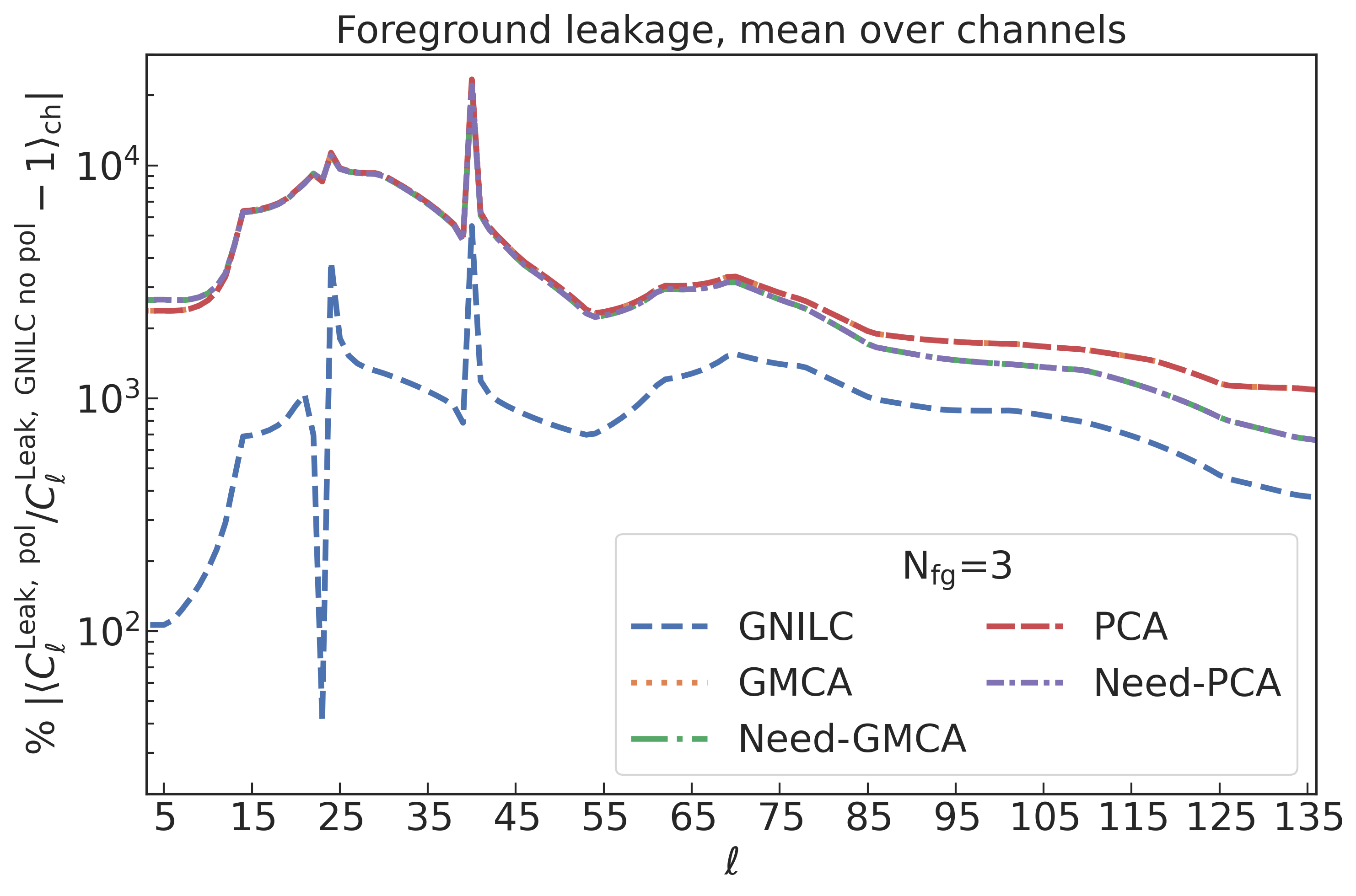}
    \caption{Percent difference (in absolute value) of the $C_{\ell}$ of the foreground leakage recovered by GNILC (blue line), GMCA (orange line), Need-GMCA (green line), PCA (red line) and Need-PCA (purple line) including the polarization leakage, with respect to the $C_{\ell}$ of the foreground leakage recovered by GNILC (blue line) without the polarization leakage. The telescope beam is Gaussian and varies across the frequency channels.} Overall, the foreground leakage is higher when the polarization leakage is included for all the methods. However, GNILC is less effected by residuals then the other cleaning methods. PCA and GMCA, both standard- and Need-, shows comparable results.
    \label{fig:abs_diff_fg_leak_pol_no_pol}
\end{figure}

As discussed in previous works \cite{Carucci-2020,Cunnington-2021}, the polarization leakage adds complexity to the data and thus, overall, the cleaning methods struggle to remove it. Thus, in general, a larger $N_{\rm fg}$ is required to clean the data.
To set $N_{\rm fg}$, we run Need-PCA with $N_{\rm fg}=$3, 6, 18 and the results are shown in Fig.~\ref{fig:cl_need_pca_sync_ff_ps_pol_Nfg_3_6_18}.
In contrast with what is found in literature, i.e. \cite{Carucci-2020}, we obtain that the optimal value does not change, i.e. we set $N_{\rm fg}=3$. This is because the most of the polarization leakage is in the Galactic plane, which is masked in our case. \\
We obtain that, in the presence of the polarization leakage, all the methods recover the $\ClRes$ with the same precision, within the $10\%$ for $\ell \gtrsim$ 30. Moreover, we find that all the methods are not affected by the presence of such systematic, as they recover the signal as in the case without the polarization leakage. In fact, the polarization leakage is mainly localized in the Galactic plane and the mask we are considering in our analysis helps the cleaning methods. To check this, we run the standard PCA with full-sky maps and obtain that we need to set N$_{\rm fg} \sim 18$ to remove the most of the foreground from the data, although losing some of the \HI~in the cleaning process.  
We check also that the apparent gain of precision at large scales when the polarization leakage is added on the data is due to an higher level of foreground residuals into the \HI~signal. Fig.~\ref{fig:abs_diff_fg_leak_pol_no_pol} shows the percent differences (in absolute value) between the power spectrum of the foreground leakage for different methods, when polarization leakage is included, and that recovered by GNILC without the polarization leakage. We find that for all the methods the foreground residuals are higher when the polarization leakage is included; GNILC is less affected by them while the results for PCA and GMCA, both standard- and Need-, are similar.
%%%%%%%%%%%%%%%%%%%%%%%%%%%%%%%%%%%%%%%%%%%%%%%%%%%%%%%%%%%%%%%%%%%%%%
\subsection{Constant Gaussian beam}
\label{subsec:constant_beam_chann}
\begin{figure}[ht]
\centering
\setlength{\tabcolsep}{0.01pt}
\begin{tabular}{cc}
    \includegraphics[width=0.49\linewidth]{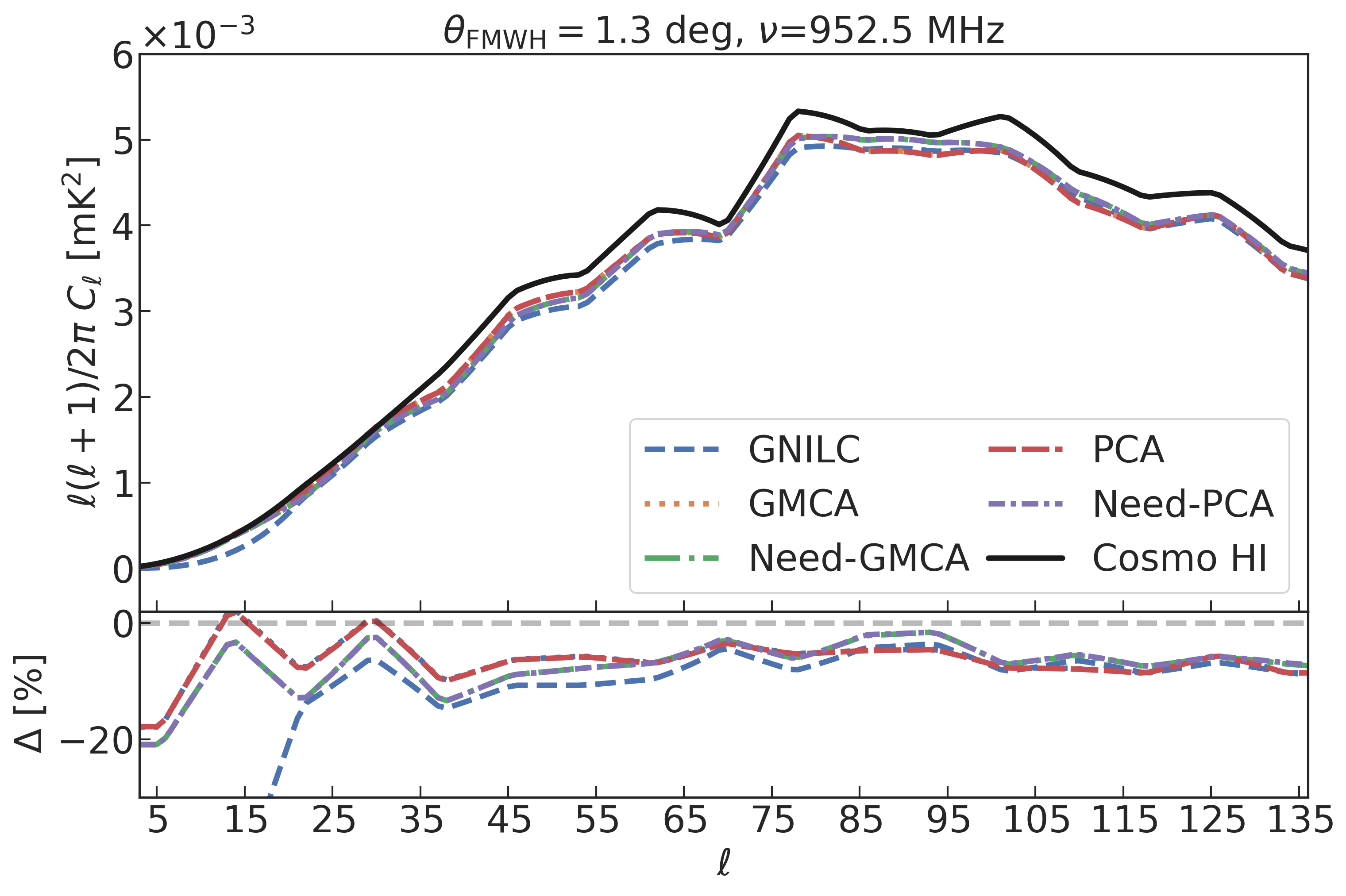}
    \includegraphics[width=0.49\linewidth]{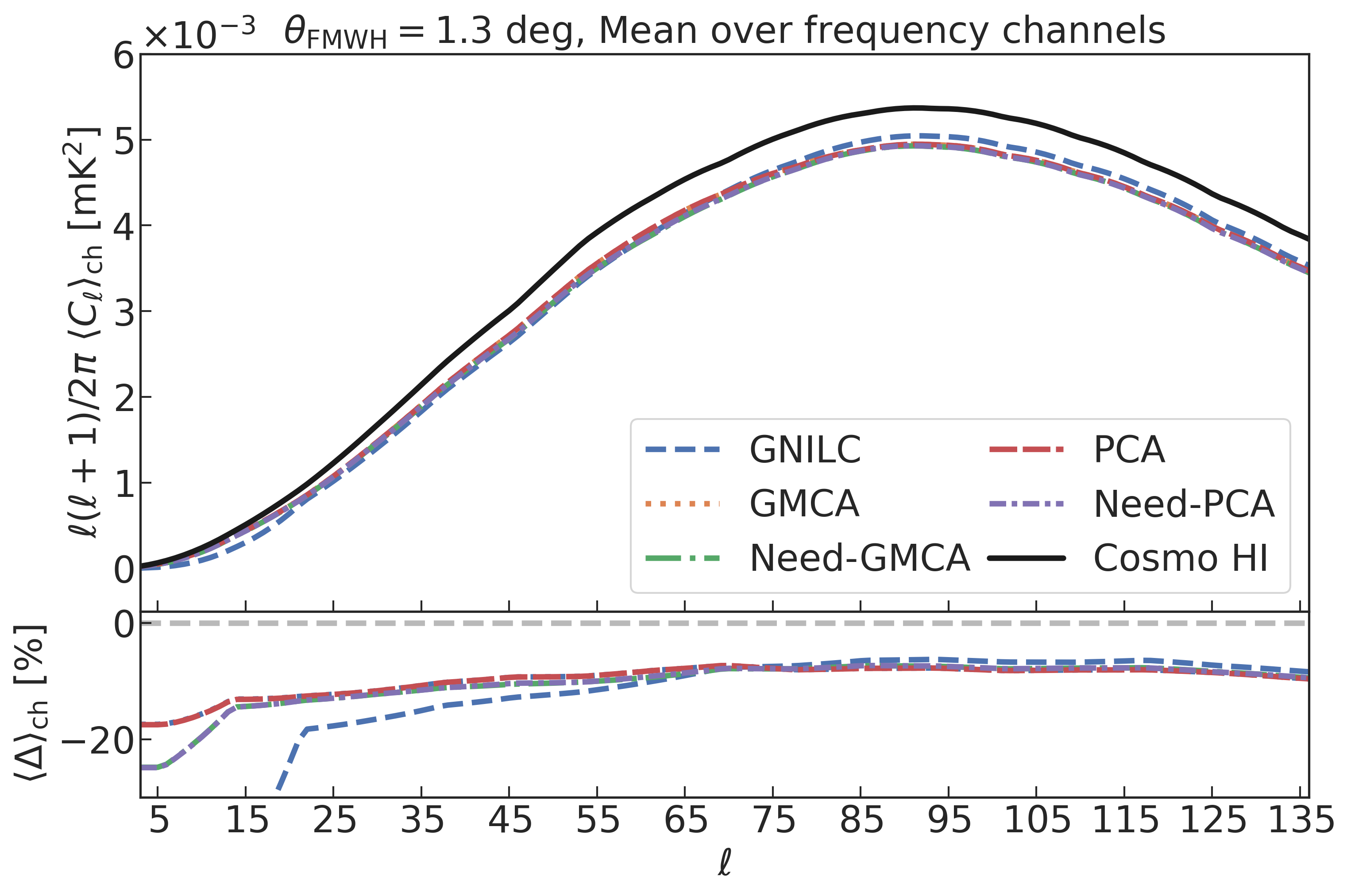}\\
    \includegraphics[width=0.49\linewidth]{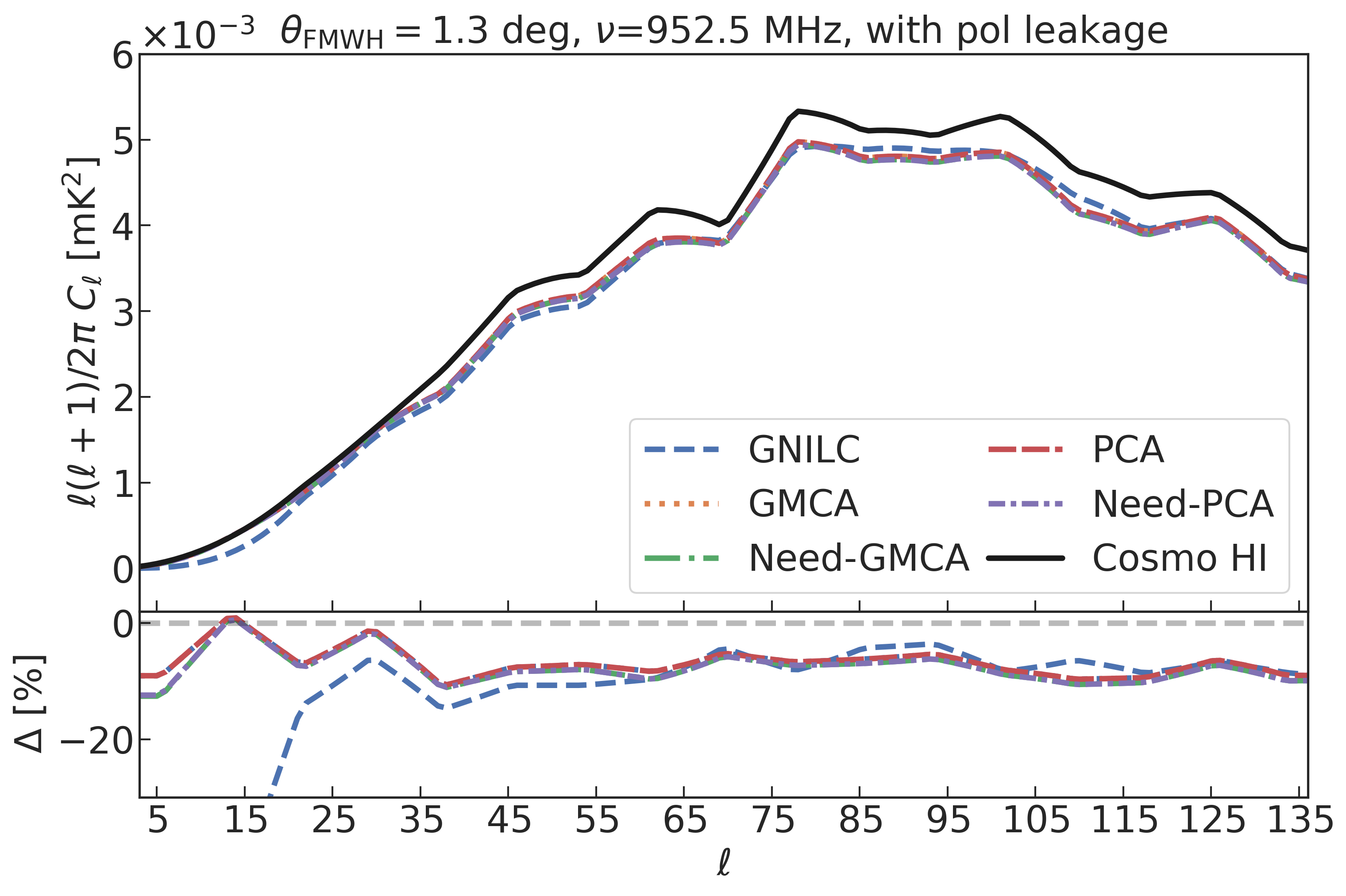}
    \includegraphics[width=0.49\linewidth]{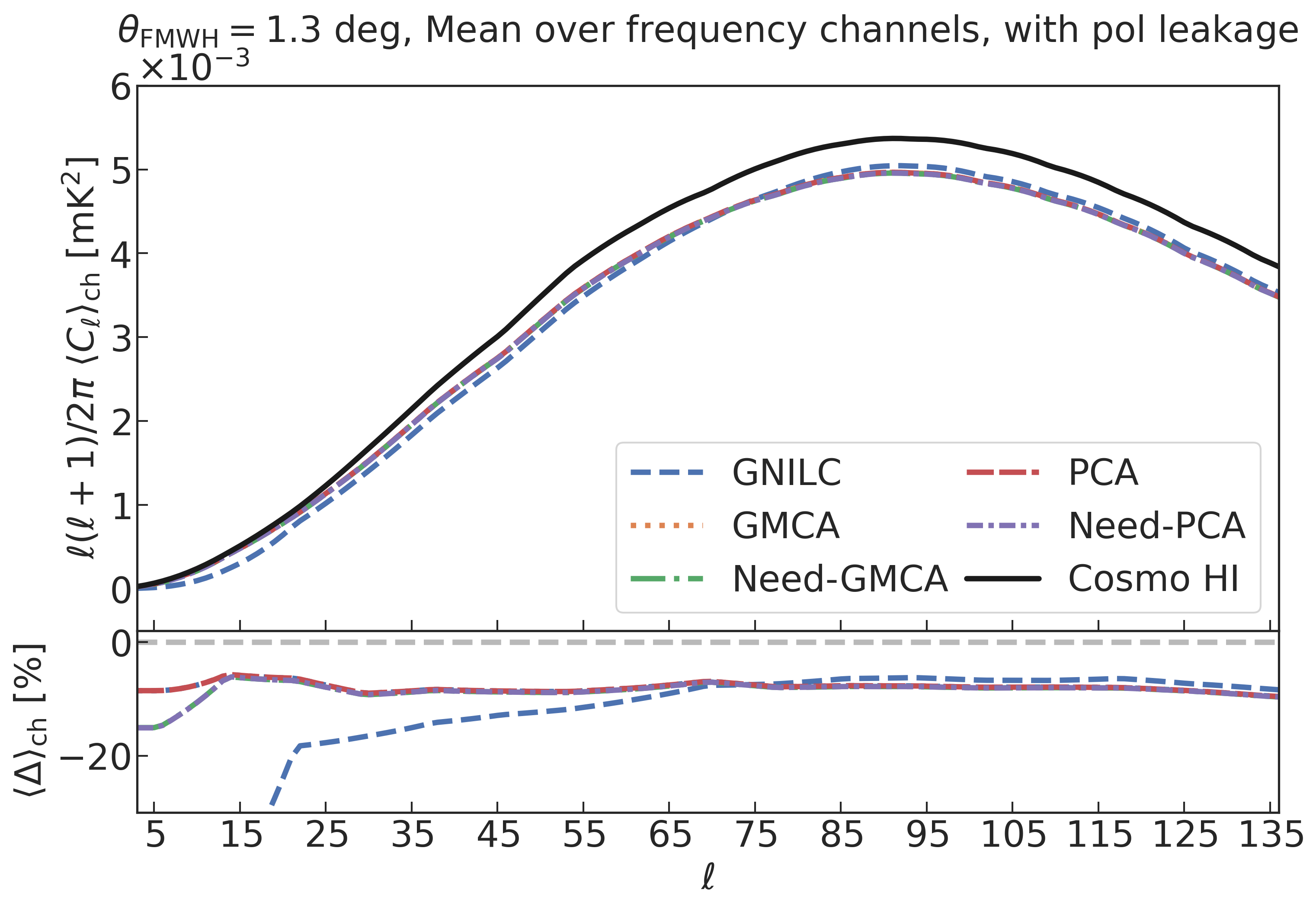}
\end{tabular}
    \caption{Comparison between the $\ClRes$ recovered with GNILC (blue line), GMCA (orange line), Needlets-GMCA (green line), PCA (red line) and Needlets-PCA (purple line). The input $\ClCosmo$ is plotted in black. The bottom panel shows the percent difference $\Delta$. The telescope beam is Gaussian and constant across the frequency channels with $\theta_{\rm FWHM}$= 1.3 deg. We set N$_{\rm fg}=$3 for PCA, Need-PCA, GMCA and Need-GMCA; GNILC computes N$_{\rm fg}$ for each scales and region of pixel according to the AIC. Left upper panel: $\ClRes$ at $\nu$=952.5 MHz; right upper panel: $\ClRes$ averaged over the frequency channels. In the bottom panel are shown the same quantities adding the polarization leakage to the analysis.}
        \label{fig:panel_summary_costant_beam}
\end{figure}

We set $\theta_{\rm FWHM}=$1.3 deg, corresponding to Eq.~\eqref{eq:theta_fwhm} computed at the lowest frequency channel, i.e., the largest beam size in the band. In contrast with the case of a frequency-varying Gaussian beam, the maps share the same resolution. We set also for this analysis N$_{\rm fg}$=3. 
The results in $\ClRes$ are shown in Fig.~\ref{fig:panel_summary_costant_beam}. We obtain $\Delta$ within 10\%. When adding the polarization leakage, we obtain the same precision in the recovery of the angular power spectrum.
%%% panel
We recover the $\ClPCA$ with the same precision of in case of frequency-dependent telescope beam, removing the same number of sources. In contrast of what is written in \cite{Carucci-2020}, we do not need to set an higher number of sources to remove to reach the same level of cleaning. We find that it is not necessary to have the maps at the same resolution, as it would cause a loss of the smallest angular information available in the observed maps \cite{Carucci-2020}.

%%%%%%%%%%%%%%%%%%%%%%%%%%%%%%%%%%%%%%%%%%%%%%%%
%%%%%%%%%%%%%%%%%%%%%%%%%%%%%%%%%%%%%%%%%%%%%%
\subsection{\textit{Cosine} beam model}
\label{subsec:cosine_beam_chann}
\begin{figure}[ht]
    \centering
    \includegraphics[width=0.7\linewidth]{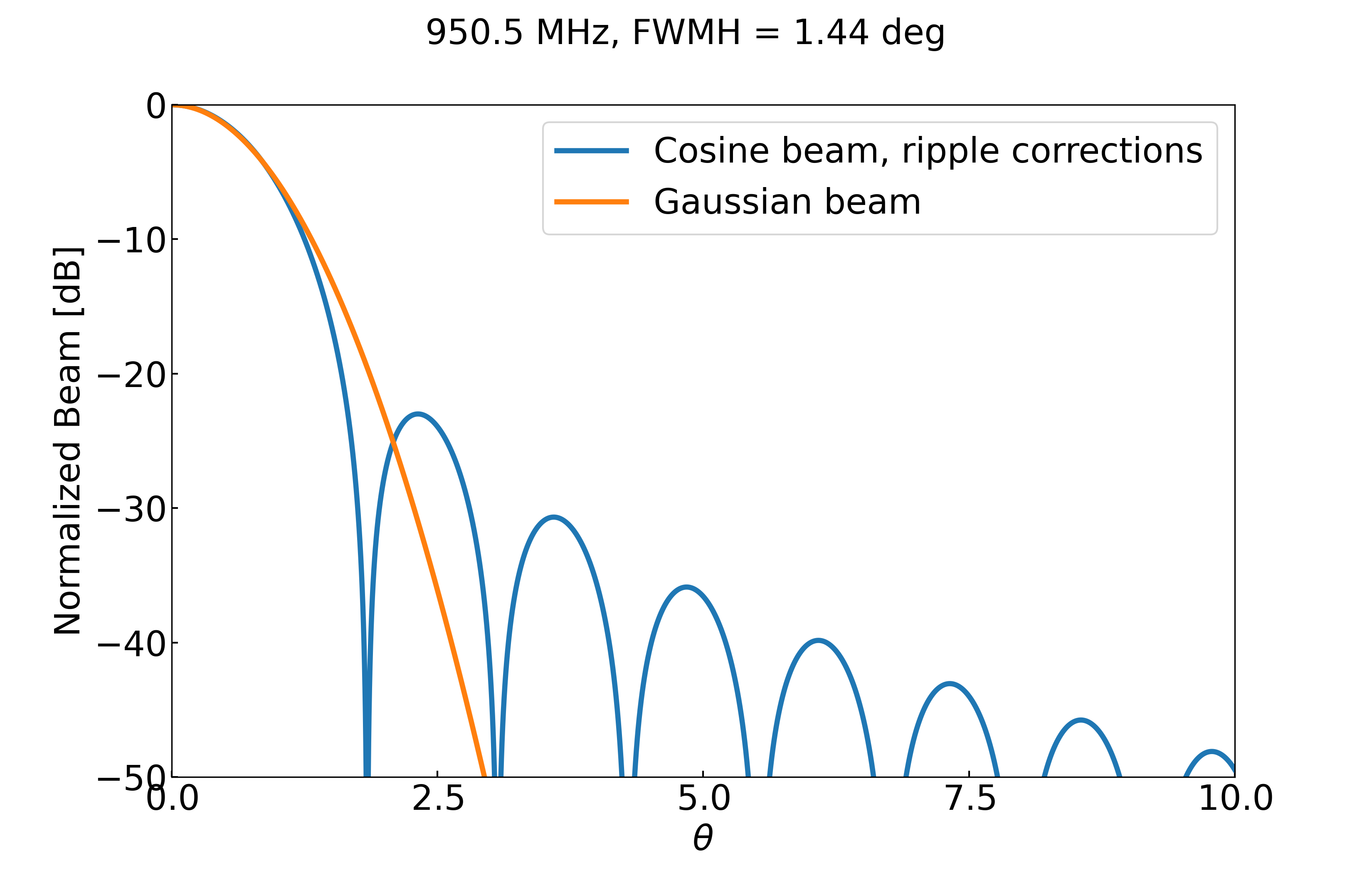}
    \caption{A comparison at 950 MHz of the beam models: the \textit{cosine} model \cite{Matshawule-2021} (orange line), with the \textit{ripple} corrections, and the Gaussian model (blue line).}
    \label{fig:beams_cosine_gaussian}
\end{figure}

\begin{figure}[ht]
\centering
\setlength{\tabcolsep}{0.01pt}
\begin{tabular}{cc}
    \includegraphics[width=0.49\linewidth]{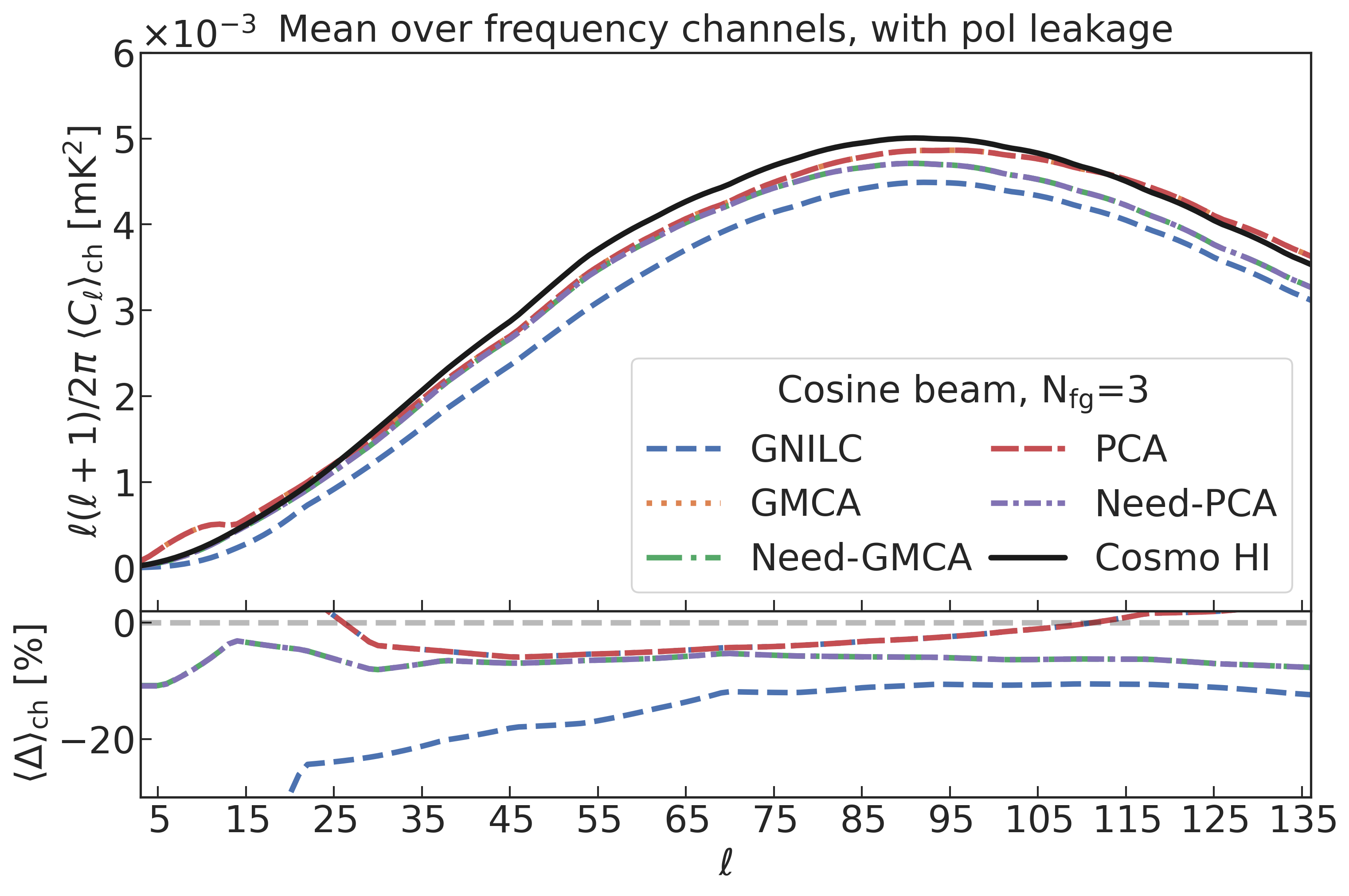}
    \includegraphics[width=0.49\linewidth]{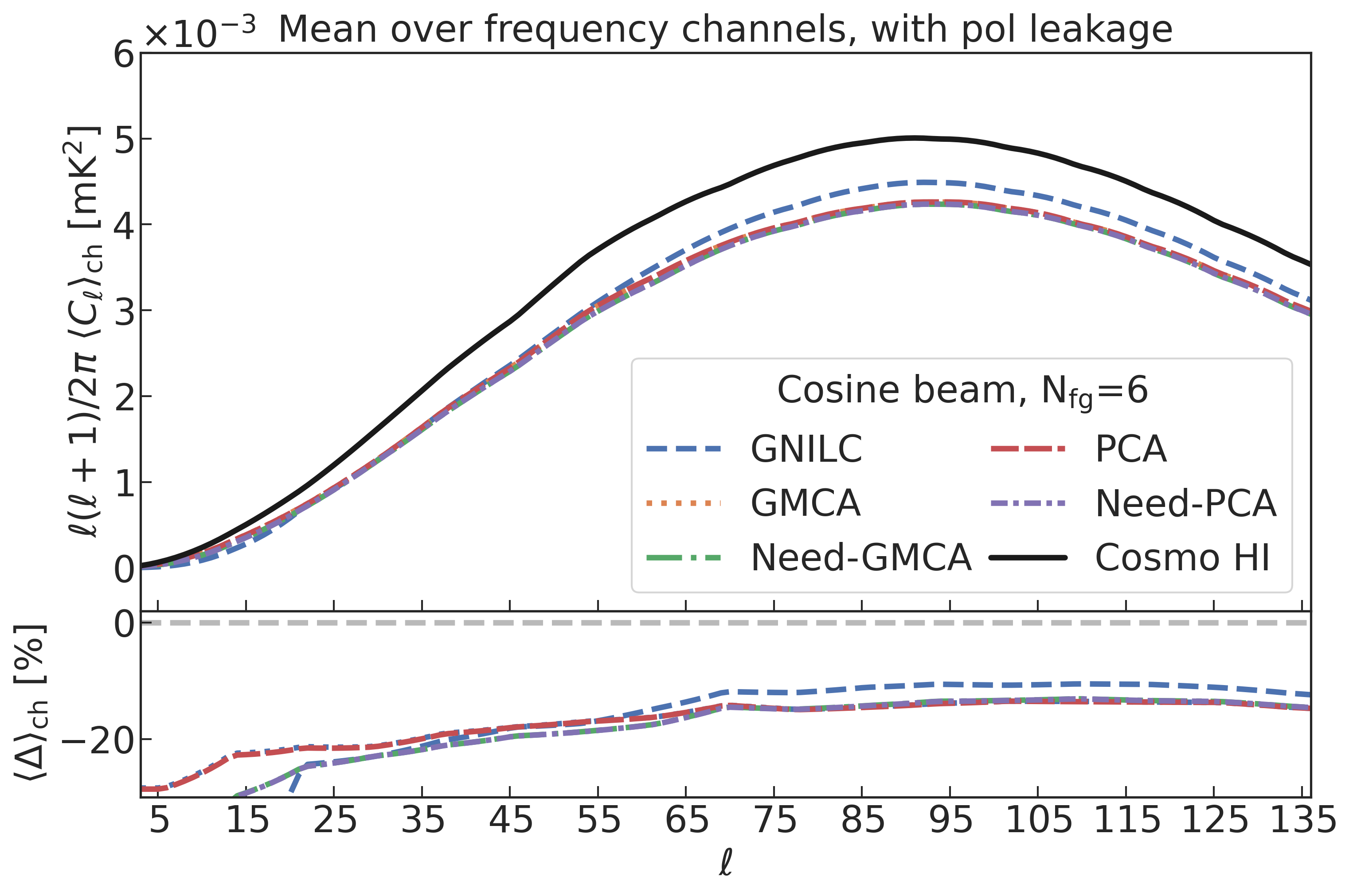}
\end{tabular}
    \caption{Comparison between the $\ClRes$ recovered with GNILC (blue line), GMCA (orange line), Needlets-GMCA (green line), PCA (red line) and Needlets-PCA (purple line). The input $\ClCosmo$ is plotted in black. The bottom panel shows the percent difference $\Delta$. The telescope beam is modelled as a cosine-tapered field illumination function, following Eq.~\eqref{eq:cosine_beam}. The resolution is frequency dependent, calculated as Eq.~\eqref{eq:theta_beam_ripple} and setting the values of the parameters listed in Table~\ref{tab:cosine_beam_ripple}. The left panel shows the results for PCA and GMCA with N$_{\rm fg}$=3 and it has to be compared with the right bottom panel of Fig. \ref{fig:panel_summary_SKA_beam}. The right panel shows the results with N$_{\rm fg}$=6.} 
    \label{fig:cl_cosine_beam_Nfg3_Nfg6}
\end{figure}

\begin{figure}[ht]
    \centering
    \includegraphics[width=0.7\linewidth]{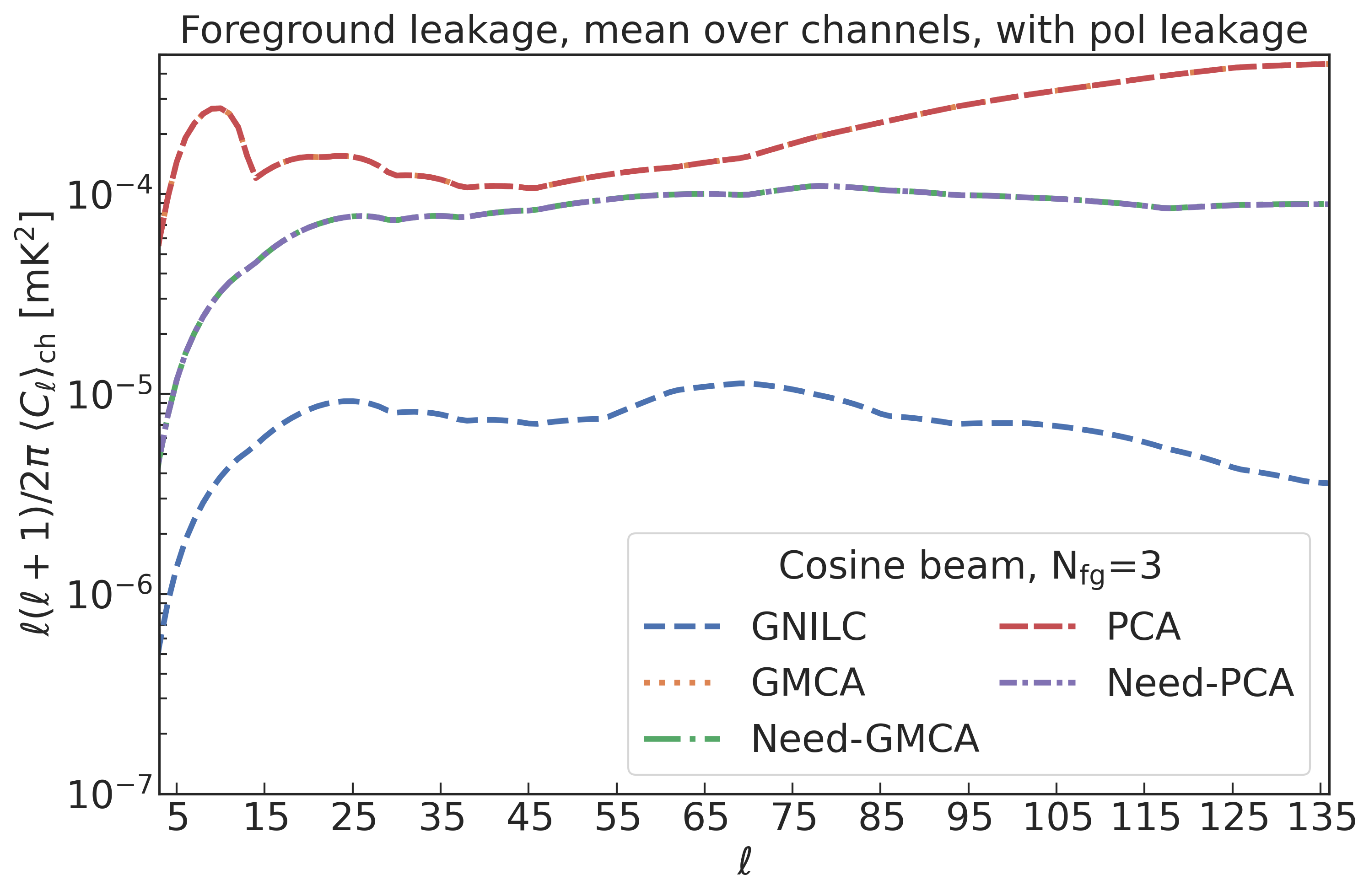}
    \caption{Comparison between the $C_{\ell}$ of the foreground leakage recovered by GNILC (blue line), GMCA (orange line), Need-GMCA (green line), PCA (red line) and Need-PCA (purple line), respectively, obtained by averaging over the frequency channels. We set N$_{\rm fg}$=3 for the standard PCA, standard GMCA and Need-PCA and Need-GMCA. The telescope beam is modelled as a cosine-tapered field illumination function, following Eq.~\eqref{eq:cosine_beam}. The resolution is frequency dependent, calculated as Eq.~\eqref{eq:theta_beam_ripple} and setting the values of the parameters listed in Table~\ref{tab:cosine_beam_ripple}. For low $\ell$, GNILC shows lower foreground leakage than the other methods, since it changes dynamically the number $N_{\rm fg}$ depending on scales and regions. The results of PCA and GMCA, in the standard and needlet versions, almost perfectly overlap.}
    \label{fig:fg_leak_mean_ch_GNILC_PCA_GMCA_cosine}
\end{figure}

We perform the analysis by implementing the telescope beam modelled as a cosine-tapered field (or cosine-squared power) illumination function \cite{Matshawule-2021}:
\begin{equation}
\label{eq:cosine_beam}
    B_C(\nu, \theta) =  \left[ \frac{\cos(1.189 \theta \pi /\theta_{\rm FWHM})}{1-4(1.189\theta/\theta_{\rm FWHM})} \right]\,.
\end{equation}
Here $\theta$ is the azimuthal angle, $\nu$ is the frequency of observation and $\theta_{\rm FWHM}$ is the angular resolution, which is approximately given by Eq.~\eqref{eq:theta_fwhm}. Following the measurements of the MeerKAT beam  as shown in \cite{Matshawule-2021} , we correct $\theta_{\rm FWHM}$ with low-level frequency-dependent \textit{ripple}. The ripple is caused by the interaction of the primary and secondary reflector of MeerKAT and it can be relevant in the foreground cleaning. They fit the smooth $\theta_{\rm FWMH}$ with a 8-th degree order polynomial and superimpose a sinusoidal oscillation with period \textit{T} and amplitude \textit{A} arc-minutes,
\begin{equation}
\label{eq:theta_beam_ripple}
    \theta_{\rm FWHM,\,r} = \frac{c}{\nu D}\big( \sum^{8}_{d=0}a_{d}\nu^{d}+A\sin\big(\frac{2\pi\nu}{T}\big)\big)\,.
\end{equation}
The values of the parameters are summarized in Table \ref{tab:cosine_beam_ripple}.
\begin{table}[ht]
\centering
  \caption{Numerical values of the coefficients of Eq.~\eqref{eq:theta_beam_ripple}, following \cite{Matshawule-2021}.}
  \begin{tabular}{c c c }
  \hline
 A [arc-min] & T [MHz]  & $a_n$ $\{n=0,...,8\}$ \\
  \hline
 & & $\{6.7\mathrm{e}{3}, -50.3, 0.16,$\\ 
 0.1  & 20 & $-3.0\mathrm{e}{-4},3.5\mathrm{e}{-7}, -2.6\mathrm{e}{-10},$\\
 & & $1.2\mathrm{e}{-13}, -3.0\mathrm{e}{-17}, 3.4\mathrm{e}{-21} \}$ \\
  \hline
  \label{tab:cosine_beam_ripple}
\end{tabular}
\end{table}
Fig.~\ref{fig:beams_cosine_gaussian} shows the sidelobes of the \textit{cosine} beam in comparison with the gaussian beam at frequency 950 MHz.\\
The results of the analysis with the \textit{cosine} beam and the \textit{ripple} corrections are shown in Fig.~\ref{fig:cl_cosine_beam_Nfg3_Nfg6}. We include the contribution of the polarization leakage. We find that standard PCA and standard GMCA do not properly reconstruct the \HI~signal at the larger scales with N$_{\rm fg}=3$, as shown in the left panel. Increasing the value of N$_{\rm fg}$ to 6 reduces the foreground leakage in the $\ClRes$, but removes part of the signal along with the foreground. On the other hands, Need-PCA and Need-GMCA are not affected by the different beam model. Indeed, the needlet decomposition helps the component separation and the cleaning algorithm by exploiting the 2D information of the maps. The GNILC method tends to remove more components than in the analysis of the Gaussian beam, yielding to a loss of the \HI~signal especially at the larger scales. Fig.~\ref{fig:fg_leak_mean_ch_GNILC_PCA_GMCA_cosine} shows the angular power spectrum of the foreground leakage, as in Fig.~\ref{fig:fg_leak_mean_ch_GNILC_PCA_GMCA}, when the \textit{cosine} beam is implemented. Overall, the foreground leakage is higher than in the Gaussian beam case, as the sidelobes of the \textit{cosine} beam add complexity to the frequency structure of the data, as explained also in \cite{Matshawule-2021}, making difficult the cleaning process. Again, GNILC is less affected by the foreground leakage than the other methods, but it tends to remove more \HI~signal along with the foreground.\\
To conclude, we have tested that the deconvolution of the maps to a common resolution should increase the performances of the cleaning methods. However, at the moment the beam of radio telescope is not well known and an accurate deconvolution is challenging. For this reason, we decide to follow a more "realistic" approach, leaving the maps as they are. Discussions about the resmoothing of the maps are found in \citep{Spinelli-2021, Matshawule-2021,Cunnington-2023,Carucci-2025}.
%%%%%%%%%%%%%%%%%%%%%%%%%%%%%%%%%%%%%%%%%%
\section{Conclusions}
\label{sec:conclusions}
The aim of this work is to assess the performance of different foreground removal methods in recovering the \HI~signal from simulated data. We use a set full-sky sky maps in the frequency range $\nu \in \left[ 900.5 - 1004.5 \right]$ MHz, corresponding to the redshift range $z \in \left[ 0.41 - 0.58 \right]$. The simulations include the cosmological signal, galactic and extragalactic foregrounds, and a polarization leakage component. To mimic observations with the SKAO-MID telescope in the AA4 configuration, each map is smoothed with a Gaussian beam, affected by thermal noise and masked over 50\% of the Galactic plane. The data-cube consists of 105 \textsc{HEALPix} maps with a frequency resolution of 1 MHz. \\
We implement a novel PCA algorithm based on spherical wavelets known as needlets. Unlike spherical harmonics, needlets are double localized in both harmonic and pixel space, making them well-suited for spherical data analysis. We apply PCA with needlets (Need-PCA), GMCA, GMCA with needlets (Need-GMCA), and GNILC to our dataset. Our main findings are as follows:
\begin{itemize}
    \item When polarization leakage is excluded, all methods recover the angular power spectrum with approximately 10\% precision for $\ell \gtrsim 30$, averaged over the frequency channels, as shown in the upper panels of Fig.~\ref{fig:panel_summary_SKA_beam}. For Need-PCA, PCA, Need-GMCA, and GMCA, we fix the number of foreground components to N$_{\rm fg}$=3.  In contrast, GNILC estimates N$_{\rm fg}$ locally at each needlet scale and spatial region using a statistical criterion. GNILC, as a \textit{semi-blind} method, requires \textit{prior} information on the \HI~power spectrum, while the others are \textit{blind} and make no assumptions about the cosmological signal. We find same results for \textit{blind} methods without any assumptions on the cosmological signal.
    
    \item With respect to the other methods considered in this work, GNILC is affected by lower foreground residuals at large angular scales, i.e. $\ell \lesssim 30$, since it varies dynamically the number of sources to remove, $N_{\rm fg}$, for different scales and regions of the sky.
    
    \item All methods are insensitive to polarization leakage, yielding the same results as in the leakage-free case, and fixing the same N$_{\rm fg}=$3, as shown in the bottom panels of Fig.~\ref{fig:panel_summary_SKA_beam}. This is because the leakage primarily affects the Galactic plane, which is masked in our analysis. Therefore, we conclude that polarization leakage–––modelled as in \cite{Alonso-2014}---has a negligible impact on \HI~recovery in SKAO-MID AA4-like surveys.
    
    \item We repeat the analysis using a constant Gaussian beam across all frequency channels, set to $\theta_{\rm FMWH}=$1.3 deg (the largest beam size in the band). The results are consistent with those obtained using a frequency-dependent beam, indicating that resolution differences across frequencies do not significantly affect the cleaning process.

    \item We smooth the simulated maps with a telescope beam modelled following Eq.~\eqref{eq:cosine_beam}. The beam resolution varies along the frequency channels and includes the \textit{ripple} corrections. This model is a realistic description of the MeerKAT telescope beam, a more exhaustive discussion is found in \cite{Matshawule-2021}. Since beam model adds complexity to the frequency structure of the data, standard PCA and GMCA do not properly remove the foreground from the data when N$_{\rm fg}$ is set to 3. The foreground leakage affects the $\ClRes$ at the larger scales and the resulting spectrum shows higher power at those scales. On the contrary, Need-PCA and Need-GMCA show the same results as in the analysis with the Gaussian beam. In this case, the needlets helps the component separation and the cleaning algorithm by exploiting the 2D information of the maps. Finally, GNILC underestimates the $\ClRes$, because it removes more components from the data, losing fraction of the \HI~signal.
\end{itemize}
In conclusion, out findings provide strong support for the application of these techniques for the analysis of SKAO-MID IM future data.
%%%%%%%%%%%%%%%%%%%%%%%%%%%%%%%%%%%%%%%%%%%%%%%%%%%%%%%%%%%%%%%%%%%%%%

\acknowledgments
We warmly thank Domenico Marinucci for precious discussions and insightful suggestions. BDC and CC thank Matteo Calabrese for very useful hints. BDC is supported by the Italian Ministry of University and Research (MUR) PRIN 2022 ``EXSKALIBUR – Euclid-Cross-SKA: Likelihood Inference Building for Universe's Research'', Grant No. 20222BBYB9, CUP C53D2300083 0006, from the European Union -- Next Generation EU, and partially supported by the INAF grant for fundamental research 2023 (No. 1.05.23.05.18) ``GEMS: Galaxy survey Exploration via Modelling of multi-scale Structures''. IPC is supported by the European Union within the Next Generation EU programme [PNRR-4-2-1.2 project No. SOE\textunderscore0000136, RadioGaGa]. MR acknowledges support by the Spanish Ministry of Science and Innovation (MCIN) and the Agencia Estatal de Investigación (AEI) through the project grants PID2022-139223OB-C21 and PID2022-140670NA-I00.

\bibliography{biblio}
\bibliographystyle{JHEP}

\appendix

\section{Need-PCA with different \boldmath{$j_{\rm max}$}}
\label{appendix:std_need_jmax12}
\begin{figure}[ht]
    \centering
    \includegraphics[width=0.6\linewidth]{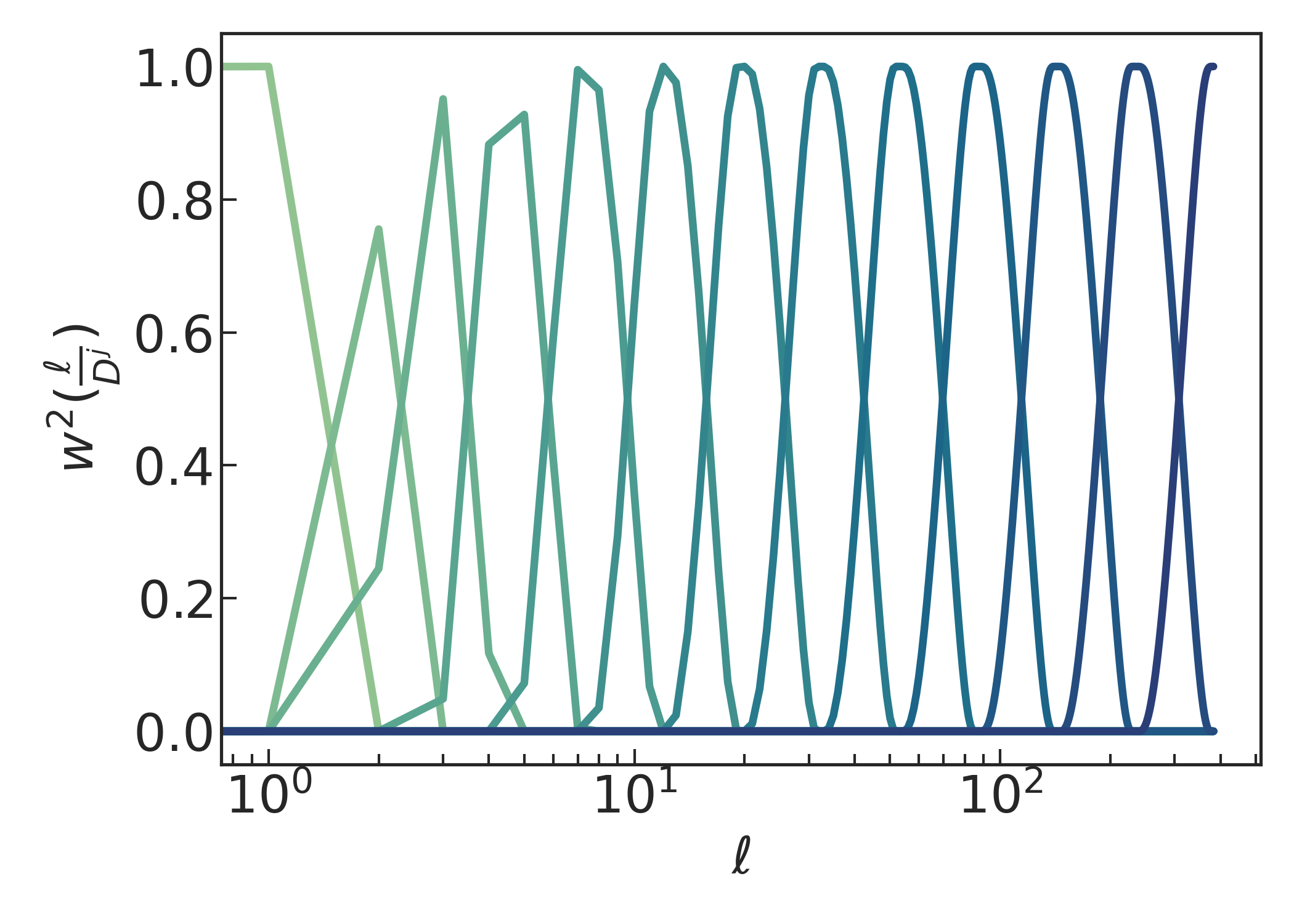}
    \caption{weight functions for the standard needlets with $j_{\rm max}=12$, $\ell_{\rm max}=383$.}
    \label{fig:window_standard_need_jmax12_lmax767}
\end{figure}

\begin{figure}[ht]
\centering
\setlength{\tabcolsep}{0.01pt}
\begin{tabular}{cc}
    \includegraphics[width=0.49\linewidth]{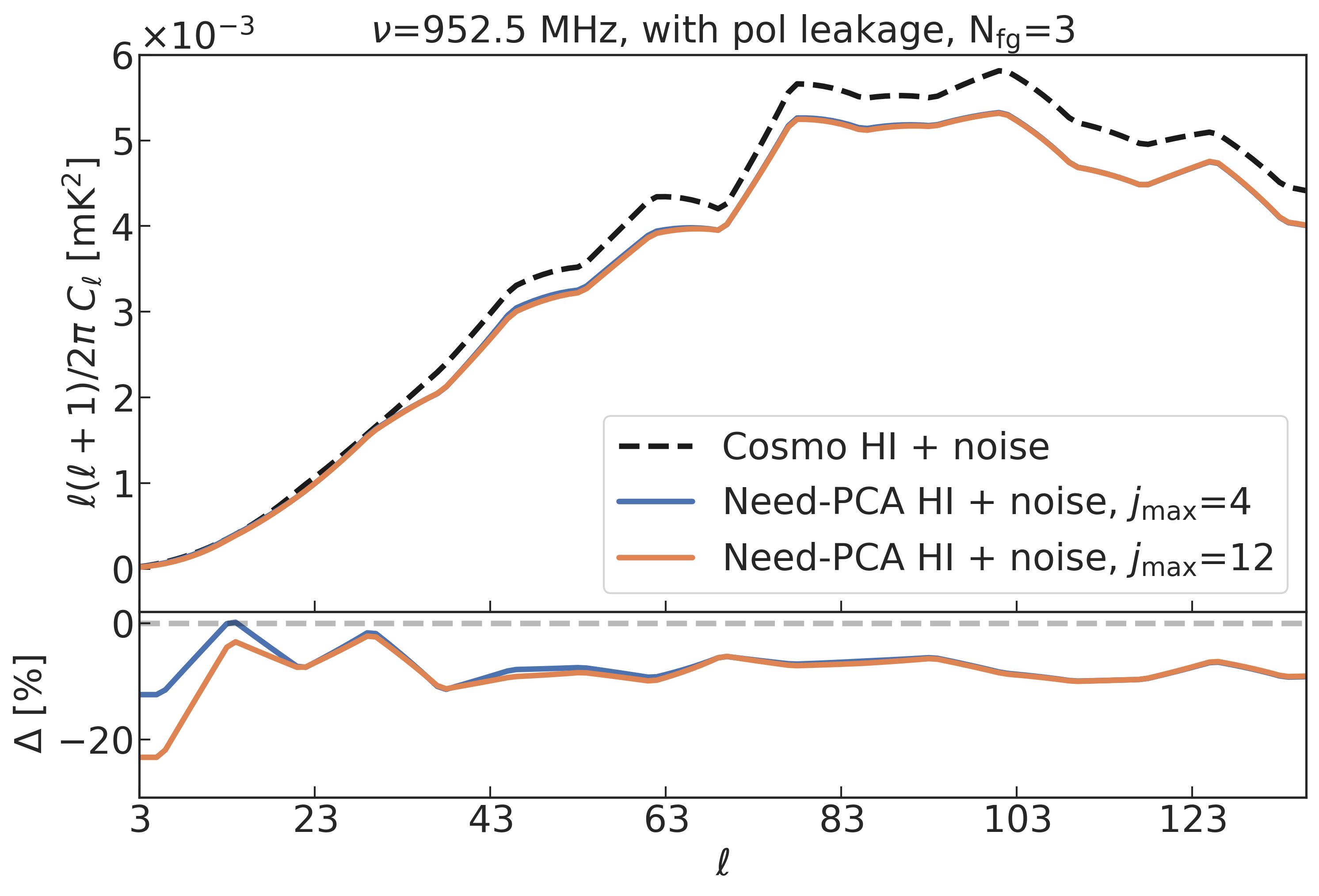}
    \includegraphics[width=0.49\linewidth]{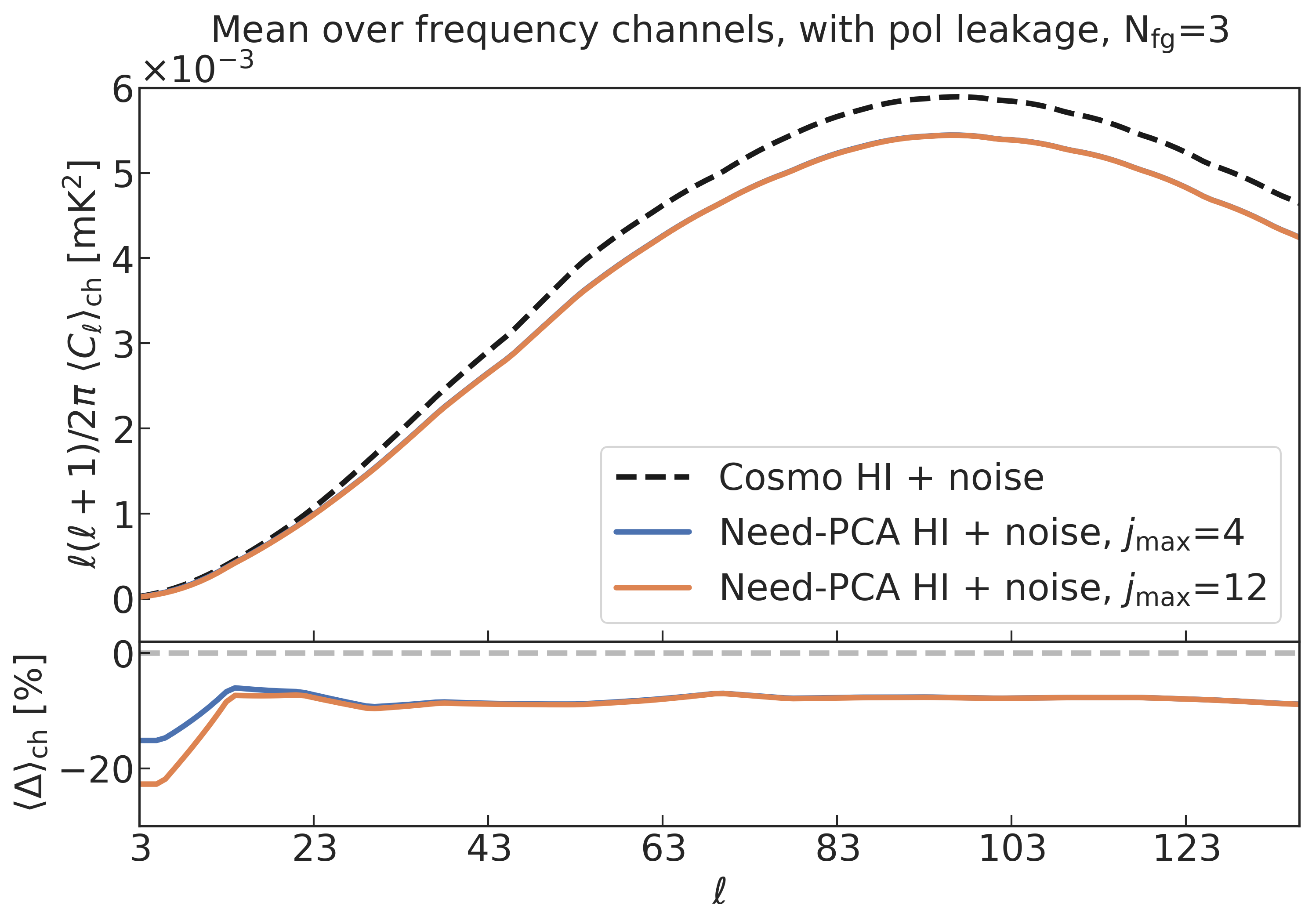}
\end{tabular}
    \caption{Angular power spectrum of the \HI~signal recovered by Need-PCA with different $j_{\rm max}$; the dotted black line is the input power spectrum, the blue line is the recovered one with the $j_{\rm}$=4 and the orange line is the $\ClPCA$ recovered with $j_{\rm}$=12. The telescope beam is Gaussian and varies across the frequency channels.}Left: $C_{\ell}^{\rm \HI}$ at frequency $\nu=952.5$ MHz, right panel: mean over the frequency channels. The bottom panel show the percent difference.
    \label{fig:cl_std_jmax_need_pca_sync_ff_ps}
\end{figure}

We test Need-PCA methods setting $j_{\rm max}$=12 and N$_{\rm fg}$=3. The shape of the weight functions are shown in Fig.~\ref{fig:window_standard_need_jmax12_lmax767}.
Higher values of $j_{\rm max}$ correspond to lower values of the parameter \textit{D}; this corresponds to a tighter localization in the harmonic space but a broader localization in the pixel space. The results of the recovered angular power spectrum with different $j_{\rm max}$ are in Fig.~\ref{fig:cl_std_jmax_need_pca_sync_ff_ps}, showing that overall the $\ClPCA$ is recovered with the same precision by the two analysis, except for the larger scales that are better recovered by the case with $j_{\rm max}$=4.

%%%%%%%%%%%%%%%%%%%%%%%%%%%%%%%%%%%%%%%%%%%%%%%%%%%%%%%%%

\section{Mexican Needlets} 
\label{appendix:mex_need}
\begin{figure}[ht]
    \centering
    \includegraphics[width=0.6\linewidth]{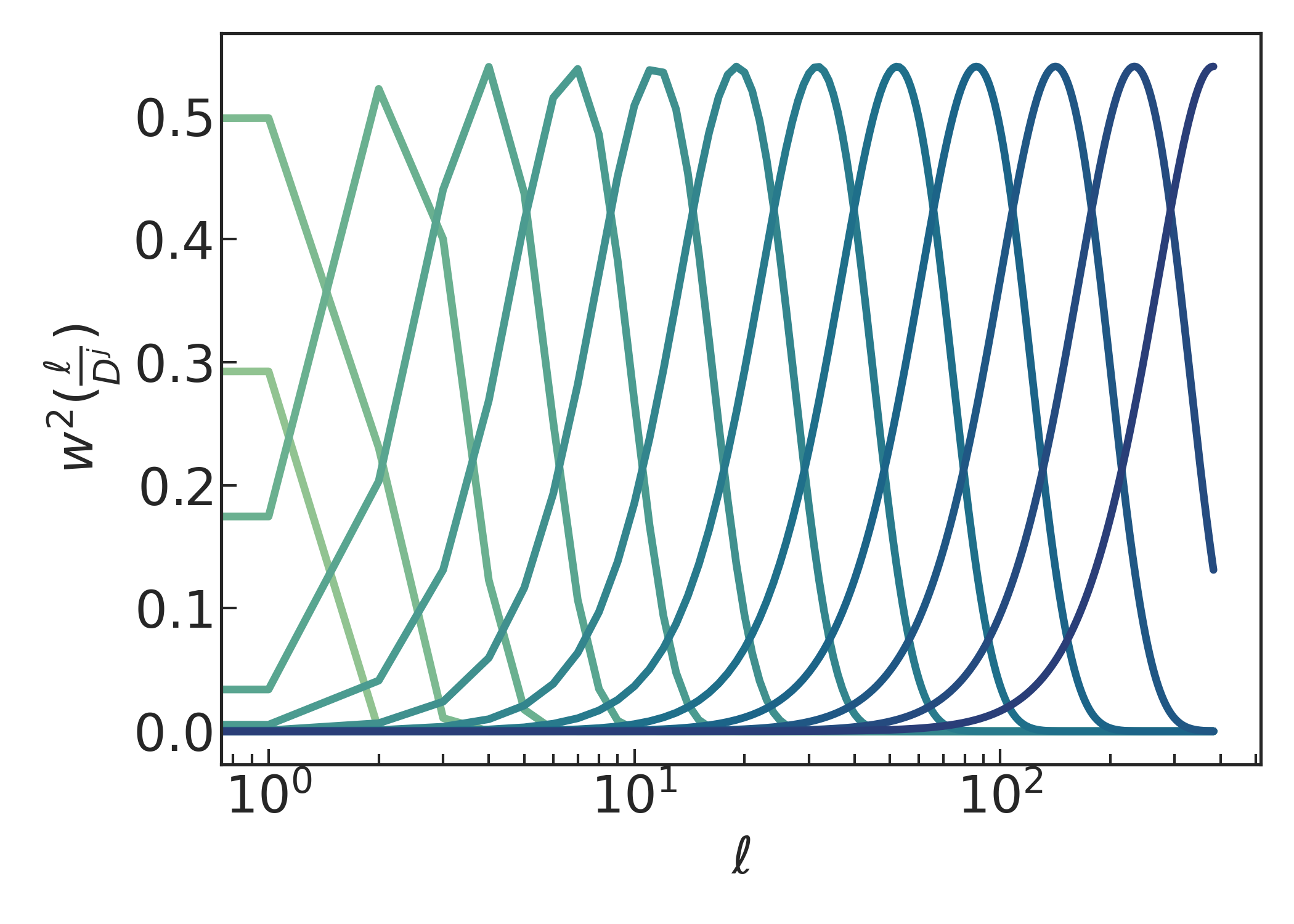}
    \caption{Weight functions for the Mexican needlets with $j_{\rm max}=$12, $\ell_{max}=767$.}
    \label{fig:window_mex_jmax12_lmax383}
\end{figure}

\begin{figure}[ht]
\centering
\setlength{\tabcolsep}{0.01pt}
\begin{tabular}{cc}
    \includegraphics[width=0.49\linewidth]{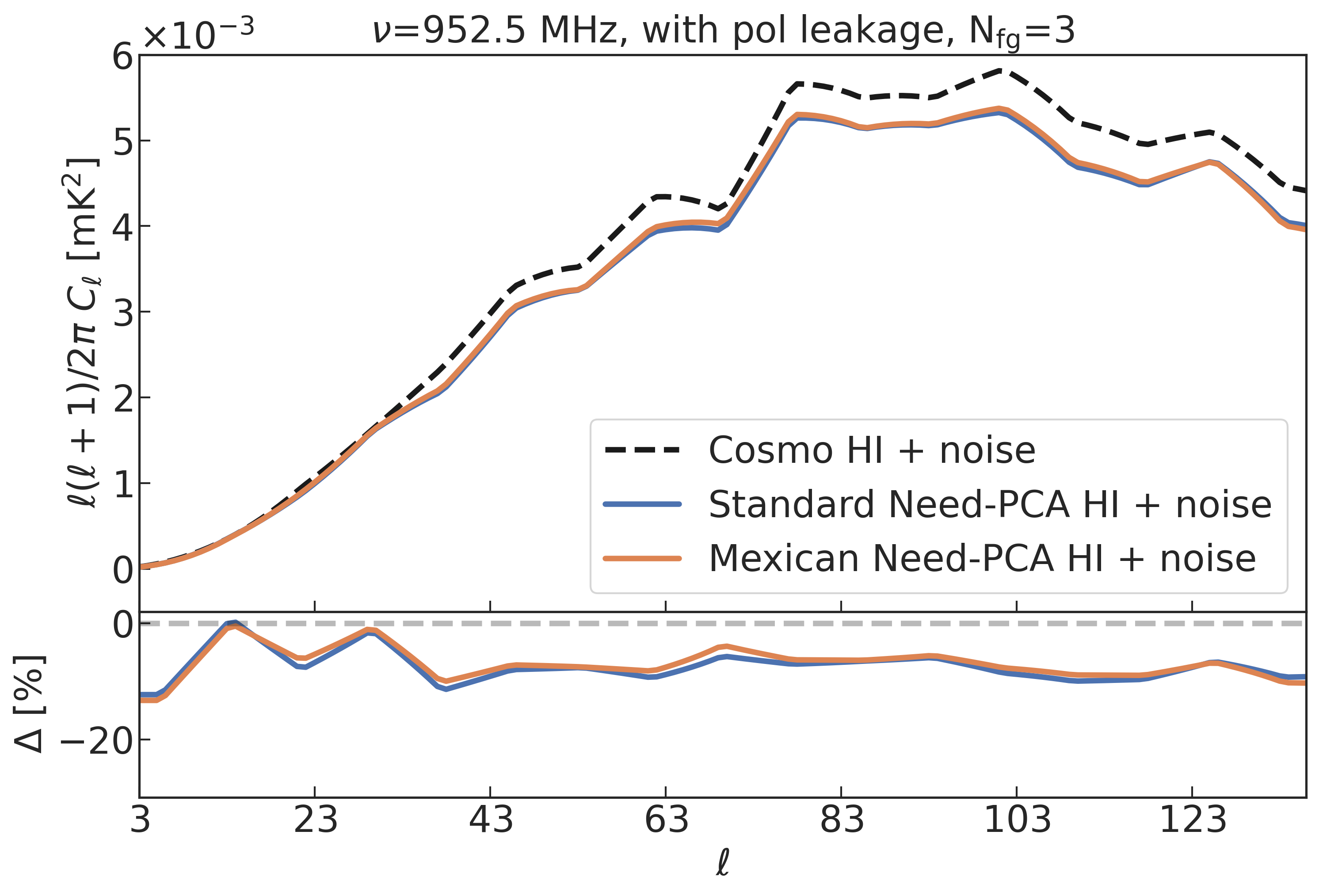}
    \includegraphics[width=0.49\linewidth]{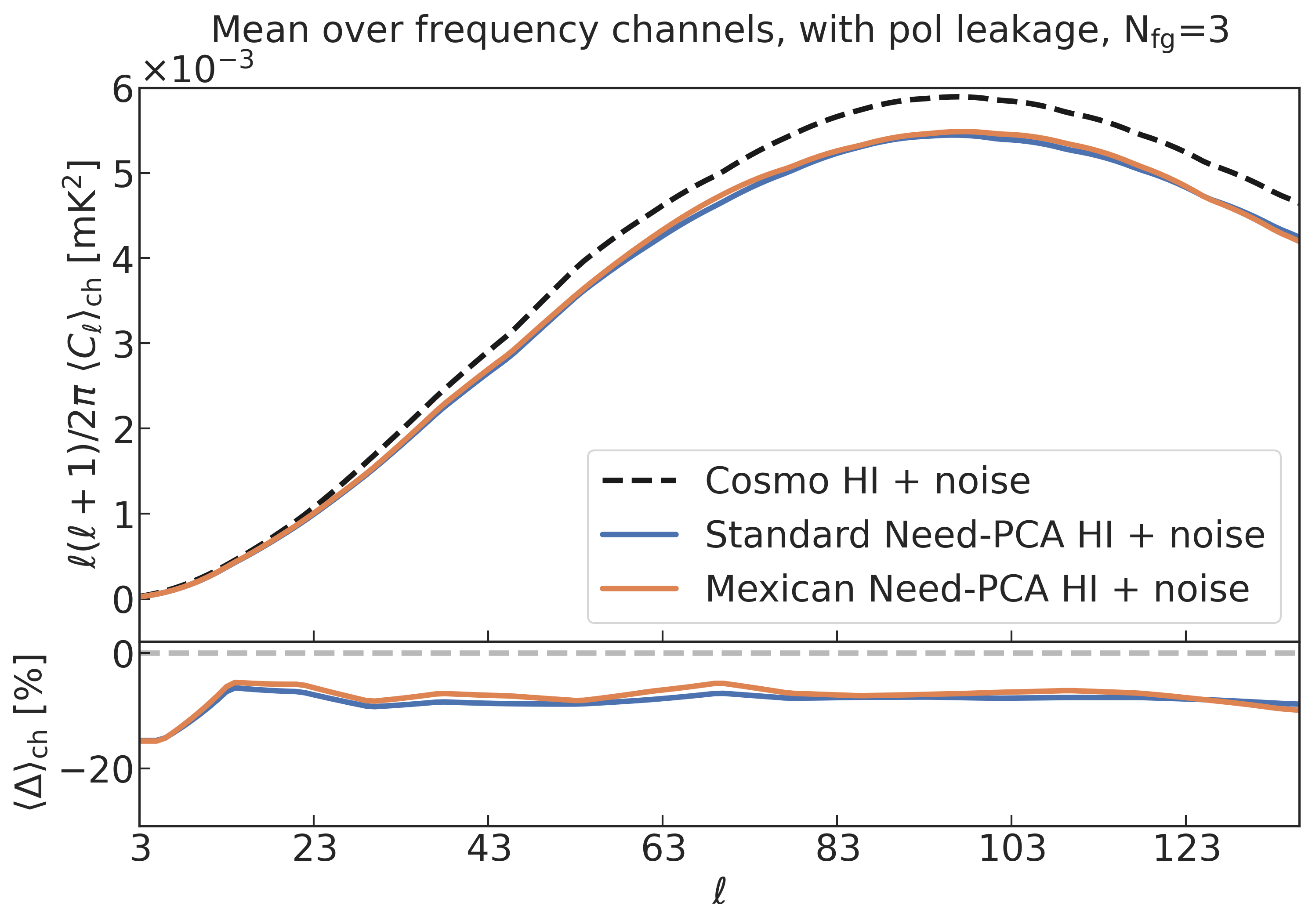}
\end{tabular}
    \caption{Angular power spectrum of the \HI~signal recovered by Need-PCA using the standard and the Mexican needlets; the dotted black line is the input power spectrum, the blue line is the recovered one with the standard needlets and the orange line is the $\ClPCA$ recovered using the Mexican needlets. The telescope beam is Gaussian and varies across the frequency channels.} Left: $C_{\ell}^{\rm \HI}$ at frequency $\nu=952.5$ MHz, right panel: mean over the frequency channels. The bottom panel show the percent difference.
    \label{fig:cl_mex_std_need_pca_sync_ff_ps}
\end{figure}

We test Need-PCA with Mexican needlets which have the weight functions shown in Fig.~\ref{fig:window_mex_jmax12_lmax383}. The main difference between standard and Mexican is that the weight function of the Mexican do not have a compact support in harmonic space and thus do not have an exact reconstruction formula.
 We compare the standard needlets with $j_{\rm max}=$4 and the Mexican needlets with $j_{\rm max}=12$. The recovered $\ClPCA$ for the two cases is shown in Fig.~\ref{fig:cl_mex_std_need_pca_sync_ff_ps}. The $\ClPCA$ is recovered with the same precision in both cases, with a difference of the order of the sub-percent.

%%%%%%%%%%%%%%%%%%%%%%%%%%%%%%%%%%%%%%%%%%%%%%%%%%%%%%%%%
\section{GNILC with different priors}
\label{app:gnilc_prior}
\begin{figure}[ht]
\centering
\setlength{\tabcolsep}{0.01pt}
\begin{tabular}{cc}
    \includegraphics[width=0.49\linewidth]{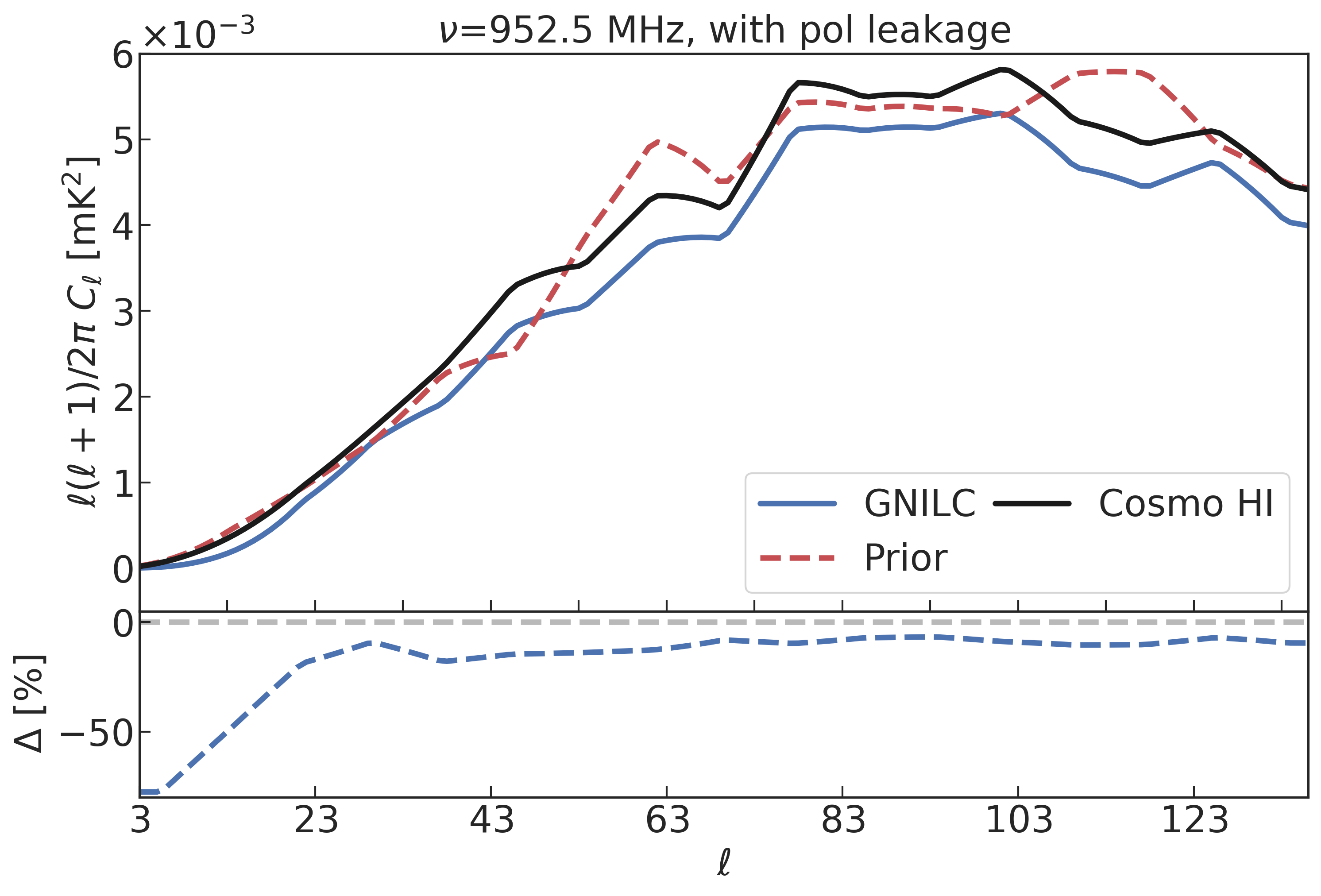}
    \includegraphics[width=0.49\linewidth]{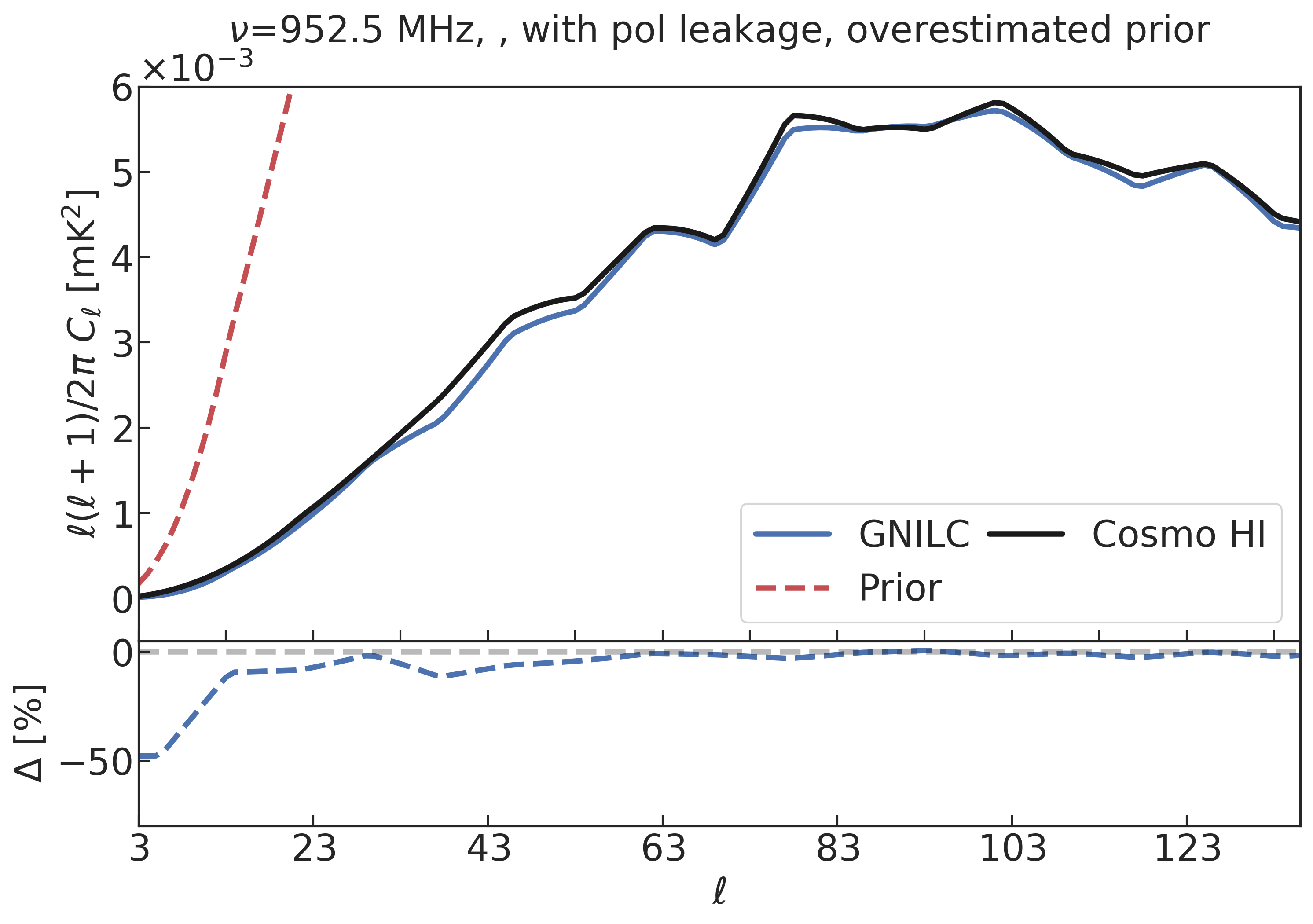}\\
    \includegraphics[width=0.49\linewidth]{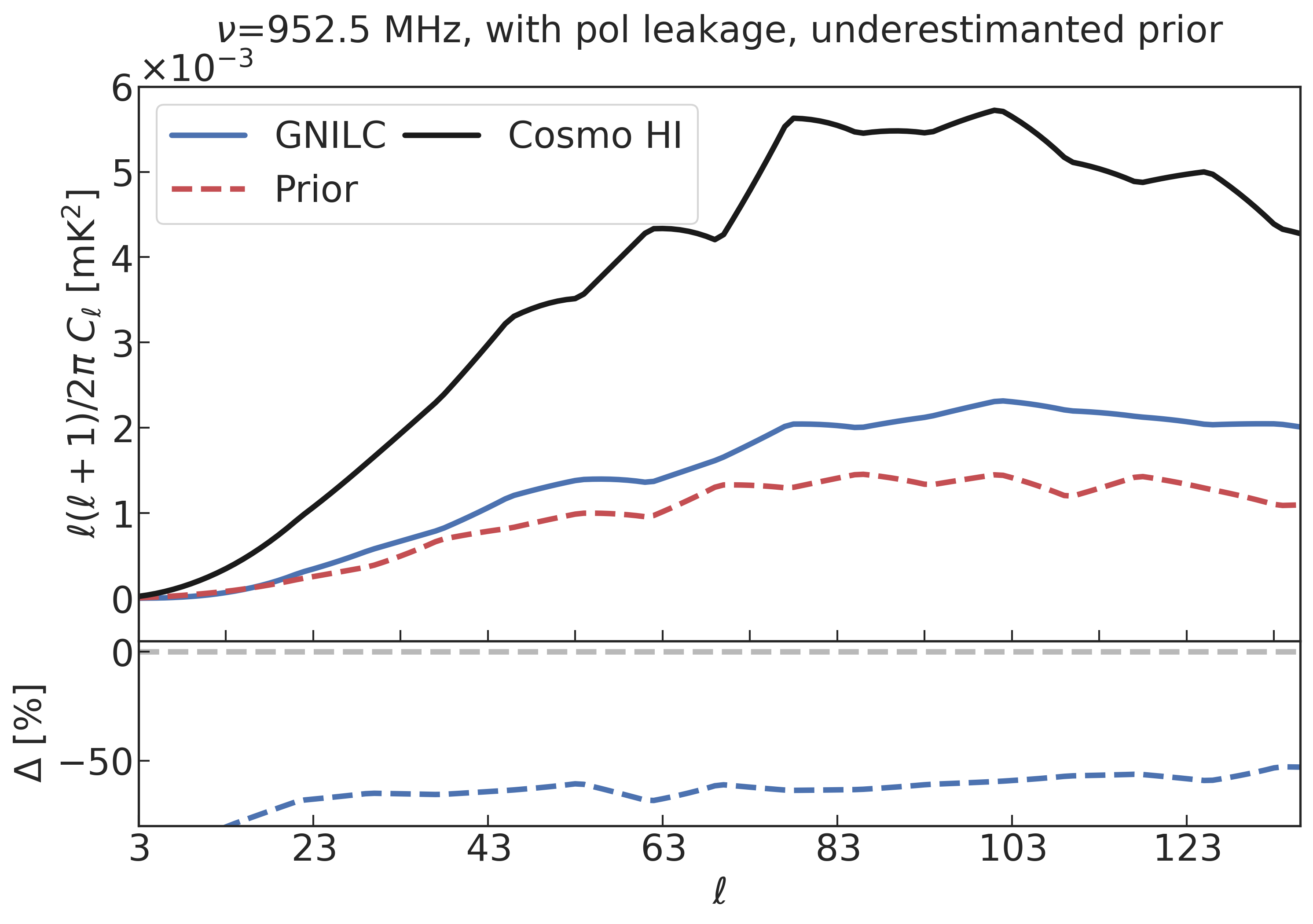}
\end{tabular}
    \caption{Angular power spectrum of the \HI~signal recovered by GNILC, $C_{\ell}^{\rm GNILC}$ with different priors, at the frequency channel $\nu=$952.5 MHz. The telescope beam is Gaussian and varies across the frequency channels. The upper left panel shows the results when implementing the correct prior, the upper right panel when the prior considered is overestimated by a factor 2.5, and the bottom panel when the prior is underestimated by a factor 2.}
    \label{fig:GNILC_prior}
\end{figure}

To run GNILC, we use a smooth prior extracted from the \HI~plus noise simulations for each frequency channels. Then, we synthetize the prior maps with \textsc{synfast} and smooth them with the proper beam width. We test the effects of having an overestimated and an underestimated smooth prior. In particular, we run the code with a smooth prior overestimated by a factor 2.5 and a smooth prior underestimated by a factor 2, following \cite{Olivari-2016}. The results in Fig.~\ref{fig:GNILC_prior} show that overestimating the prior does not affect the accuracy of recovering $C_{\ell}^{\HI}$. On the other hand, we are not able to properly recover the $C_{\ell}^{\HI}$ when the smooth prior is underestimated. A more detailed discussion is found in \cite{Olivari-2016}.
%%%%%%%%%%%%%%%%%%%%%%

\end{document}